\documentclass[acmlarge]{acmart}

\usepackage{booktabs} 
\usepackage{enumerate}
\usepackage{booktabs}
\usepackage{makecell}
\usepackage{multirow}
\usepackage{flushend}
\usepackage{latexsym}
\usepackage{multirow}
\usepackage{amsmath}
\usepackage{amsfonts}
\usepackage{subfigure}
\usepackage[noend]{algpseudocode}
\usepackage{graphicx}
\usepackage{color}
\usepackage{algpseudocode}  
\usepackage{amsmath}
\usepackage{rotating} 
\usepackage{bm}
\usepackage{lipsum}
\usepackage{verbatim}
\usepackage{adjustbox}
\usepackage[graphicx]{realboxes}
\usepackage{rotating}
\usepackage{array}
\usepackage[utf8]{inputenc}

\acmDOI{0000001.0000001}
    
\received[Received]{August 21, 2023}
\received[Revised]{February 24, 2024}
\received[Accepted]{April 15, 2024}
\newcommand{\revised}[1]{\textcolor{blue}{}\textcolor{black}{#1}}

\usepackage[linesnumbered,vlined,ruled,commentsnumbered]{algorithm2e}
\newlength\mylen

\makeatletter
\newcommand{\algorithmfootnote}[2][\footnotesize]{%
  \let\old@algocf@finish\@algocf@finish
  \def\@algocf@finish{\old@algocf@finish
    \leavevmode\rlap{\begin{minipage}{\linewidth}
    #1#2
    \end{minipage}}%
  }%
}

\theoremstyle{definition}

\newtheorem{defn}{Definition}[section]

\usepackage{xspace}

\newcommand{\framework}{UnDBot}

\usepackage{filecontents}

\begin{document}
\title{Unsupervised Social Bot Detection via Structural Information Theory}

\author{Hao Peng}
\authornote{This is the corresponding author.}
\affiliation{%
  \institution{Beihang University}
  \city{Beijing}
  \country{China}
  }
\email{penghao@buaa.edu.cn}
\author{Jingyun Zhang}
\affiliation{%
  \institution{Beihang University}
  \city{Beijing}
  \country{China}
  }
\email{zhangjingyun@buaa.edu.cn}
\author{Xiang Huang}
\affiliation{%
  \institution{Beihang University}
  \city{Beijing}
  \country{China}
 }
\email{huang.xiang@buaa.edu.cn}
\author{Zhifeng Hao}
\affiliation{%
  \institution{University of Shantou}
  \city{Shantou}
  \country{China}
  }
\email{zfhao@gdut.edu.cn}
\author{Angsheng Li}
\affiliation{%
  \institution{Beihang University}
  \city{Beijing}
  \country{China}
 }
\email{angsheng@buaa.edu.cn}
\author{Zhengtao Yu}
\affiliation{%
  \institution{Kunming University of Science and Technology}
  \city{Kunming}
  \country{China}
  }
\email{ztyu@hotmail.com}
\author{Philip S. Yu}
\affiliation{%
  \institution{University of Illinois at Chicago}
  \city{Chicago}
  \country{USA}
 }
\email{psyu@uic.edu}

\begin{abstract}
Research on social bot detection plays a crucial role in maintaining the order and reliability of information dissemination while increasing trust in social interactions.
The current mainstream social bot detection models rely on black-box neural network technology, e.g., Graph Neural Network, Transformer, etc., which lacks interpretability.
In this work, we present \framework{}, a novel unsupervised, interpretable, yet effective and practical framework for detecting social bots.
This framework is built upon structural information theory.
We begin by designing three social relationship metrics that capture various aspects of social bot behaviors: \emph{Posting Type Distribution}, \emph{Posting Influence}, and \emph{Follow-to-follower Ratio}.
Three new relationships are utilized to construct a new, unified, and weighted social multi-relational graph, aiming to model the relevance of social user behaviors and discover long-distance correlations between users.
Second, we introduce a novel method for optimizing heterogeneous structural entropy.
This method involves the personalized aggregation of edge information from the social multi-relational graph to generate a two-dimensional encoding tree.
The heterogeneous structural entropy facilitates decoding of the substantial structure of the social bots network and enables hierarchical clustering of social bots.
Thirdly, a new community labeling method is presented to distinguish social bot communities by computing the user's stationary distribution, measuring user contributions to network structure, and counting the intensity of user aggregation within the community.
Compared with ten representative social bot detection approaches, comprehensive experiments demonstrate the advantages of effectiveness and interpretability of \framework{} on four real social network datasets.
\end{abstract}


\authorsaddresses{
Authors' addresses: 
H. Peng, School of Cyber Science and Technology, Beihang University, No. 37 Xue Yuan Road, Haidian District, Beijing, 100191, China, and State Key Laboratory of Public Big Data, Guizhou University, No. 2708, South Section of Huaxi Avenue, Huaxi District, Guiyang City, Guizhou Province, 550025, China, and Guangxi Key Lab of Multi-source Information Mining \& Security, Guangxi Normal University, Guilin 541004, China, and Yunnan Key Laboratory of Artificial Intelligence, Kunming University of Science and Technology, Kunming 650500, China; email: \path{penghao@buaa.edu.cn}; 
J. Zhang, School of Cyber Science and Technology, Beihang University, No. 37 Xue Yuan Road, Haidian District, Beijing, 100191, China; email: \path{zhangjingyun@buaa.edu.cn};
X. Huang, School of Cyber Science and Technology, Beihang University, No. 37 Xue Yuan Road, Haidian District, Beijing, 100191, China; email: \path{huang.xiang@buaa.edu.cn};
Z. Hao, College of Science, University of Shantou, No. 243, University Road, Shantou, 515063, China; email: \path{haozhifeng@stu.edu.cn};
A. Li, School of Computer Science and Engineering, Beihang University, No. 37 Xue Yuan Road, Haidian District, Beijing, 100191, China; email: \path{angsheng@buaa.edu.cn};
Z. Yu, Faculty of Information Engineering and Automation, and Yunnan Key Laboratory of Artificial Intelligence, Kunming University of Science and Technology, Kunming 650500, China; email: \path{ztyu@hotmail.com};
P. S. Yu, Department of Computer Science, University of Illinois at Chicago, Chicago 60607, IL; email: \path{psyu@uic.edu}.
}


\maketitle

\renewcommand{\shortauthors}{H. Peng et al.}

\section{Introduction}\label{sec: introduction} 
Social bots are usually controlled by programs, pretending to be humans to publish harmful and low-credibility information~\cite{mendoza2024detection,shao2018spread, pozzar2020threats}, and even manipulate or guide public behavior in social networks~\cite{keller2019social,weng2022public} to reduce social trust and disrupt the orderly dissemination of information.
For example, spreading misinformation about COVID-19 has triggered an ``information epidemic''~\cite{himelein2021bots,duan2022algorithmic}.
Nowadays, the development of Artificial Intelligence Content Generator technology~\cite{zhou2023comprehensive,zhang2023complete} also makes the competition between social bot detection and anti-detection more intense~\cite{le2022socialbots,yang2023fedack,wu2020using,zeng2024adversarial}.
It has been shown that the more human-like a bot account behaves, the more likely users are to interact with it~\cite{wischnewski2022agree}, which makes it harder for social bots to be detected from social networks.
Even in public emergencies, the significant volume of low-credibility information and objectionable content spread by social bots has the potential to manipulate public emotions and disrupt the trajectory of Internet public opinion~\cite{stella2018bots,cheng2020dynamic, Allem2018could}.
Studies also have shown that the content disseminated by social bots primarily adopts a critical, correct, and questioning tone~\cite{suarez2022assessing,chen2021neutral}, aiming to polarize netizens' speech and disrupt the orderly dissemination of information~\cite{pastor2022profiling}.
Therefore, the research of detecting social bots is of great significance to protecting the orderly dissemination of information and maintaining social trust~\cite{ferrara2016rise,kobis2021bad, Makovi2023trust}.\par

Effective and reliable social bot detection approaches need adequate modeling, representation, and analysis of social user behaviors~\cite{wu2023botshape}.
However, accomplishing such a task is challenging due to the varied and dynamic interactive behaviors of social bots~\cite{cresci2020decade,cresci2019detecting,ng2023botbuster}.
Consequently, relying solely on social bot detection methods based on manual feature engineering~\cite{knauth2019language,beskow2019its,kantepe2017preprocessing} and traditional machine learning classifiers, such as Logistic regression~\cite{beskow2019its}, K-means~\cite{miller2014twitter}, SVM~\cite{efthimion2018supervised}, naive Bayesian~\cite{fazil2017identifying}, Random forest~\cite{yang2020scalable}, etc., practical detection accuracy becomes bottlenecked.
Expressly, limited by the simple and discrete low-dimensional manual features, such as \emph{username length}, \emph{the number of followers}, \emph{the number of tweets}, \emph{the number of likes}, etc., the above model's performance falls short of the ideal~\cite{feng2021satar}.
Deep learning-based social bot detection models have made significant advancements in learning feature embeddings from metadata, including Graph Neural Network (GNN)-based~\cite{liu2018heterogeneous,wang2019fdgars,feng2021botrgcn,guo2021social,yang2022rosgas,ali2019detect,breuer2020friend,arin2023deep} and Transformer-based~\cite{feng2022heterogeneity,heidari2020using,martin2021deep} models. 
The former enables improving social bot detection performance by understanding semantic relationships in the neighborhood through transmitting information between users and learning embedded user characteristics.
The latter aggregates user influence through multiple relationships and learns critical discriminative features for bot detection through self-attention mechanisms. 
Moreover, they are capable of better understanding contextual information.
Additionally, the federated knowledge distillation-based bot detection model~\cite{yang2023fedack} is employed for the cross-platform and cross-language social bots.
While the aforementioned data-driven representative models are deemed sufficient, they necessitate a substantial number of labeled social user samples, posing a risk to their generalization ability.
The behavior of social bots evolves in a realistic environment, making the labeling process challenging and rendering supervised/semi-supervised models less effective for out-of-sample data ~\cite{yang2021generalized}.\par

Existing unsupervised social bot detection models~\cite{chen2017hunting,chavoshi2016debot,miller2014twitter,mazza2019rtbust,anwar2020bot,mannocci2022mulbot} typically rely on identifying time series, simple clustering methods, or explicit behavioral markers that are exclusive to social bots.
These markers include \emph{highly synchronized}, \emph{repeated tweets}, \emph{URL shortening services}, \emph{retweeting behaviors}, etc.
However, contemporary social bots have exhibited enhanced intelligence and proficiency in disguising their objectives and concealing their authentic identity. 
Therefore, the behavioral indicators for detecting bots are only sometimes apparent or discernible.
Moreover, the behavioral markers are always low-order and discrete, often describing superficial behaviors that undermine the accuracy of detecting social bots.
Therefore, it is essential to develop effective, practical, interpretable, and unsupervised social bot detection models by utilizing user behavior data to their fullest potential to govern social bots effectively.\par

The significance of social network structural features in social bot detection cannot be underestimated, as they offer valuable insights into user interaction patterns and information dissemination.
The local social network contains valuable information that is crucial for identifying potential social bots~\cite{dehghan2023detecting}.
Despite this, structure-based detection methods are not yet prevalent, and the primary strategy for social bot detection is random walk propagation of user labels~\cite{jia2017random,wang2018structure}.
To create directed graphs, social network modeling primarily relies on direct social relationships between users, such as following, commenting, liking, and sharing.
However, most downstream tasks after graph modeling still rely on neural networks~\cite{lo2023xg,li2024graph}, which have black-box characteristics~\cite{samek2017explainable,liang2021explaining} and lack interpretability due to the absence of a direct causal logical relationship between input and output features.
In addition, developing interpretable social bot detection models can significantly enhance the detection process's reliability by improving our understanding of how the model works. 
Therefore, it is highly imperative to design a new unsupervised social bot detection framework based on user behavior data and graph structure with high effectiveness and interpretability.\par

In this work, we propose ~\framework, an effective, practical, \underline{Un}supervised, and interpretable \underline{D}etection framework of social \underline{Bot}s based on structural information theory.
Our framework seeks to reveal the significant structure of social bot networks, thereby achieving hierarchical clustering from graph to tree and detection in an unsupervised manner.
Firstly, we construct a multi-relational graph from a social bot's perspective, representing users as nodes and depicting heterogeneous social behavior commonalities as edges.
This approach enables us to define new types of social relationships that represent the similarity of users in their behavioral characteristics, such as posting type distribution, posting influence, and follow-to-follower ratio. 
Unlike traditional graphs that rely on direct user interactions like following and retweeting, this multi-relational graph gives more weight to the hidden commonalities of social behaviors among users.
Secondly, we present a new heterogeneous structural entropy optimization method to partition social users into distinct communities.
By aggregating the various types of relationships in the multi-relational graph of social users, we extend the structural entropy~\cite{li2016structural} from the simple graph to the multi-relational graph.
A new encoding tree is constructed and optimized to minimize the multi-relational graph's structural entropy, from which hierarchical community partitions of social users are provided.
Thirdly, we propose a novel community labeling method combining community influence and cohesion to identify social bot communities.
We employ Multirank~\cite{ng2011multirank} to calculate the co-ranking of vertices and multi-relational edges in the graph, obtaining a stationary distribution to quantify the influence of social user communities. 
The entropy of community nodes on the encoding tree is also utilized to quantify community cohesion. 
Combining community influence and cohesion distinguishes social bot communities from normal human communities.\par

We conduct extensive experiments on four datasets, Cresci-2015~\cite{cresci2015fame}, Cresci-2017~\cite{cresci2017social}, Pronbots-2019, and Botwiki-2019~\cite{yang2020scalable}, to demonstrate the effectiveness, interpretability, and efficiency of \framework{}.
\revised{More additional human accounts and tweets are added to the Pronbots-2019 and Botwiki-2019 datasets to make the detection experiments more realistic.}
First, the comparative experimental results indicate that \framework{} demonstrates superior overall performance compared to existing unsupervised social bot detection models and models based on unsupervised network representation learning. 
It significantly enhances the accuracy of social bot detection.
Second, a series of ablation experiments demonstrate the necessity and rationality of the graph modeling introduced in the \framework{}.
Each type of edge contributes to the performance improvement of \framework{}, and the proposed multi-relational graph significantly enhances accuracy compared to other modeling approaches.
Third, the experiment of time analysis further illustrates the balance between accuracy and efficiency of \framework{}.
Finally, visualizations for the model effect showcase the interpretability of \framework{}.
All codes and datasets of this work are publicly available at GitHub~\footnote{\url{https://github.com/SELGroup/UnDBot}}.

The main contributions of this work are summarized as follows:
\begin{itemize}
    \item An unsupervised and interpretable social bot detection framework is proposed with high accuracy that decodes the significant structural features of the network using structural information theory.

    \item A new, unified, and weighted social multi-relational graph is devised based on \revised{social bot} behavioral similarity, which breaks through the traditional approach solely on direct user interactions and better models the activity of social \revised{bot} users.

    \item A new heterogeneous structural entropy optimization method is proposed that aggregates different types of edges by assigning personalized weights to edges to achieve hierarchical community partitioning of social users.

    \item A new community labeling method is developed that involves integrating community influence (measured by stationary distribution) and community cohesion (measured by node entropy) to identify social bot communities with higher accuracy.

    \item A series of comparative and analytical experiments demonstrate that \framework{} achieves high detection accuracy and comprehensively analyzes the model's interpretability.
\end{itemize}

The structure of this paper is as follows: \revised{Section~\ref{sec: background} outlines the background and preliminaries of our work.} 
In Section~\ref{sec: methodology}, we describe the technical details of the proposed framework, named \framework. 
Section~\ref{sec: Experimental Setup} presents the experimental setup, and Section~\ref{sec: Results And Discussion} discusses the experiment's results. 
Section~\ref{sec: Related Work} provides an overview of related works. 
Finally, we conclude the paper in Section~\ref{sec: conclusion}.
\section{Background and Preliminaries}
\label{sec: background}
In this section, we first summarize the problems and challenges of social bot detection.
\revised{Then we introduce the structural information theory used in our social bot detection framework and elaborate on the basic concept of structural entropy.}
The comprehensive list of the primary symbols used throughout this paper is presented in Table~\ref{tab: notation}.

\begin{table*}[b]
\aboverulesep=0ex
\belowrulesep=0ex
\caption{Forms and interpretations of notations.} 
\centering
\begin{tabular}{r|l} 
\toprule
\textbf{Symbol} & \textbf{Definition} \\
\hline
$\mathcal{G}; \mathcal{G}^r$ & Homogeneous Graph; Multi-relational Graph (heterogeneous graph).\\

$\mathcal{V};\mathcal{E}; \mathcal{E}_k$ & Vertex set; Edge set of Homogeneous Graph; Edge set under $k$-th relationship.\\

$E$ & The connected user pairs in graph $\mathcal{G}^r$.\\

$R$ & Total number of relationships.\\

$v; N$ & Vertex in graph; Total number of vertices. \\

$e_{i,j}^k$ & Edge between vertex $i$ and vertex $j$ under $k$-th relationship.\\

$w_{i,j}^k$ & Weight of edge $e_{i,j}^k$ between vertex $i$ and vertex $j$ under the $k$-th relationship.\\

\hline

$\mathcal{T}; \lambda$ & Encoding tree; The root node of the encoding tree. \\

$ \alpha;T_\alpha$ & Node on encoding tree; Label of node $\alpha$. \\

$\alpha_i; \alpha^-$ & $i$-th child node of node $\alpha$; Parent node of node.\\

$d_i; g_\alpha$ & Degree of vertex $i$; Number of cutting edges of node $\alpha$. \\

$vol(\mathcal{G}^r); vol(\alpha)$ & Volume of Graph $\mathcal{G}$; Volume of node $\alpha$.\\

$H^\mathcal{T}(\mathcal{G})$ & The structural entropy of $\mathcal{G}$ under encoding tree $\mathcal{T}$.\\

$H_k(\mathcal{G})$ & The $k$-dimensional structure entropy.\\

$H(\mathcal{G}^r;\alpha)$ & The structural entropy of node $\alpha$ on an encoding tree.\\

$Mg(\mathcal{T};\alpha,\beta)$ & Merging operator between node $\alpha$ and node $\beta$.\\

$\mathcal{T}_{mg}$ & Encoding tree after Merging operator.\\

$\Delta_{\mathcal{G}^r}^{Mg}(\mathcal{T};\alpha,\beta)$ & Difference of Structural entropy after merging node $\alpha$ and $\beta$.\\

$p$ & The maximum scale ratio for parallel merge operators.\\
\hline

$Pt; Inf; ff$  & Distribution of posting types; Influence of posting; Following-followers ratio.\\

$d_{i,j}$ & Manhattan distance between posting type distribution between user $i$ and user $j$.\\

$\Delta_{i,j} $ & Deviation ratio of influence between user $i$ and user $j$.\\

$\xi$ &  Threshold of feature similarity in modeling.\\

\hline

$\mathcal{A}$ & A three-dimensional tensor of multi-relational graph $\mathcal{G}^r$.\\

$\mathcal{O}$ & Tensor of transition probabilities to reach different vertices.\\

$\mathcal{S}$ & Tensor of transition probabilities through different edges.\\

$o_{i,j,k}$ &  The possibility of reaching vertex $v_j$, from vertex $v_i$ through $k$-th edge.\\

$s_{i,j,k}$ &  The possibility of going through $k$-th edge, from vertex $v_i$ to vertex $v_j$.\\

$\rho$ & The stop threshold in random walk.\\ 

$\mathcal{X}; \mathcal{Y}$ &  Stationary distribution of users; stationary distribution of edges.\\

$x_{comm}^\alpha$ & The community influence.\\

$Ev(\alpha)$ & Evaluation score of the community represented by node $\alpha$.\\

$\theta$ & Threshold of evaluation index for communities.\\

$\pi$ & Weighted parameters of influence and cohesion.\\
\bottomrule
\end{tabular}
\label{tab: notation}
\end{table*}

\subsection{Problem and Challenges}
\label{sec: challenges}
Intuitively, we model with multi-relational graphs for practical tasks to transform the social bot detection into an unsupervised vertex classification problem.
The vertices in the graph can be hierarchically clustered using a two-dimensional structural entropy minimum algorithm.
Once the social user community is formed, the next task is to classify the community.
Overall, to achieve effective, unsupervised, and interpretable user vertices classification, social bot detection faces the following 3 main challenges.

\noindent
\textbf{Challenge 1: How to model social user networks to serve social bot clustering task? } 

Most of the current graph modeling approaches create edges between users based on their direct interactions on social networks, such as liking, commenting, retweeting, favoriting, following/being followed, and mentioning.
Although these techniques can intuitively capture the interactions among social users and facilitate neighbor aggregation of the graph neural network to learn high-order user features, they have minimal impact on detecting social bots from a network structure perspective.
Given the increasing intelligence of social bots, their interactions with actual human users are becoming more commonplace.
Many social bots infiltrate typical human social networks, which are hard to detect via social network structure information modeled solely by interaction relationships~\cite{ferrara2016rise}.
Hence, it is necessary to define new types of social relationships and model a social user network from the hidden behavior commonality of social bots.

\noindent
\textbf{Challenge 2: How to achieve adaptive hierarchical clustering of social users?} 

Currently, two types of clustering algorithms are used in social bot detection: feature-based clustering algorithms and network structure-based community detection algorithms~\cite{chen2017people}, such as K-nearest neighbors, density clustering, maximum flow minimum cut theory, etc. 
However, these traditional models exhibit a singular level of user clustering, wherein the predetermined number of clusters can potentially influence the efficacy of clustering. 
Deep clustering methods based on neural networks lack interpretability.
Additionally, the distance measurement method employed may not be entirely suitable.
The structural information theory provides an adaptive hierarchical community division of social users by building a structural entropy encoding tree and decoding the essential structure of social networks.
However, the current structural entropy optimization strategy~\cite{li2016structural} is aimed at homogeneous networks and does not consider the heterogeneity of user relationships in social network environments. 
Therefore, an effective structure entropy optimization strategy under multi-relational graphs is needed to achieve adaptive hierarchical clustering of social users.

\noindent
\textbf{Challenge 3: How to identify social user communities and implement binary classification?}

Theoretically, the encoding tree of a multi-relational graph can only achieve the effect of clustering users. 
It cannot directly separate users into two large social bots and human communities. 
Moreover, the number of communities formed on the encoding tree depends on the structure of the multi-relational graph.
Although humans and social bots occupy separate communities on the encoding tree, it is necessary to study the differences between different types of user communities and employ appropriate discrimination methods to explicitly identify social bot communities in the absence of anchor markers or any labels.
\revised{Unsupervised and interpretable discriminative behavioral features play a more important role in realistic social bot detection tasks.}

\subsection{\revised{Structural Information Theory}}
\label{subsec: structural information theory}
Structural information theory~\cite{li2016structural} was originally proposed in 2016 for measuring the structural information contained within a graph. 
Specifically, this theory aims to calculate the structural entropy of the homogeneous graph $\mathcal{G}=(\mathcal{V},\mathcal{E})$, which reflects its uncertainty when undergoing hierarchical division. 
\revised{In our work, the hierarchical partitions are represented by a tree structure known as the encoding tree.}
We introduce encoding trees and $k$-dimensional structural entropy below.

\textbf{Encoding tree}.
\revised{Similar to the previous study \cite{zeng2023effective},} an encoding tree of a graph $\mathcal{G}$ is defined as a rooted tree $\mathcal{T}$ with the following properties:
\begin{enumerate}
\item For each node $\alpha$ in the encoding tree $\mathcal{T}$, there is a subset $T_{\alpha} \in \mathcal{V}$ of vertices in the graph $\mathcal{G}$ corresponding to it.
\item For the root node $\lambda$ in the encoding tree, $T_\lambda=\mathcal{V}$.
\item The children of node $\alpha$ are denoted as $\alpha_{i}$ and sorted from left to right as $i$ increases. 
The parent node of $\alpha_{i}$ is denoted as $\alpha_{i}^-=\alpha$.
\item If node $\alpha$ has $L$ children, then the vertex subset $T_{\alpha_{i}}$ of child nodes is mutually exclusive, and $T_{\alpha}=\cup T_{\alpha_{i}}$.
\item Each leaf node $v$ in the tree, $v$ corresponds to a single vertex in the vertex set $\mathcal{V}$ in the graph $\mathcal{G}$.
\end{enumerate}

The $k$-dimensional encoding tree means that the tree's height is $k$ (the height of the root node is 0).
Intuitively, the encoding tree embodies the hierarchical community division of graph vertices, the parent node is a large community, and the child nodes are small communities in the large community.

\textbf{Structure entropy}.
The structural information of the homogeneous graph $\mathcal{G}$ determined by the encoding tree $\mathcal{T}$ is defined as:
\begin{eqnarray} 
H^\mathcal{T}(\mathcal{G})=-\sum_{\alpha \in \mathcal{T},\alpha \neq \lambda}\frac{g_\alpha}{vol(\mathcal{G})}log\frac{vol(\alpha)}{vol(\alpha^-)},
\end{eqnarray}
where $vol(\mathcal{G})$ is the sum of the degrees of all vertices in the graph $\mathcal{G}$. 
$vol(\alpha)$ is the volume of $T_{\alpha}$ and is the sum of the degrees of all vertices in the vertex subset $T_{\alpha}$. 
$g_{\alpha}$ is the sum of weights of all edges from vertex subset $T_{\alpha}$ to vertex subset $\mathcal{V}/ T_{\alpha}$, which can be understood as the total weight of the edges from the vertices outside the vertex subset $T_{\alpha}$ to the vertices inside the vertex $T_{\alpha}$, or the total weight of the cut edges.
$\frac{g_\alpha}{vol(\mathcal{G})}$ represents the probability that the random walk enters $T_{\alpha}$.
The structural entropy $H(\mathcal{G})$ of graph $\mathcal{G}$ is the minimum $H^{\mathcal{T}}(\mathcal{G})$. 
Let $\mathcal{T}_k$ be encoding trees whose height is not greater than $k$, then the $k$-dimensional structural entropy of $\mathcal{G}$ is defined as follows:
\begin{eqnarray} 
H_k(\mathcal{G})=min H^{\mathcal{T}_k}(\mathcal{G}).
\end{eqnarray}
Furthermore, one-dimensional structural entropy is special as there are only root nodes and leaf nodes in the encoding tree of one layer.
All the vertices in the graph $\mathcal{G}$ belong to a large community $\lambda$ under the one-dimensional encoding tree, which is unique in terms of community division so that the one-dimensional structural entropy can be directly expressed as:
\begin{eqnarray} 
H_1(\mathcal{G})=-\sum_{i=1}^{n}\frac{d_i}{vol(\mathcal{G})}log\frac{d_i}{vol(\mathcal{G})},
\end{eqnarray}
where $d_i$ is the sum of weights of all edges connected to vertex $v_i$ in graph $\mathcal{G}$ and is called the degree of vertex $v_i$. 
One-dimensional structural entropy measures the uncertainty of graph $\mathcal{G}$ without layering.
\section{Methodology}
\label{sec: methodology}
\begin{figure}[h]
    \centering
    \vspace{-0.5cm}
    \includegraphics[width=1\textwidth]{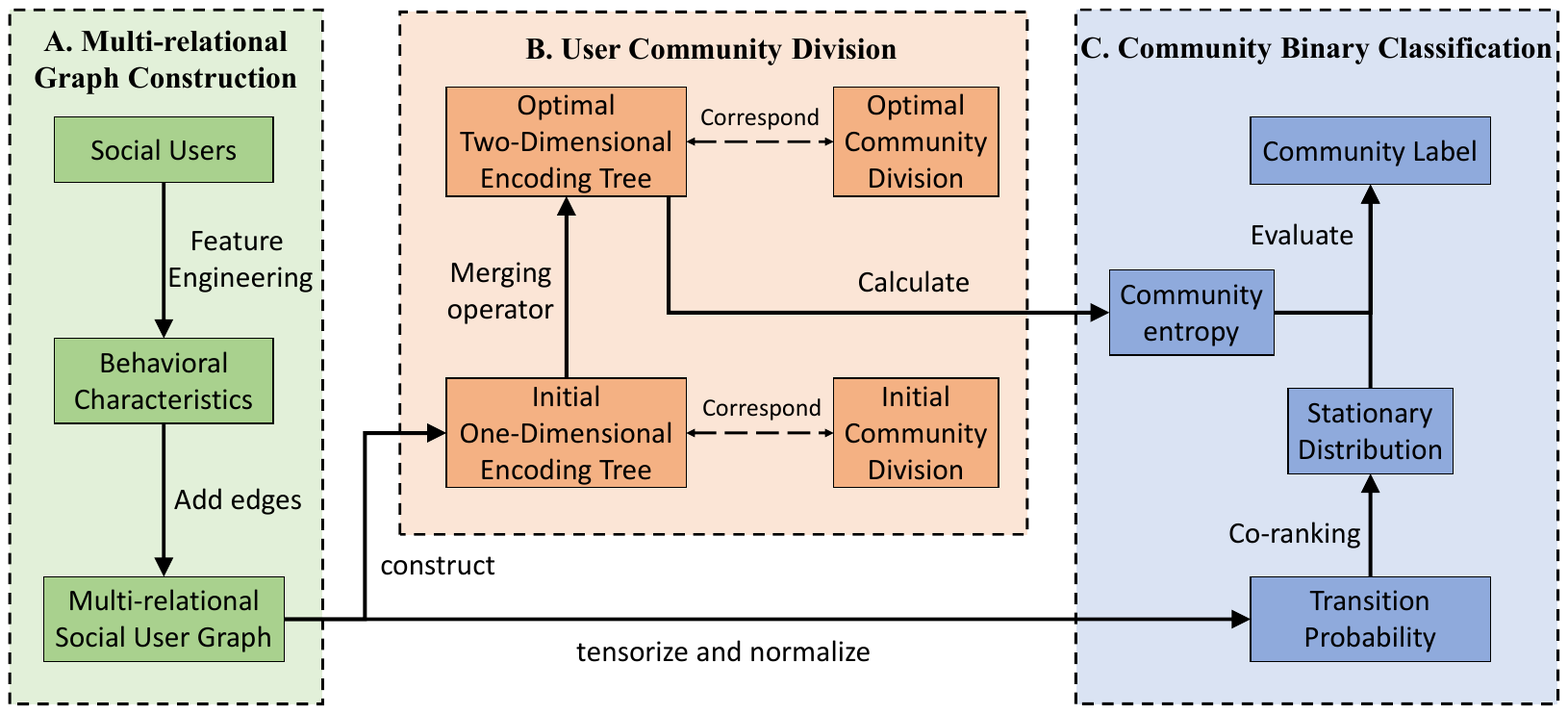}
    \vspace{-4.0cm}
    \caption{The overall framework of ~\framework .}
    \label{fig: overall-framwork}
    \vspace{-0.4cm}
\end{figure}

In this section, we will introduce the social bot detection framework \framework{} based on the structural information theory and the multi-relational graph.
As shown in Figure~\ref{fig: overall-framwork}, \framework{} consists of three key modules: Multi-relational Graph Construction, User Community Division, and Community Binary Classification.
\textbf{(1) Multi-relational Graph Construction.} To begin with, social users are constructed as a multi-relational graph based on the similarity of bot behavioral characteristics. (Section~\ref{sec: Multi-relational Graph Construction})
\textbf{(2) User Community Division.} Following the graph construction process, an optimal two-dimensional encoding tree is created based on the principle of structural entropy minimization.
Additionally, social users are allocated to different subtrees on the encoding tree, resulting in community division. (Section~\ref{sec: User Community Division})
\textbf{(3) Community Binary Classification.} For each community, the stationary distribution and community entropy are utilized to quantify community influence and cohesion, which are then used for binary classification \revised{to bot or human accounts}. (Section~\ref{sec: Binary Classification of Communities})

\subsection{\revised{Multi-relational social user graph}}\label{sec: Multi-relational Graph Construction}
In this subsection, we first define the presented multi-relational social user graph and compare it to the relationships used in other models.
Subsequently, a detailed description of constructing a multi-relational graph in \framework{} will be provided.

\begin{defn}
    \textbf{Multi-relational graph in the social bot detection task.}\par
    The multi-relational graph in social bot detection, denoted as $\mathcal{G}^r=\{\mathcal{V},{\mathcal{E}_{k}}|_{k=1}^{R}\}$, is defined in terms of $\mathcal{V}$ and $\mathcal{E}_{k}$.
    $\mathcal{V}$ represents the set of users $\{v_{1}, \dots, v_{N}\}$, and $N$ denotes the total number of users in the social network.
    $\mathcal{E}_{k}$ represents a set of edges $e_{i, j}^{k} = (v_{i}, v_{j}, w_{ij}^k)\in \mathcal{E}_{k}$ with weight between users under $k$-th relationship.
    $R$ is the total number of different relationships, and weight $w_{ij}^k$ is the similarity under $k$-th relationship. \par

    When constructing a multi-relational graph for a social bot detection task, relationships are always direct interactions between users on social networks, such as following, replying, retweeting, mentioning, and liking, as shown in Figure~\ref{fig:multi-relational graph}(b).
    This modeling method may be suitable for neighbor aggregation in architectures based on graph neural networks, but it is not beneficial for structure-based detection models.
    Due to the intelligence of social bots, their interactions with humans are becoming more and more frequent. 
    It is difficult to directly identify a social bot in a complex network through interactive relationships or basic user information such as their ID, profile, username length, and the number of tweets they have made.
    \revised{Because the social bot will disguise itself by interacting with normal human users, there are also rich connections between social bots and human users in terms of following, replying, etc.}
    Nevertheless, the existence of social bots is always for certain purposes, like spreading malicious news, guiding public opinion, or acting as fake fans. 
    These purposes are always achieved through social behaviors such as \revised{tweeting, following, commenting, retweeting, and liking}.
    \revised{Traditional social behaviors-based multi-relationship graph modeling in practice makes it difficult to achieve effective bot detection performance.}
    
    As shown in Figure~\ref{fig:multi-relational graph}(a), we abandon the traditional multi-relational graph modeling method and focus on the commonality of social behaviors from the perspective of social bots to construct a multi-relational graph, thereby transforming the social bot detection task into a vertex binary classification problem.
    The connections we selected are closely related to social bots, as shown in Table~\ref{tab: relationships}.
    \revised{We divide social behaviors into posting type (tweeting, retweeting, and commenting), posting influence (retweeting, liking, commenting), and follow-to-follower ratio (following).}
    \revised{The above social behaviors include active and passive behaviors.}
    \revised{For example, retweeting and commenting in the posting type are the actions of the user, while retweeting and commenting in the posting influence are received by the user, that is, the actions of other users towards the user.}
    \revised{The connections formed include} having the same \emph{posting type distribution} (U-T-U), having the same \emph{posting influence} (U-I-U), and having the same \emph{follow-to-follower ratio} (U-F-U).
    \revised{Meanwhile, the closeness of different types of connections can be formalized as the weight of the edges. 
    Intuitively, the more prominent the social bot behavioral characteristics, the richer the information around the suspected nodes in the graph.}
    
    \begin{table}[h]
        \centering
        \aboverulesep=0ex
        \belowrulesep=0ex
        \caption{\revised{Relationships in ~\framework .}}
        \begin{tabular}{m{5cm}<{\centering}|m{5cm}<{\centering}|m{4cm}<{\centering}}
            \toprule
            \textbf{Relation} & \textbf{Illustration} & \revised{\textbf{Social Behaviors}} \\
            \hline
            \makecell{\underline{U}ser-\emph{posting \underline{T}ype distribution}-\underline{U}ser\\ (\emph{U-T-U})}  & It connects two users who have the same distribution of posting types. & \revised{tweeting, retweeting, and commenting} \\
            \hline
            \makecell{\underline{U}ser-\emph{posting \underline{I}nfluence}-\underline{U}ser\\ (\emph{U-I-U})} & It connects two users who have the same posting influence on social networks. & \revised{retweeting, liking, and commenting} \\
            \hline
            \makecell{\underline{U}ser-\emph{\underline{F}ollow-to-follower ratio}-\underline{U}ser \\(\emph{U-F-U})} & It connects users who have the same ratio of followings to followers. & \revised{following} \\
            \bottomrule
        \end{tabular}
        \label{tab: relationships}
    \end{table}

    \begin{figure}
        \centering
        \includegraphics[width=1\textwidth]{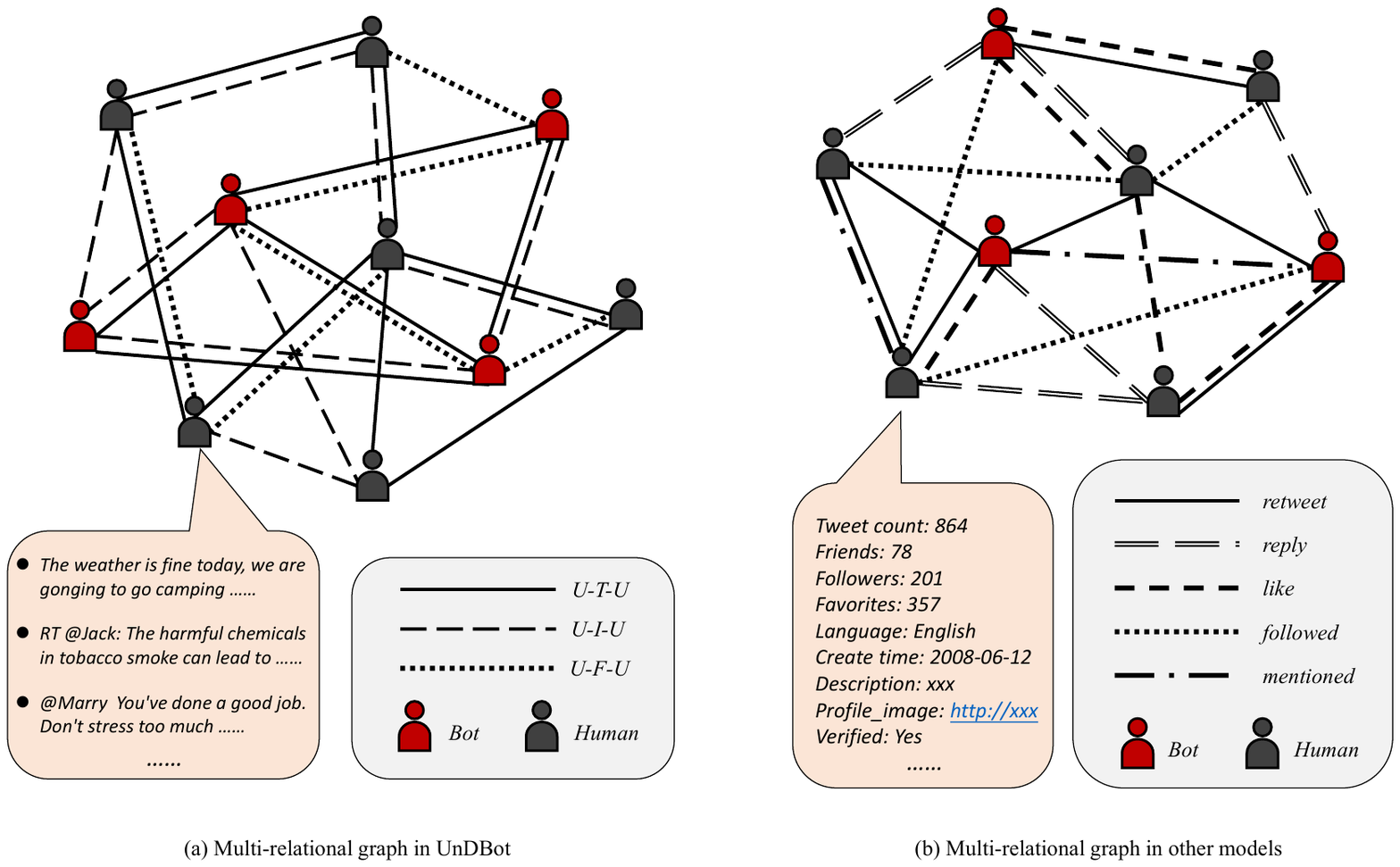}
        \vspace{-2cm}
        \caption{\revised{Multi-relational graph in social bot detection.}}
        \label{fig:multi-relational graph}
        \vspace{-0.5cm}
\end{figure}    
\end{defn}

\begin{defn}
    \textbf{Multi-relational Graph Construction.}\par
    \revised{In terms of posting behavior, the} posting type can be divided into original tweets, retweets, and comments.  
    Social spambots typically disseminate a substantial volume of spam tweets and promotional content to lure internet users. 
    Conversely, certain types of social bots engage in excessive retweeting and commenting on tweets about specific subjects to influence the trend of public opinion topics.
    Therefore, we argue that \textbf{\emph{Posting Type Distribution}} is a characteristic closely related to social bots.
    To this end, we propose a new kind of edge to aggregate the proportion information for each type of tweet.
    We utilize a vector $Pt_i=[pt_i^1,pt_i^2,pt_i^3]$ to represent the distribution of posting types for each user, where $pt_i^1$ signifies the proportion of original tweets, $pt_i^2$ indicates the proportion of retweets and $pt_i^3$ denotes the proportion of comments.
    Since the \emph{posting type distribution} is not a discrete value, we also use the Manhattan distance~\cite{singla2012comparative} to calculate the similarity between two users' posting type distribution vectors, thereby determining whether the posting type distributions are similar. 
    The distance of $Pt_i$ and $Pt_j$ and the weight are as follows:
    \begin{equation} 
        d_{i,j}=\mid pt_i^1-pt_j^1 \mid + \mid pt_i^2-pt_j^2 \mid + \mid pt_i^3-pt_j^3 \mid, \quad
        w_{ij}^1=1-d_{i,j},
        \label{Equation: dij}
    \end{equation}
    where $d_{i,j} \leq 1$ is the Manhattan distance of the \emph{posting type distribution} and $w_{ij}^1$ is the weight of edge between user $v_i$ and user $v_j$ defined by the similarity of \emph{posting type distribution}.
    The larger the value of $d_{i,j}$, the smaller the weight of the edge $w_{ij}^1$.
    When the Manhattan distance $d_{i,j}$ is less than the threshold $\xi$, we consider that the two users have similar \emph{posting type distribution}, indicating they have similar posting preferences.
    In this case, an edge $e_{i,j}^1=(v_i,v_j,w_{ij}^1)$ is added on the multi-relational graph $\mathcal{G}^r$ .\par
    
    \revised{In terms of social influence, the} influence on social networks is manifested in the number of comments, likes, and retweets received by users’ tweets.
    Generally speaking, the most influential users are some official media or social media influencers, but social bots lurk in social networks, accumulating certain influence in long-term interactions with normal users.
    Therefore, the \textbf{\emph{Posting Influence}} is also an important focus of social bot detection.
    In our work, we define the influence of a user's post $Inf_i$ as the sum of the average number of comments, likes, and retweets of original tweets.
    Unlike the \emph{posting type distribution} $Pt_i$, there is no upper limit to the $Inf_i$ defined here.
    If normalization is performed before utilizing the Manhattan distance to determine the similarity of influence between two users, it may disadvantage users with significantly fewer likes, comments, and retweets than the maximum value. 
    This is because normalization narrows the influence gap between users.
    Therefore, we use the deviation ratio of influence between users to calculate the weight as follows:
    \begin{eqnarray}
        \Delta_{i,j}=\frac{\mid Inf_i-Inf_j\mid}{max(Inf_i,Inf_j)},\quad
        w_{ij}^2=1-\Delta_{i,j},
        \label{Equation: delta ij}
    \end{eqnarray}
    where $\Delta_{i,j}$ is the deviation ratio of the influence of the two users, and $w_{ij}^2$ is the weight of the edge between user $v_i$ and user $v_j$ defined by the similarity of \emph{posting influence}.
    When the deviation ratio is less than the threshold $\xi$, we consider that the \emph{posting influences} of these two users are similar.
    In this case, an edge $e_{i,j}^2=(v_i,v_j,w_{ij}^2)$ is added on the multi-relational graph $\mathcal{G}^r$.\par
    
    Apart from tweeting, the act of following other users also has an impact on social networks.
    However, some unethical individuals or commercial organizations have recognized the commercial value of having many fans.
    ``Fake fans'' refers to followers that are artificially created through the use of robots, virtual identities, and other methods. 
    These followers typically do not interact with other accounts.
    Therefore, the \textbf{\emph{Follow-to-follower Ratio}} is also an important indicator for analyzing social user behavior.
    To fit users who do not have any followers, we define the \emph{follow-to-follower ratio} as follows:
    \begin{eqnarray}
        {ff}_i=\frac{{num \text{-} following}_i+1}{{num \text{-} follower}_i+1},
        \label{Equation: ffi}
    \end{eqnarray}
    where ${ff}_i$ is the \emph{follow-to-follower ratio}, ${num \text{-} following}_i$ and ${num \text{-} follower}_i$ refer to the number of followings and fans of user $v_i$, respectively.
    Similar to the \emph{posting influence}, there is also no upper limit to ${ff}_i$, so we use the deviation ratio to calculate the weight as follows:
    \begin{eqnarray}
        w_{ij}^3=1-\frac{\mid {ff}_i-{ff}_j\mid}{max({ff}_i , {ff}_j)}.
        \label{Equation: SF}
    \end{eqnarray}
    Here $w_{ij}^3$ is the weight of the edge between user $v_i$ and user $v_j$ defined by the similarity of \emph{follow-to-follower ratio}.
    When the deviation ratio is less than $\xi$, we consider that the \emph{follow-to-follower ratio} of these two users are similar.
    To prioritize the following behavior of users, we implement a restriction $\varphi$ on the rules for building edges: only if both the $ff_i$ and ${ff}_j$ of user pair $(i,j)$ are greater than $\varphi$, an edge $e_{i,j}^3=(v_i,v_j,w_{ij}^3)$ is added to the multi-relational graph $\mathcal{G}^r$.\par
\end{defn}

\revised{Overall, we design the above three social relationship metrics that capture various aspects of social bot behaviors: \emph{Posting Type Distribution}, \emph{Posting Influence}, and \emph{Follow-to-follower Ratio}, to model the multi-relational social user graph.}

\begin{figure}[thp]
    \centering
    \vspace{-0.5cm}
    \includegraphics[width=1\textwidth]{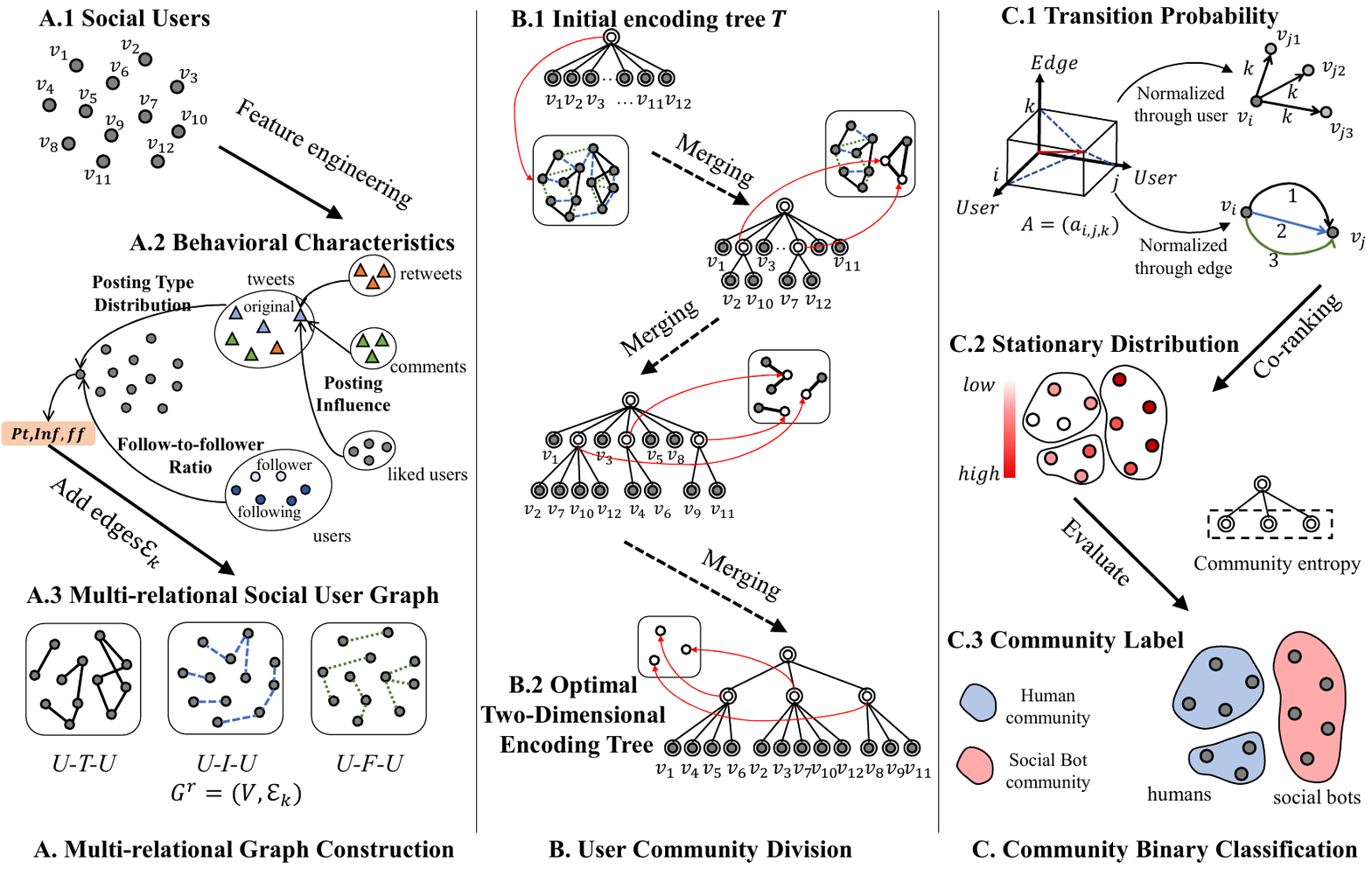}
    \vspace{-1.0cm}
    \caption{The Structural Information Theory-Based Social Bot Detection.}
    \label{fig: UnDBot_architecture}
\end{figure}

\subsection{User Community Division}
\label{sec: User Community Division}\par
The encoding tree clustering method demonstrates superior adaptability in selecting the number and size of communities on graphs~\cite{li2015discovering,li2016structural,zeng2023unsupervised,cao2024hierarchical}, compared to other clustering methods. 
However, it is important to note that this approach is restricted to homogeneous and simple graphs.
To tackle the challenge of constructing encoding trees and partitioning communities on multi-relational graphs, we first expand the concept of structural entropy to encompass multi-relational graphs defined in Section~\ref{sec: Multi-relational Graph Construction}. 
Then, we construct the optimal encoding tree for multi-relational graphs to partition the social user community effectively.

\textbf{Multi-Relational Graph Structural Entropy.}
The definition of traditional structural entropy is given in Section~\ref{subsec: structural information theory}
After constructing a multi-relational social user graph based on the similarity of social behaviors, our objective is to classify users with similar social behaviors into the same community using structural information theory.
However, the initial structural entropy analysis is conducted on the homogeneous graph. 
Therefore, we design an optimized structural entropy of the multi-relational graph $H^\mathcal{T}(\mathcal{G}^r)$ under the encoding tree $\mathcal{T}$. 
This is achieved by representing all variables of edges as the summation of the weights across the three kinds of relationships, which include the degree of vertices ($d_i$), community volume ($vol(\alpha)$), and cut edges ($g_\alpha$):
\begin{eqnarray}
    \left\{
    \begin{array}{l}
        d_i=\omega_1d_i^1+\omega_2d_i^2+\omega_3d_i^3.\\
        vol(\alpha)=\omega_1 vol^1(\alpha)+\omega_2 vol^2(\alpha)+\omega_3 vol^3(\alpha).\\
        g_\alpha=\omega_1 g^1_\alpha+\omega_2 g^2_\alpha+\omega_3 g^3_\alpha.
    \end{array}
    \right.
    \label{Equation: multi edge}
\end{eqnarray} 
The resulting structural entropy integrates the three newly defined social relationships, providing a more comprehensive depiction of the structure of the multi-relational graph.\par

\textbf{Optimized Encoding Tree.}
The definition of encoding tree is also given in Section~\ref{subsec: structural information theory}
An optimal hierarchical encoding tree is constructed during the computation of the structural entropy.
The leaf nodes of this encoding tree represent the vertices in the graph.
If two leaf nodes share the same parent node, it signifies that the corresponding graph vertices belong to the same community.
The initial one-dimensional encoding tree represents the simplest two-level structure, where the leaf nodes in the graph are directly connected to the root node, \revised{as shown in Figure~\ref{fig: UnDBot_architecture} B.1}.
The \emph{Merge operator} combines the nodes, and a greedy search strategy is employed to construct an optimal encoding tree.
The \emph{Merge operator} combines two subtrees under the same parent node, resulting in a single subtree. 
This is visualized in the graph as the merging of two small communities.
In the optimal encoding tree, the graph structure exhibits minimal uncertainty, and the nodes achieve a balanced and stable state, leading to the optimal partitioning of user vertices.
The steps of the \emph{Merge operator} are outlined as follows:
\begin{enumerate}
\item Let label $T_\alpha=\{x_1, x_2... x_M, y_1, y_2,... , y_N\}$, and merge the label of node $\beta$ into node $\alpha$;
\item For every subtree $\alpha_s, s\in \{{1, 2... , M}\}$, define $T_{\alpha_s}=\{x_s\}$, assign the level $h(\alpha_s) \xleftarrow{} h(\alpha)+1$;
\item For every subtree $\alpha_t, M + 1 \leq t \leq M + N$, define $T_{\alpha_t}=\{ y_{t-M}\}$, assign 
 the level $h(\alpha_t) \xleftarrow{} h(\alpha)+1$;
\item Delete subtree node $\beta$ and its all subtrees.
\end{enumerate}
Step (1) updates the label of node $\alpha$ to the merged one, and steps (2) and (3) define the child nodes of the merged node $\alpha$.
Recording $\mathcal{T}_{mg} (\alpha, \beta)$ as the encoding tree after $\mathcal{T}$ runs $Mg(\mathcal{T}; \alpha, \beta)$, the difference in structural entropy of the graph $\mathcal{G}^r$ determined by the two encoding trees $\mathcal{T}_{mg} (\alpha, \beta)$ and $\mathcal{T}$ is:
\begin{eqnarray} 
\begin{split}
\Delta_{\mathcal{G}^r}^{Mg}(\mathcal{T};\alpha,\beta)=\left( H^{\mathcal{T}_{mg}(\alpha,\beta)}(\mathcal{G}^r;\alpha)+\sum_{\delta^-=\alpha}H^{\mathcal{T}_{mg}(\alpha,\beta)}(\mathcal{G}^r;\delta)\right)\\
-\left( H^\mathcal{T}(\mathcal{G}^r;\alpha)+H^\mathcal{T}(\mathcal{G}^r;\beta)+\sum_{\delta^-=\alpha or \delta^-=\beta }H^\mathcal{T}(\mathcal{G}^r;\delta) \right),
\end{split}
\label{Equation: Merging opertor}
\end{eqnarray}
where $H^\mathcal{T}(\mathcal{G}^r; \alpha)$ is the structural information of subtree $\alpha$, $H^\mathcal{T}(\mathcal{G}^r; \beta)$ is the structural information of subtree $\beta$, $H^\mathcal{T}(\mathcal{G}^r; \delta)$ is the structural information of $\alpha$'s and $\beta$'s subtree $\delta$; $H^{\mathcal{T}_{mg}(\alpha, \beta)}(\mathcal{G}^r; \alpha)$ is the structural information of subtree $\alpha$ after merging subtree $\beta$ into $\alpha$; 
Similarly, $H^{\mathcal{T}_{mg}(\alpha, \beta)}(\mathcal{G}^r; \delta)$ is the structure information of the subtree $\delta$ after merging subtree $\beta$ into $\alpha$. 
\revised{Eq.~\ref{Equation: Merging opertor} calculates the change in the related subtree structure entropy before and after running the Merge operator on nodes $\alpha$ and $\beta$.}
If $\Delta_{\mathcal{G}^r}^{Mg}(T; \alpha, \beta)\le 0$, then the Merging operator runs successfully, denoted as $Mg(\mathcal{T}; \alpha, \beta)\downarrow$.
According to Eq.~\ref{Equation: Merging opertor} $\Delta_{\mathcal{G}^r}^{Mg}(\mathcal{T}; \alpha, \beta)$ is locally computable. 
The Merging operator only merges two subtrees into one subtree, and other nodes in the encoding tree do not change.
For a two-dimensional encoding tree, the structural entropy difference after the Merging operator is expanded as:
\begin{eqnarray} 
\begin{split}
\Delta_{\mathcal{G}^r}^{Mg}(\mathcal{T};\alpha,\beta)=-\frac{g_{\alpha'}}{vol(\mathcal{G}^r)}log_2\frac{vol(\alpha')}{vol(\mathcal{G}^r)}+\sum_{v_i \in \alpha'}-\frac{d_i}{vol(\mathcal{G}^r)}log_2\frac{d_i}{vol(\alpha')}\\
-\left(-\frac{g_\alpha}{vol(\mathcal{G}^r)}log_2\frac{vol(\alpha)}{vol(\mathcal{G}^r)}+\sum_{v_i \in \alpha}-\frac{d_i}{vol(\mathcal{G}^r)}log_2\frac{d_i}{vol(\alpha)}\right)\\
-\left(-\frac{g_\beta}{vol(\mathcal{G}^r)}log_2\frac{vol(\beta)}{vol(\mathcal{G}^r)}+\sum_{v_i \in \beta}-\frac{d_i}{vol(\mathcal{G}^r)}log_2\frac{d_i}{vol(\beta)}\right),
\end{split}
\label{Equation: delta merge}
\end{eqnarray}
where $g_\alpha$ is the sum of the cut edge weights of vertex subset $T_\alpha$, 
$d_i$ represents the degree of vertex $v_i$, which is the sum of the weights of its edges. 
$vol(\mathcal{G}^r)$ represents the volume of the root node and is the sum of the degrees of all vertices;
$vol(\alpha)$ represents the volume of node $\alpha$, which is the sum of degrees of vertex subset $T_\alpha$.\par

The initial one-dimensional encoding tree implies that each user vertex $v_i$ forms an independent community as illustrated in Figure~\ref{fig: UnDBot_architecture} B.1.
\revised{Next, we calculate the difference in structural entropy before and after executing the Merge operator on any two nodes, and select no more than the maximum number of pairs, denoted as $max\_operate$, from the pairs of nodes whose structural entropy is reduced the most to run the Merge operator. }
Continue this process until no pair of nodes is found that allows the Merge operator to execute successfully. 
The value of $max\_operate$ for each iteration is determined by the Equation $ceil((current\_num-1)*p)$, where $current\_num$ represents the current number of communities/nodes, and $p$ is a hyperparameter between 0 and 1 that controls the speed of parallel operations for the Merge operator.
The operator ceil is a mathematical function that rounds a given number up to the nearest integer.
The final optimal two-dimensional encoding tree partitions users into distinct communities, as illustrated in Figure~\ref{fig: UnDBot_architecture} B.2. 
Like information entropy, which measures information uncertainty, structural entropy quantifies the uncertainty resulting from graph partitioning.
A lower value of structural entropy indicates reduced uncertainty and greater stability in the graph structure, thereby achieving the most favorable division of user nodes.\par

\subsection{Community Binary Classification}
\label{sec: Binary Classification of Communities}\par
Up to this point, we have successfully partitioned social users on the multi-relational graph into distinct communities.
This subsection introduces a method for quantifying community influence in a stationary distribution. 
We propose a community labeling strategy by combining community influence and cohesion.

\textbf{Stationary Distribution.}
MultiRank~\cite{ng2011multirank} is a framework designed for the co-ranking of vertices and edges in multi-relational graphs. 
It assesses the significance of users and relationships when computing the probability distribution for multi-relational data.
Users on the multi-relational graph are connected through different relationships, so we represent the multi-relational graph $\mathcal{G}^r$ as a three-dimensional tensor $\mathcal{A} = (a_{i,j,k})$, where $a_{i,j,k}$ is the weight between nodes $v_i$ and $v_j$ under $k$-th relationship.
When user vertices $v_i$ and $v_j$ have no edge under $k$-th relationship, the $(i,j,k)$ term is zero.
\revised{And this step is called tensorization and corresponds to 'tensorize' in Figure~\ref{fig: overall-framwork}}:
\begin{eqnarray}
    \mathcal{A}=(a_{i,j,k})=\left\{
    \begin{array}{l}
        1, \quad \exists e_{i,j}^k.\\
        0, \quad \nexists e_{i,j}^k.
    \end{array}
    \right.
    \label{Equation: A}
\end{eqnarray}
To obtain the influence of social users on the entire multi-graph through co-ranking, we normalize the tensor $\mathcal{A}$ based on vertices and relationships separately, as shown in Figure~\ref{fig: UnDBot_architecture} C.1 of the UnDBot architecture \revised{which corresponds to 'normalize' in Figure~\ref{fig: overall-framwork}}. 
This process results in two tensors, $O$ and $S$:
\begin{eqnarray}
    \left\{
    \begin{array}{l}
        \mathcal{O}=(o_{i,j,k})=\left\{
        \begin{array}{l}
            \frac{a_{i,j,k}}{\sum_{j=1}^N a_{i,j,k}}, \quad \exists e_{i,j}^k,\\
            \frac{1}{N},\forall j \in \{1,2,...,N\}, \quad \nexists e_{i,j}^k,
        \end{array}
        \right.\\
        \mathcal{S}=(s_{i,j,k})=\left\{
        \begin{array}{l}
            \frac{a_{i,j,k}}{\sum_{k=1}^3 a_{i,j,k}},\quad \exists e_{i,j}^{k}.\\
            \frac{1}{3},\forall k \in \{1,2,3\}, \quad \nexists e_{i,j}^{k}.
        \end{array}
        \right.
    \end{array}
    \right.
    \label{Equation: OS}
\end{eqnarray}
Here $o_{i,j,k}$ represents the possibility of reaching vertex $v_j$ from vertex $v_i$ and $k$-th edge. 
And $s_{i,j,k}$ represents the possibility of going through $k$-th edges from vertex $v_i$ to vertex $v_j$.
In particular, if the vertex $v_i$ does not have any edge under the $k$-th relationship ($\sum_{j=1}^N a_{i,j,k}=0$), or if the vertex pair $(v_i, v_j)$ has no edge under any relationship ($\sum_{k=1}^3 a_{i,j,k}=0$), the average value is used instead.
The tensors $\mathcal{O}$ and $\mathcal{S}$ represent the transition probability of a random walk on the multi-relational graph $\mathcal{G}^r$.
We use $t-1$ to represent the state of the random walk at the previous moment and use $t$ to represent the state after a random walk.
Then the possibility $Prob_O[j;t]$ of reaching user $v_j$ and the possibility $Prob_S[k;t]$ of passing the $k$-th edge at the current moment are formalized by the following Equation:
\begin{eqnarray} 
    \left\{
    \begin{split}
        Prob_O[j;t]=\sum_{i=1}^{N} \sum_{k=1}^{3} o_{i,j,k} \times Prob_O[i;t-1] \times Prob_S[k;t],\\
        Prob_S[k;t]=\sum_{i=1}^{N} \sum_{j=1}^{N} s_{i,j,k} \times Prob_O[i;t-1] \times Prob_O[j;t].
    \end{split}
    \right.
    \label{Equation: possibility}
\end{eqnarray}
\revised{The possibility of reaching user $v_j$ at time $t$ ($Prob_O[j;t]$) is expressed as the accumulation of the possibility of reaching user $v_i$ at the previous moment $t-1$ ($Prob_O[i;t-1]$) and passing edge $k$ from user $v_i$ at the current moment $t$ ($Prob_S[k;t]$). }
\revised{The possibility of passing the $k$-th edge at time $t$ ($Prob_S[k;t]$) is expressed as the accumulation of the possibility of reaching user $v_i$ at the previous moment $t-1$ ($Prob_O[i;t-1]$) and reaching user $v_j$ at the current moment $t$ ($Prob_O[j;t]$). }
After the continuous random walk, the possibility of arriving at user $v_j$ and passing $k$-th edge reaches a steady state, i.e., $Prob_O[j;t] \approx Prob_O[j;t-1]$. 
Denoting $\overline{x_j} = \lim_{t\to\infty}Prob_O[j;t]$ and $\overline{y_k}=\lim_{t\to\infty}Prob_S[k;t]$, Eq.~\ref{Equation: possibility} is formalized as follows when the time zone is infinite:
\begin{eqnarray}
    \left\{
    \begin{split}
        \overline{x_j}=\sum_{i=1}^{N} \sum_{k=1}^{3} o_{i,j,k} \times \overline{x_i} \times \overline{y_k},\\
        \overline{y_k}=\sum_{i=1}^{N} \sum_{j=1}^{N} s_{i,j,k} \times \overline{x_i} \times \overline{x_j}.
    \end{split}
    \right.
    \label{Equation: xy}
\end{eqnarray}
We represent the steady state as a stationary distribution in the tensor form $\mathcal{X}=[\overline{x_1},\overline{x_2},…,\overline{x_N}]$ and $\mathcal{Y}=[\overline{y_1},\overline{y_2},\overline{y_3}]$.
The random walk process is simulated to iterate the initial distribution tensor until the two tensors tend to be stable.
The stationary distribution $\mathcal{X}$ can thus be obtained from the system of equations: 
\begin{equation}
    \left\{
    \begin{split}
        \mathcal{X}=\mathcal{X}\mathcal{O}\mathcal{Y}^\top, \\ \mathcal{Y}=\mathcal{X}\mathcal{S}\mathcal{X}^\top.
    \end{split}
    \right.
    \label{Equation: XY}
\end{equation}
\revised{Constantly update tensor $\mathcal{X}$ and tensor $\mathcal{Y}$ through the equation, and a stabilize threshold $\rho$ is set to stop updating when $|\mathcal{X}_{new}-\mathcal{X}_{old}|+|\mathcal{Y}_{new}-\mathcal{Y}_{old}|<\rho$.}

\textbf{Community Label.}
The stationary distribution $\mathcal{X}$ represents the distribution of influence that each user vertex has on the entire network, as depicted in Figure~\ref{fig: UnDBot_architecture} C.2. 
The average influence of users quantifies community influences $x_{comm}^\alpha$: 
\begin{eqnarray}
    \begin{array}{c}
    x_{comm}^\alpha=avg(\overline{x_i},v_i \in T_\alpha),
    \end{array}
    \label{Equation: influence}
\end{eqnarray}
where $x_{comm}^\alpha$ is the community influence, $T_\alpha$ is the community represented by node $\alpha$, and $avg$ represents a mathematical function that calculates the average value.
In this work, since the multi-relational graph constructed in section~\ref{sec: User Community Division} is based on the similarity of behavior between users, the distribution of vertex influence $\mathcal{X}$ is interpreted as the similarity between each user and other users.
The degree of community influence is interpreted as the degree of similarity of user behavior within the community.\par

In addition to community influence, internal cohesion within a community is an important indicator for analyzing the community.
The entropy $H(\mathcal{G}^r;\alpha)$ of the community node on the encoding tree quantifies community cohesion:
\begin{eqnarray}
    \begin{array}{c}
    H(\mathcal{G}^r;\alpha)=-\frac{g_\alpha}{vol(\mathcal{G}^r)}log_2\frac{vol(\alpha)}{vol(\mathcal{G}^r)}.
    \end{array}
    \label{Equation: cohesion}
\end{eqnarray}
The higher the similarity among users within a community, the more interconnected edges exist within the community, resulting in a higher entropy of community nodes and stronger cohesion of the community, resembling a social bot community.
We combine community influence and community cohesion to define the community label function as follows:
\begin{eqnarray}
    Ev(\alpha)=(1-\pi) \frac{x^\alpha_{comm}}{x^\lambda_{comm}}+\pi \frac{H(\mathcal{G}^r;\alpha)}{\sum_{\beta^-=\lambda}H(\mathcal{G}^r;\beta)}.
    \label{Equation: E}
\end{eqnarray}
If the evaluation index $Ev(\alpha)$ of the community $\alpha$ is greater than the threshold $\theta$, the community $\alpha$ is judged as a social bot community. 
Otherwise, it is a human community, as shown in Figure~\ref{fig: UnDBot_architecture} C.3.

\begin{algorithm}[htbp]
    \SetAlgoLined
    \caption{The overall process of ~\framework.}\label{algorithm}
    \KwIn{Tweet information of users: $Pt$; Influence Information of tweets: $Inf$;
    following Information of users: $ff$; 
    threshold of the similarity for three features in modeling: $\xi$; 
    the stop threshold in random walk: $\rho$; the community evaluation threshold: $\theta$; the parameters for parallel Merge Operators: $p$; weighted parameters of influence and cohesion: $\pi$.
    }
    \BlankLine
    Construct the multi-relational graph $\mathcal{G}^r$ via Eq.~\ref{Equation: dij}, Eq.~\ref{Equation: delta ij} and Eq.~\ref{Equation: SF};

    $comms \leftarrow$ the initial division, $current\_num \leftarrow $ number of users;; \tcp{Each user forms a community}
    
    Initialize $edge$ with user pairs that connected in $\mathcal{G}^r$, $merge\_all \leftarrow False$;
    
    \While{ not $merge\_all$}{
        $max\_operate \leftarrow ( current\_num -1)*p $;\tcp{The maximum number of operator runs}
        
        \For{$(u,v)$ in $edge$}{
            $dH(u,v) \leftarrow \Delta_\mathcal{G}^{Mg}(\mathcal{T};{comm}_u,{comm}_v ) $ via Eq.~\ref{Equation: delta merge};
        }
        $op\_edge \leftarrow (u,v)$ with $dH(u,v)>median(dH>0)$;\tcp{Select edges that can be merged}
        
        \If{len($op\_edge$)>$max\_operate$}
        {
            $op\_edge \leftarrow (u,v)$ with top $max\_operate$ $dH$;
        }
        \If{len($op\_edge$)=0}
        {
            $merge\_all=True$;
        }
        
        $current\_num \leftarrow current\_num-$len($op\_edge$);\tcp{Update the number of communities}
        
        update $edge$ and $comms$;\tcp{Running Merge Operator on $op\_edge$}
    }
    
    Initialize $\mathcal{A} \leftarrow$Eq.~\ref{Equation: A}, $\mathcal{O},\mathcal{S}\leftarrow$Eq.~\ref{Equation: OS}; \tcp{Normalized}
    
    Initialize the distribution tensor $\mathcal{X}_{old}$ and $\mathcal{Y}_{old}$;
    
    \While{True}{
        $\mathcal{X}_{new} \leftarrow \mathcal{X}_{old} \mathcal{O} \mathcal{Y}_{old}^\top$, $\mathcal{Y}_{new} \leftarrow \mathcal{X}_{old} \mathcal{S} \mathcal{X}_{old}^\top$ via Eq.~\ref{Equation: XY};\\
        \eIf{$ \mid \mid {\mathcal{X}_{new}-\mathcal{X}_{old}} \mid \mid +\mid \mid \mathcal{Y}_{new}-\mathcal{Y}_{old} \mid \mid < \rho$}
        {
            Break; \tcp{The distribution tensor tends to be stable}
        }
        {
            $\mathcal{X}_{old} \gets \mathcal{X}_{new}$, $\mathcal{Y}_{old} \gets \mathcal{Y}_{new}$; \tcp{Update}
        }
    }

    $\mathcal{X} \gets \mathcal{X}_{old}$;

    \For{$comm$ in $comms$}{
        Calculate $x_\alpha$ and $H_\alpha$ via Eq.~\ref{Equation: influence} and Eq.~\ref{Equation: cohesion}; \tcp{Calculate community influence and cohesion}
        
        $E(comm) \leftarrow (1-\pi)x_\alpha+\pi H_\alpha$ via Eq.~\ref{Equation: E}; \tcp{Calculate the evaluation score}
	\eIf{$E(comm)>\theta$}
        {
            Users in $comm$ are social bots;
        }
        {
            Users in $comm$ are humans;
        }
    }  
\end{algorithm}

\subsection{Put them together}\par
Algorithm~\ref{algorithm} outlines the overall detection process of the proposed ~\framework. 
The process includes the construction of a multi-relational graph, the user community division using structural entropy, and a community labeling process that utilizes stationary distribution and community node entropy.
We identify three new types of social relationships based on the definition in Section~\ref{sec: Multi-relational Graph Construction}. 
Using these relationships, we construct a social user multi-relational graph based on the similarity of social behavior (Line 1).
We use the supplemented multi-relational graph-based structural information theory to construct a two-dimensional encoding tree with structural entropy minimization. 
To speed up, we run the \emph{Merge operator} in parallel that select no more than $max\_operate$ edges (Lines 5-12) to merge in each round until the \emph{Merge operator} fails (Lines 13-15). 
This process partitions user vertices into communities.
We then translate the multi-relational graph as a three-dimensional tensor (Line 19), as described in Section~\ref{sec: Binary Classification of Communities}, and calculate the stationary distribution using co-ranking (Lines 20-28). 
Once we obtain the user influence distribution vector $\mathcal{X}$ (Line 29), we calculate the influence $x_\alpha$ and cohesion $H_\alpha$ of each community to distinguish the social bot community from the human community (Lines 30-38).

\subsection{Time Complexity}
\revised{The entire \framework{} model is divided into three parts: multi-relational graph construction, user community division, and community binary classification.}
\revised{Among them,} modeling the social user multi-relational graph takes $O(N^2)$ \revised{,where $N$ is the total number of social users}. 
Constructing the structural entropy encoding tree takes $O(|E|log\frac{N-1}{1-p})$, \revised{where $|E|$ is the number of connected user pairs in graph $MG$}. 
Specifically, during each round of execution, the time complexity of calculating the difference in structural entropy for each pair of nodes after merging is $O(|E_i|)$, where $E_i$ represents the current inter-community edges. 
The maximum number of rounds of execution is $log_{1-p}\frac{1-p}{N-1}$, where $p$ is a hyperparameter controlling the parallel operation of Merge operators and can be regarded as a constant.
So the maximum time complexity of \revised{user community division} is $O(|E|*log_{1-p}\frac{1-p}{N-1})=O(|E|(1+logN))$. 
The MultiRank module (lines 21-30 in Algorithm~\ref{algorithm}) takes $O(k(N+3))$. 
$k$ is the number of iterations related to the graph structure in the MultiRank calculation process.
Generally, $k$ is a small number, so the time complexity of community binary classification is simplified to $O(N)$.
Consequently, the total time complexity of \framework{} is $O(N^2+N)+O(|E|(1+logN))=O(N^2+|E|(1+logN))$.
\section{Experimental Setup}
\label{sec: Experimental Setup}

\subsection{Software and Hardware}
We implement all models with Python 3.10.
All experiments are executed on a Linux server with a 128-core Intel Xeon Platinum 8336C CPU, 503GB of RAM, and an NVIDIA A800-SXM4-80GB. 
As for baselines, we utilize open-source implementation of node2vec from the library PecanPy~\footnote{\url{https://github.com/krishnanlab/PecanPy}}, as well as the codes provided by the authors for other baselines.

\subsection{Datasets}

\begin{table}[h]
    \caption{Statistics of datasets.}
    \aboverulesep=0ex
    \belowrulesep=0ex
    \centering
    \begin{tabular}{c|cccccc}
        \toprule
        Datasets & Human & Bot & User & Tweet & Edges & Feature Similarity\\
        \midrule
        \emph{Cresci-2015} & 1,950 & 3,351 & 5,301 & 2,827,757 & \makecell{
        \emph{U-F-U}: 726,323\\ \emph{U-T-U}: 2,970,033\\ \emph{U-I-U}: 412,741} & \makecell{0.3735\\0.2791\\0.2960}\\
        \midrule
        \emph{Cresci-2017} & 3,474 & 9,263 & 12,737 & 6,637,616 & \makecell{\emph{U-F-U}: 2,279,920\\ \emph{U-T-U}: 16,652,890\\ \emph{U-I-U}: 1,026,120} & \makecell{0.3215\\0.3323\\0.3133}\\
        \midrule
        \emph{Pronbots-2019} & \revised{1,481} & 17,882 & \revised{19,363} & \revised{231,224} & \makecell{\emph{U-F-U}: \revised{66,979}\\ \emph{U-T-U}: \revised{131,853,041}\\ \emph{U-I-U}: \revised{15,015}} & \makecell{\revised{0.2775} \\ \revised{0.7091} \\ \revised{0.7604} }\\
        \midrule
        \emph{Botwiki-2019} & \revised{65} & 698 & \revised{763} & \revised{357,851} & \makecell{\emph{U-F-U}: \revised{68}\\ \emph{U-T-U}: \revised{176,210} \\ \emph{U-I-U}: \revised{5,561} } & \makecell{\revised{0.2119} \\ \revised{0.6145} \\ \revised{0.1778} }\\
        \bottomrule
    \end{tabular}
    \label{tab: datasets}
\end{table}

We employ four publicly available social user datasets to assess the performance of the models in the context of social bot detection.
The original \emph{Cresci-2015} and \emph{Cresci-2017} datasets consist of user tweet data and user attribute data. 
As \framework{} requires user tweet information in the dataset to analyze posting type and posting influence distribution, the experiments conducted in this study on the \emph{Cresci-2017} dataset exclude users who have no tweet data. 
\revised{Furthermore, we collect tweets from users in the \emph{Botwiki-2019} and \emph{Pronbots-2019} datasets and add human users to build datasets required for the experiment to make it more consistent with real scenarios.}
Table~\ref{tab: datasets} presents various statistical information about the datasets we utilized and the number of connected edges for different relationships on the datasets under the graph construction method of \framework{}. 
Additionally, we calculated the average feature similarity based on three characteristics of user nodes, as defined in Section~\ref{sec: Multi-relational Graph Construction}. 
The detailed descriptions of these four datasets are as follows:
\begin{itemize}
    \item \textbf{\emph{Cresci-2015}} ~\cite{cresci2015fame}. 
    The Cresci-2015 dataset contains both benign Twitter users and fake accounts on Twitter.
    Among them, datasets \emph{E13} and \emph{TFP} contain 1481 and 469 active accounts from different social classes and backgrounds, manually verified as benign accounts by social scientists.
    Datasets \emph{FSF}, \emph{INT}, and \emph{TWT} are fake accounts purchased from different online marketplaces, containing 1169, 1337, and 845 users, respectively.
    
    \item \textbf{\emph{Cresci-2017}} ~\cite{cresci2017paradigm}. 
    The Cresci-2017 dataset contains benign users and three types of fake users on Twitter.
    The dataset \emph{genuine} contains 3474 benign accounts that have been manually verified.
    The dataset \emph{social spambot1} contains 991 automated accounts related to the mayoral election of Rome. 
    These social bots are similar to real accounts in profile, tweeting behavior, and friending behavior.
    The dataset \emph{social spambot2} contains 3457 social spambots related to the \emph{Talnts} mobile application.
    The dataset \emph{social spambot3} contains 464 social spambots related to advertising products for sale on amazon.com.
    The dataset \emph{fake followers} contains 3351 fake follower accounts purchased from different online marketplaces.
    The dataset \emph{traditional spambot1} contains 1000 traditional spambots from ~\cite{yang2013empirical}.
    
    \item \textbf{\emph{Pronbots-2019}}~\cite{yang2019arming}. 
    The \revised{original} Pronbots-2019 dataset contains 17882 Twitter bots related to advertising scam sites.
    The dataset was first shared by Andy Patel (github.com/r0zetta/pronbot2) and collected for training Botometer models.
    \revised{In this experiment, to make the dataset more consistent with real scenarios without destroying the structure of the original \emph{Pronbots-2019} dataset, we add 1481 human users from \emph{E13} of \emph{Cresci-2015}.}

    \item \textbf{\emph{Botwiki-2019}} ~\cite{yang2020scalable}. 
    The Botwiki-2019 dataset contains 698 self-identified bots from botwiki.org, which is an open catalog that includes bots designed for social media platforms, messaging apps, websites, and other online platforms. 
    The dataset retains only active users of the Twitter platform.
    \revised{In this experiment, to make the dataset more consistent with real scenarios without destroying the structure of the original \emph{Botwiki-2019} dataset, we add 65 human users from \emph{TFP} of \emph{Cresci-2015}.}
\end{itemize}

\subsection{Baselines and Variations}
\label{sec: baselines and variations}
\textbf{Baselines.}
To verify the effectiveness of the proposed \framework{} for social bot detection tasks, we compare it with various unsupervised machine learning-based and network embedding-based baselines.
All models based on unsupervised network embedding are constructed on the multi-relational graph described in Section~\ref{sec: Multi-relational Graph Construction}, followed by binary classification using K-means.
To demonstrate the effectiveness of multi-relational graph modeling, we also conduct experiments on homogeneous graphs for comparison.
The homogeneous graph consists of two types: one is constructed based on the action of following other users, which is contained in the original dataset Cresci-2015; the other follows anomaly detection~\cite{samariya2023comprehensive} and constructs the graph based on attribute similarity.

\begin{itemize}
    \item \textbf{K-means}~\cite{miller2014twitter}.
    K-means is a classic clustering method. 
    After predefining the number of clusters and the center point, users with similar characteristics are grouped into spherical clusters based on Euclidean distance. 
    The improved streaming algorithm~\cite{miller2014twitter} detects abnormal users (social bots) unsupervised by calculating the shortest distance between newly arrived users and core clusters.
    \item \textbf{DNA}~\cite{cresci2016dna}.
    This social user behavior modeling method is based on digital DNA technology, which simulates user behavior and interaction through strings. 
    The model uses the similarity of DNA sequences to discover user groups with highly similar behaviors and unsupervised identify social bot groups.
    
    \item \textbf{DeepWalk}~\cite{perozzi2014deepwalk}.
    DeepWalk is an unsupervised structure-preserving network embedding learning algorithm. 
    It uses the random walk method of depth-first traversal to sample neighbor nodes in the graph and utilizes the co-occurrence relationship between nodes in the graph to learn the vector representation of nodes.

    \item \textbf{LINE}~\cite{tang2015line}.
    LINE uses breadth-first sampling of neighbor nodes and defines the first-order and second-order similarity of nodes.
    The first-order and second-order similarities are optimized separately to learn vector representations of nodes.

    \item \textbf{Node2vec}~\cite{grover2016node2vec}.
    Node2vec is an extension of DeepWalk, incorporating a biased walk that comprehensively considers breadth-first and depth-first for domain sampling. 
    The objective is to optimize the likelihood of encountering neighboring nodes within the specified constraints of each vertex.

    \item \textbf{SDNE}~\cite{wang2016structural}.
    SDNE is an extension of LINE that utilizes an auto-encoder structure to simultaneously optimize first- and second-order similarity. 
    The learned node vector preserves the local and global structure.

    \item \textbf{GraRep}~\cite{cao2015grarep}.
    GraRep is a network embedding learning model that maps the k-order information of nodes to distinct subspaces, thereby computing k models individually, with each model capturing information of different orders.
    Embeddings learned by each model are concatenated as an embedding that captures all k-level information of a node.
    It excels in distinguishing the neighbors of different orders of nodes in representation learning and can be extended to capture neighbors of any order.

    \item \textbf{DNGR}~\cite{cao2016deep}.
    DNGR is a graph representation learning model that uses the random surfing model to extract the graph structure information directly.
    This allows it to more accurately and directly learn the graph structure information of weighted graphs. 
    Recovering vertex representations from Positive Pointwise Mutual Information (PPMI) matrices can capture potentially complex, non-linear relationships between different vertices.

    \item \textbf{HOPE}~\cite{ou2016asymmetric}.
    HOPE is a graph representation learning model based on matrix decomposition. 
    It leverages the relationship between graphs, matrices, and discrete Fourier transforms to effectuate the mapping of nodes from high-dimensional spaces to low-dimensional spaces. 
    This process preserves the integrity of high-order similarity relationships among nodes.
    
    \item \textbf{GAE}~\cite{zhang2020deep}.
    GAE is an unsupervised neural network for graph-structured representation learning. 
    This work uses GCN as its encoder to learn low-dimensional features of nodes and graphs. 
    The decoder reconstructs the original features from the embedded features and is trained to preserve useful structural features in a low-dimensional space for node classification.    
\end{itemize}\par

\noindent
\textbf{Variations.}
We generate several variants of the full \framework{} model to understand how each module works within the overall detection framework and to assess better how each module individually contributes to detection performance improvements. 
Because the modeling in the first step directly affects the effect of structural entropy user community division, the key part of \framework{} lies in the construction of multi-relational graphs and the evaluation indicators in community marking. 
We selectively enable or disable some of these modules for ablation studies. 
The details of these changes are described below:
\begin{itemize}
    \item \textbf{UnDBot-FT}.
    This variant only uses the \emph{posting type distribution} and \emph{posting influence} to construct the multi-relational graph. 
    When calculating the structural entropy, the weight ratio of the two types of edges is $1:1$.
    Only these two relationships are used to quantify community influence when calculating the stationary distribution.
    The community evaluation index weighs community influence and community node entropy.
    \item \textbf{UnDBot-FI}.
    This variant only uses the \emph{posting type distribution} and \emph{following-follower ratio} to construct the multi-relational graph. 
    When calculating the structural entropy, the weight ratio of the two types of edges is $1:1$. 
    Only these two relationships are used to quantify community influence when calculating the stationary distribution. 
    The community evaluation index weighs community influence and community node entropy.
    \item \textbf{UnDBot-TI}.
    This variant only uses the \emph{posting influence} and \emph{following-follower ratio} to construct the multi-relational graph. 
    When calculating the structural entropy, the weight ratio of the two types of edges is $1:1$. 
    Only these two relationships are used to quantify community influence when calculating the stationary distribution.
    The community evaluation index weighs community influence and community node entropy.
    \item \textbf{UnDBot-G}.
    This variant replaces the multi-relational graph $MG$ with a homogeneous graph $G$. 
    For datasets with primitive graph structures, the following relation in the dataset is used for modeling, while for other datasets, attribute similarity is used for modeling.
    \item \revised{\textbf{UnDBot-mg}.}
    \revised{This variant replaces the multi-relational graph $MG$ with another multi-relation graph $mg$. $mg$ is a multi-relational graph constructed from simplified user behavior similarities. It only contains the user's direct active behavior, including the number of tweets (\emph{U-twe-U}), the number of friends (\emph{U-fri-U}), and the number of likes (\emph{U-fav-U}). The specific construction method is the same as $MG$.}
\end{itemize}

\subsection{Model Running}
\label{sec: setting}
We use the following settings: similarity threshold (\revised{$\xi$=0.01 for \emph{Pronbots-2019} and} $\xi$=0.1 \revised{for other datasets}) of \emph{posting type distribution}, \emph{posting influence}, and \emph{follow-to-follower ratio} for multi-relational graph construction.
For user community division, we adopt the ratio of three edge weights ($1:1:1$) in the multi-relational graph structure entropy and the maximum scale ratio for parallel merge operators (\revised{$p=0.05$ for \emph{Pronbots-2019} and }$p=0.15$ \revised{for other datasets}).
For community binary classification, we set the stabilize threshold ($\rho$=0.004) for the distribution tensor, the weight of community influence (1-$\pi$), the weight of community entropy (\revised{$\pi=0.6$ for \emph{Botwiki-2019} and} $\pi=0.4$ \revised{for other datasets}), and the threshold of evaluation index for communities ($\theta=0.60$ for \emph{Pronbots-2019}, $\theta=0.55$ for \emph{Botwiki-2019}, and $\theta=1$ for other datasets ).
For baselines, we uniformly set the epochs in network embedding learning (50), the number of clusters in K-means (2), and the embedding size (128).

\subsection{Evaluation Metrics}
As the social bot detection task is essentially a binary classification problem, and the number of benign accounts and malicious accounts in the data set is relatively balanced, we use the ACCURACY rate to evaluate the overall performance of the classifier:
\begin{equation}
    ACC=\frac{TP+TN}{TP+TN+FP+FN},
\end{equation}
where $TP$ is True Positive, $TN$ is True Negative, $FP$ is False Positive, $FN$ is False Negative.
In addition, to measure the Pertinency Factor of the social bot detection model, we use the Precision rate as an evaluation index, that is, the actual proportion of social bots in the samples judged as social bots:
\begin{equation}
    Precision=\frac{TP}{TP+FP}.
\end{equation}
To assess the comprehensiveness of social bot detection, we use the Recall rate evaluation index, that is, the proportion of correctly identified in the social bot samples.
\begin{equation}
    Recall=\frac{TP}{TP+FN}.
\end{equation}
Furthermore, we employ a balanced F-score for a more comprehensive measure of effectiveness.
\begin{equation}
    F1=\frac{2 \times Precision \times Recall}{Precision+Recall}.
\end{equation}
\section{Results And Discussion}
\label{sec: Results And Discussion}

\begin{sidewaystable}[thp]
    \setlength{\belowcaptionskip}{16.cm}
    \aboverulesep=0ex
    \belowrulesep=0ex
    \centering
    \renewcommand\arraystretch{1.3}
    \caption{Comparison of the average $ACC$, $Precision$, $Recall$, and $F1$ across different methods for social bot detection (unit:\%). The best results are bolded, and the second-best results are underlined.}
    \label{tab: results}
    \setlength{\tabcolsep}{1mm}{
    \begin{tabular}{c|cccc|cccc|cccc|cccc}
        \toprule
        
        \multirow{2}*{Method} & \multicolumn{4}{c|}{\multirow{1}*{\textbf{Cresci-2015}}} & \multicolumn{4}{c|}{\multirow{1}*{\textbf{Cresci-2017}}} & \multicolumn{4}{c|}{\multirow{1}*{\textbf{Pronbots-2019}}} & \multicolumn{4}{c}{\multirow{1}*{\textbf{botwiki-2019}}}\\

       \cline{2-17}

        & {\textbf{ACC}} & {\textbf{P}} & {\textbf{R}} & \multicolumn{1}{c|}{\textbf{F1}} & {\textbf{ACC}} & {\textbf{P}} & {\textbf{R}} & \multicolumn{1}{c|}{\textbf{F1}} & {\textbf{ACC}} & {\textbf{P}} & {\textbf{R}} & \multicolumn{1}{c|}{\textbf{F1}} & {\textbf{ACC}} & {\textbf{P}} & {\textbf{R}} & {\textbf{F1}}\\
        
        \hline
        \textbf{K-means} & 76.51 & 73.22 & 99.07 & 84.21 & 73.62 & 76.86 & 93.31  & 84.29 & \revised{80.28} & \revised{97.71} & \revised{80.53} & \revised{88.30} & \revised{77.85} & \revised{95.84} & \revised{79.23} & \revised{86.75} \\
        
        \textbf{DNA} & 36.22 & 0.87 & 32.95  & 1.69 & 33.09 & 14.39 & 93.71 & 23.34 & \revised{7.80} & \revised{66.67} & \revised{0.34} & \revised{0.67} & \revised{82.83} & \revised{96.85} & \revised{83.95} & \revised{89.95} \\

        \textbf{GAE} & 74.83 & 72.09 & 98.21 & 83.15 & 73.80 & 95.58 & 76.17 & 84.53 & \revised{53.68} & \revised{94.99} & \revised{52.62} & \revised{67.73} & \revised{64.74} & \revised{96.52} & \revised{63.75} & \revised{76.79} \\

        \hline

        $\textbf{DeepWalk}_{MG}$ & \underline{87.75} & 92.93 & 87.25  & \underline{90.00} & \underline{90.33} & 89.38 & 98.39 & \underline{93.67} & \revised{91.49} & \revised{99.98} & \revised{90.80} & \revised{95.17} & \revised{82.18} & \revised{90.72} & \revised{89.68} & \revised{90.20}  \\

        $\textbf{LINE}_{MG}$ & 55.30 & 62.01 & 78.38  & 69.01 & 72.72 & 72.72 & \textbf{99.99} & 84.20 & \revised{92.36} & \revised{92.36} & \revised{\textbf{100}} & \revised{\underline{96.03}} & \revised{71.17} & \revised{97.05} & \revised{70.63} & \revised{81.76} \\

        $\textbf{Node2vec}_{MG}$ & 73.85 & 89.25 & 66.67 & 76.32 & 90.32 & 89.38 & 98.38 & 93.66 & \revised{91.24} & \revised{99.50} & \revised{90.97} & \revised{95.05} & \revised{82.18} & \revised{90.72} & \revised{89.68} & \revised{90.20} \\

        $\textbf{SDNE}_{MG}$ & 74.54 & 98.32 & 60.76  & 75.09 & 66.59 & 83.45 & 72.04 & 75.37 & \revised{91.31} & \revised{99.69} & \revised{90.87} & \revised{95.08} & \revised{71.69} & \revised{\textbf{100}} & \revised{69.05} & \revised{81.69} \\

        $\textbf{GraRep}_{MG}$ & 80.04 & \underline{98.89} & 69.20 & 81.43 & 61.63 & \textbf{99.70} & 47.38 & 64.24 & \revised{91.51} & \revised{\textbf{100}} & \revised{90.81} & \revised{95.18} & \revised{84.67} & \revised{99.15} & \revised{83.95} & \revised{90.92} \\

        $\textbf{DNGR}_{MG}$ & 83.06 & \textbf{98.92} & 74.00 & 84.67 & 73.71 & 73.59 & 99.72 & 84.67 & \revised{\underline{92.54}} & \revised{99.56} & \revised{92.33} & \revised{95.81} & \revised{82.18} & \revised{90.61} & \revised{89.83} & \revised{90.22} \\

        $\textbf{HOPE}_{MG}$ & 51.01 & 67.60 & 43.21  & 52.72 & 60.21 & 87.40 & 52.92 & 65.92 & \revised{81.66} & \revised{\textbf{100}} & \revised{80.14} & \revised{88.98} & \revised{82.83} & \revised{90.67} & \revised{90.54} & \revised{90.61} \\

        \hline
        $\textbf{DeepWalk}_{G}$ & 76.08 & 79.80 & 82.88 & 81.41 & 62.71 & 73.52 & 79.41 & 76.33 & \revised{63.14} & \revised{99.47} & \revised{60.41} & \revised{75.17} & \revised{77.72} & \revised{99.81} & \revised{75.79} & \revised{86.16} \\

        $\textbf{LINE}_{G}$ & 64.80 & 66.67 & 88.61 & 76.09 & 58.80 & 96.78 & 47.23 & 63.48 & \revised{92.34} & \revised{92.35} & \revised{\underline{99.99}} & \revised{96.02} & \revised{61.07} & \revised{98.08} & \revised{58.59} & \revised{73.36} \\

        $\textbf{Node2vec}_{G}$ & 64.34 & 63.94 & 99.99 & 78.00 & 50.42 & 71.77 & 57.06 & 63.57 & \revised{66.99} & \revised{99.23} & \revised{64.76} & \revised{78.37} & \revised{82.44} & \revised{100} & \revised{80.80} & \revised{89.38} \\

        $\textbf{SDNE}_{G}$ & 68.63 & 66.84 & \textbf{100} & 80.12 & 72.20 & 77.46 & 89.88 & 82.96 & \revised{87.48} & \revised{91.96} & \revised{94.72} & \revised{93.32} & \revised{53.87} & \revised{86.34} & \revised{58.88} & \revised{70.01} \\

        $\textbf{GraRep}_{G}$ & 67.65 & 66.15 & \textbf{100} & 79.62 & 57.48 & 71.78 & 72.39 & 72.08 & \revised{50.04} & \revised{\textbf{100}} & \revised{45.90} & \revised{62.92} & \revised{66.71} & \revised{\textbf{100}} & \revised{63.61} & \revised{77.76} \\

        $\textbf{DNGR}_{G}$ & 60.24 & 64.72 & 81.58 & 72.17 & 75.80 & 75.82 & \underline{99.96} & 86.23 & \revised{92.36} & \revised{92.37} & \revised{99.98} & \revised{\underline{96.03}} & \revised{\underline{90.96}} & \revised{91.55} & \revised{\textbf{99.28}} & \revised{\underline{95.26}} \\

        $\textbf{HOPE}_{G}$ & 66.67 & 65.47 & \textbf{100} & 79.14 & 50.40 & 75.92 & 50.65 & 60.76 & \revised{50.46} & \revised{94.03} & \revised{49.50} & \revised{64.85} & \revised{53.34} & \revised{\textbf{100}} & \revised{48.99} & \revised{65.77} \\
        \hline
        
        \textbf{~\framework} & \textbf{89.12} & 96.11 & 86.27 & \textbf{90.93} & \textbf{93.46} & \underline{98.07} & 92.83 & \textbf{95.38} & \revised{\textbf{92.68}} & \revised{92.75} & \revised{99.88} & \revised{\textbf{96.18}} & \revised{\textbf{93.18}} & \revised{97.78} & \revised{\underline{94.70}} & \revised{\textbf{96.22}} \\

        \bottomrule
    \end{tabular}
    }
\end{sidewaystable}

This section conducts several experiments to evaluate the performance of ~\framework. 
We mainly answer the following questions:
\begin{itemize}
    \item Q1:
    How do different models perform under different datasets, i.e., the algorithm's effectiveness (Section~\ref{sec: effectiveness})?
    \item Q2:
    How do the multi-relational graph ($MG$), each individual edge within \framework{}, and the hyperparameters contribute to the overall effectiveness (Section~\ref{sec: ablation})?
    \item Q3:
    How does \framework{} work in terms of interpretability (Section~\ref{sec: interpretability})?
    \item Q4:
    How efficient different baselines and \framework{} can attain (Section~\ref{sec: efficient})?
    \item Q5:
    How to explore the detection result and gain deeper insights into the distinctions in human perspective (Section~\ref{sec: case study}).
\end{itemize}

\subsection{Model Effectiveness}
\label{sec: effectiveness}
In this section, we conduct experiments on four publicly available social bot detection datasets (Cresci-2015, Cresci-2017, Pronbots-2019, and Botwiki-2019) to evaluate the effectiveness of our social bot detection model \framework{}.
Table~\ref{tab: results} presents the test accuracy (ACC), precision (P), recall (R), and F1 score of the unsupervised baselines and our model \framework{}.
All baselines that rely on unsupervised graph representation learning are evaluated on the social user multi-relational graph (models denoted as $baseline_{MG}$) and the homogeneous graph (models denoted as $baseline_{G}$ ).
Our model demonstrates a generally superior level of effectiveness compared to the baseline methods.
\framework{} outperforms all baselines on the evaluation indices ACC and F1 for the Cresci-2015, Cresci-2017, Pronbots-2019, and botwiki-2019 datasets.
Although it does not achieve the highest precision and recall on three datasets, it still has an advantage over most baselines and achieves a balance between precision and recall. 
Overall, \framework{} significantly outperforms most unsupervised social bot detection methods regarding detection accuracy, precision, recall, and F1.
This demonstrates the effectiveness of \framework{} in the social bot detection scenario.
Compared with existing unsupervised social bot detection methods (K-means and DNA), the detection accuracy on the four datasets is increased by at least 12.61$\%$, 19.66$\%$, \revised{12.40$\%$ and 10.36$\%$} respectively.
Compared to methods based on unsupervised graph representation learning, the detection accuracy is improved by at least 1.37$\%$, 3.13$\%$, \revised{0.14$\%$, and 2.23$\%$} on the four datasets respectively.\par

For the baselines, the first two methods in Table~\ref{tab: results} are existing unsupervised models for the social bot detection task, while the third method is an unsupervised model based on autoencoders.
\emph{DNA} consistently performs the worst, even lower than 50$\%$, except in the botwiki-2019 dataset.
This is because \emph{DNA} encodes social user behavior into sequences of behavior strings using genetic techniques and identifies social bots based on the presence of common behavior substrings.
This method only encodes behaviors related to tweeting and does not consider other social behaviors of users. 
Additionally, quantifying behavior similarity based on the longest common behavior substring is prone to false positives due to the influence of the order of action behaviors.
\emph{K-means} treats social bot identification as anomaly detection and labels users whose feature properties deviate from the core population as social bots.
Although \emph{K-means} demonstrates higher detection accuracy compared to \emph{DNA}, it is sensitive to the initial clustering centers and struggles to handle noisy data effectively.
Furthermore, clustering social users solely based on directly obtained user features easily results in highly disguised social bots being wrongly assigned to human clusters, thus affecting accuracy. 
Therefore, the accuracy of \emph{K-means}  is only around 75$\%$ in the four dataset scenarios.
\emph{GAE} combines GCNs and autoencoders to learn low-dimensional feature embeddings for unlabeled samples. 
However, its performance is influenced by the choice of autoencoder and graph structure, and the overall effect is not as good as \emph{K-means}.
In comparison, the advantage of \framework{} lies in its ability to analyze user behavior characteristics while considering the network structure. 
Unlike \emph{K-means} and \emph{DNA}, which directly determine individual social bots based on the numerical values of raw feature values or action sequences, \framework{} identifies social bot communities through behavioral similarity within the entire social network. 
It comprehensively considers the relationships with other users, enhancing detection effectiveness.\par

\begin{figure}[thp]
   \centering
    \subfigure[${DeepWalk}_{MG}$ on Cresci15.]{
        \label{fig: DeepWalk_cresci15_X}
        \begin{minipage}[t]{0.32\linewidth}
            \centering
            \includegraphics[width=1\linewidth]{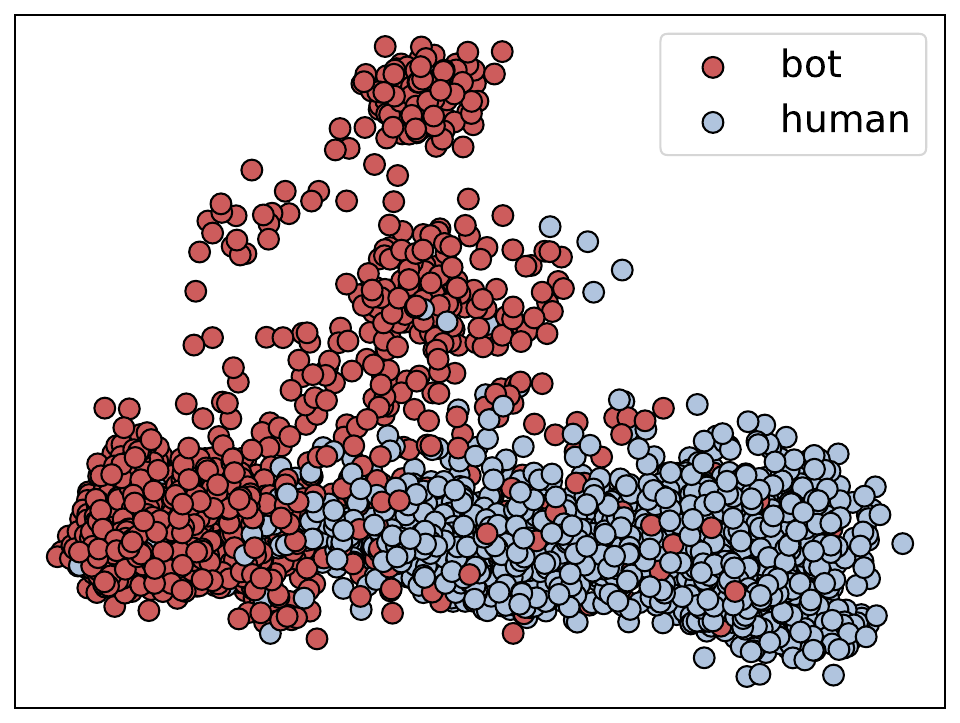}
        \end{minipage}%
    }
    \subfigure[${LINE}_{MG}$ on Cresci15.]{
        \label{fig: LINE_cresci15_X}
        \begin{minipage}[t]{0.32\linewidth}
            \centering
            \includegraphics[width=1\linewidth]{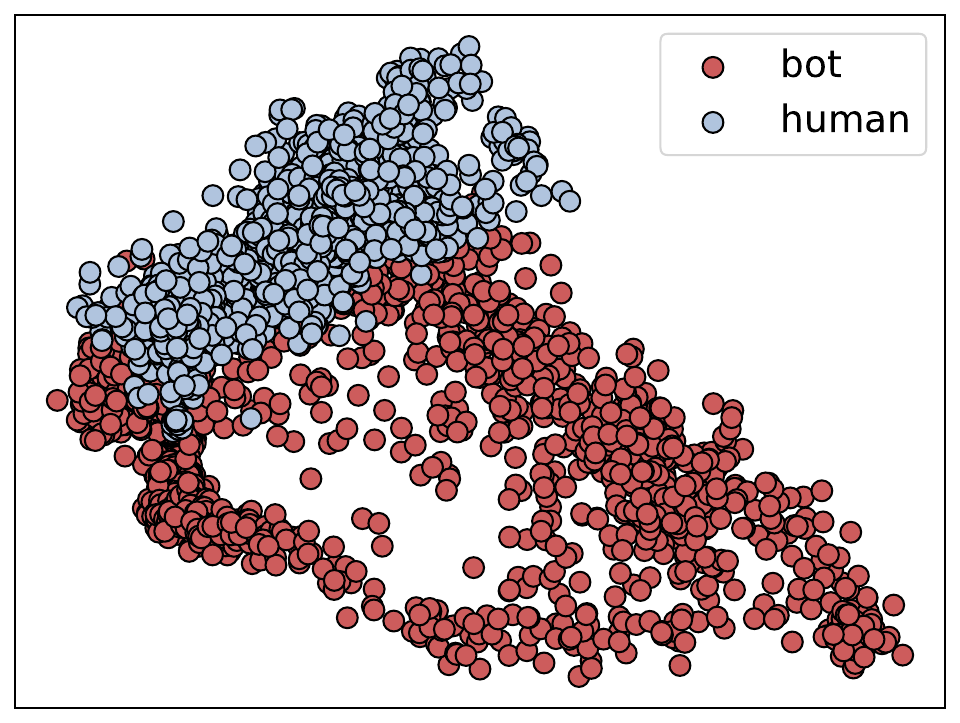}
        \end{minipage}%
    }
    \subfigure[${SDNE}_{MG}$ on Cresci15.]{
        \label{fig: SDNE_cresci15_X}
        \begin{minipage}[t]{0.32\linewidth}
            \centering
            \includegraphics[width=1\linewidth]{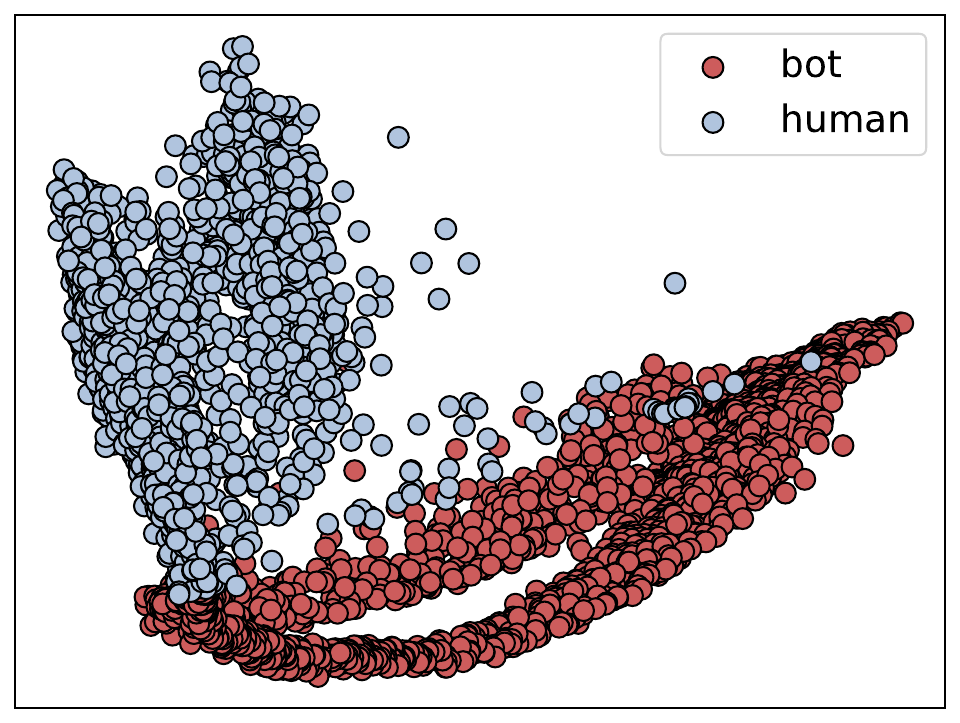}
        \end{minipage}%
    }

    \subfigure[${GraRep}_{MG}$ on Cresci15.]{
        \label{fig: GraRep_cresci15_X}
        \begin{minipage}[t]{0.32\linewidth}
            \centering
            \includegraphics[width=1\linewidth]{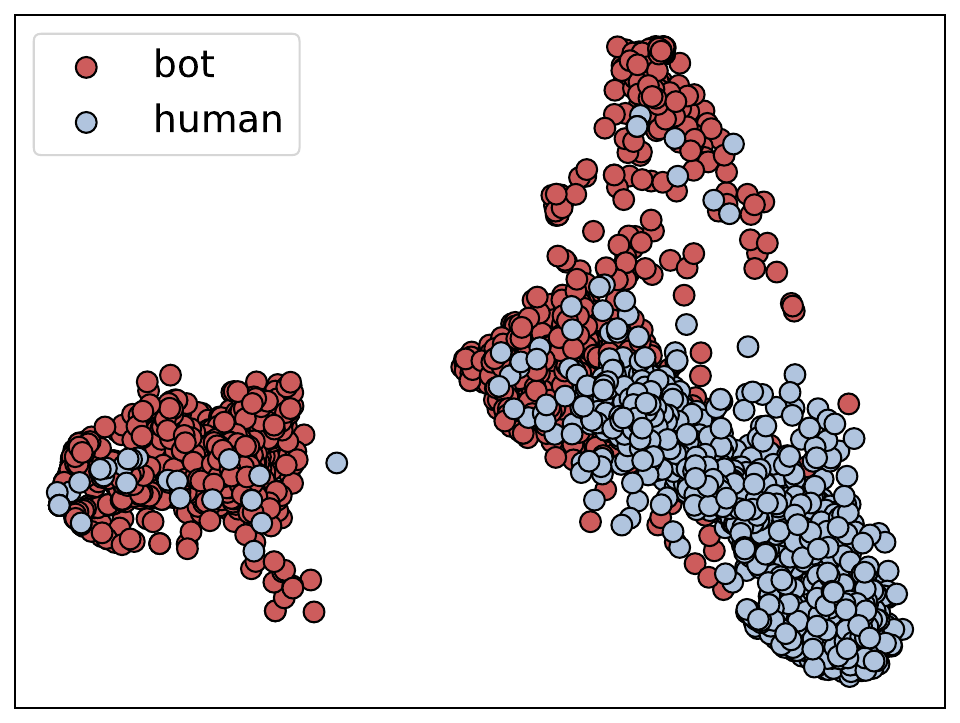}
        \end{minipage}%
    }
    \subfigure[${DNGR}_{MG}$ on Cresci15.]{
        \label{fig: DNGR_cresci15_X}
        \begin{minipage}[t]{0.32\linewidth}
            \centering
            \includegraphics[width=1\linewidth]{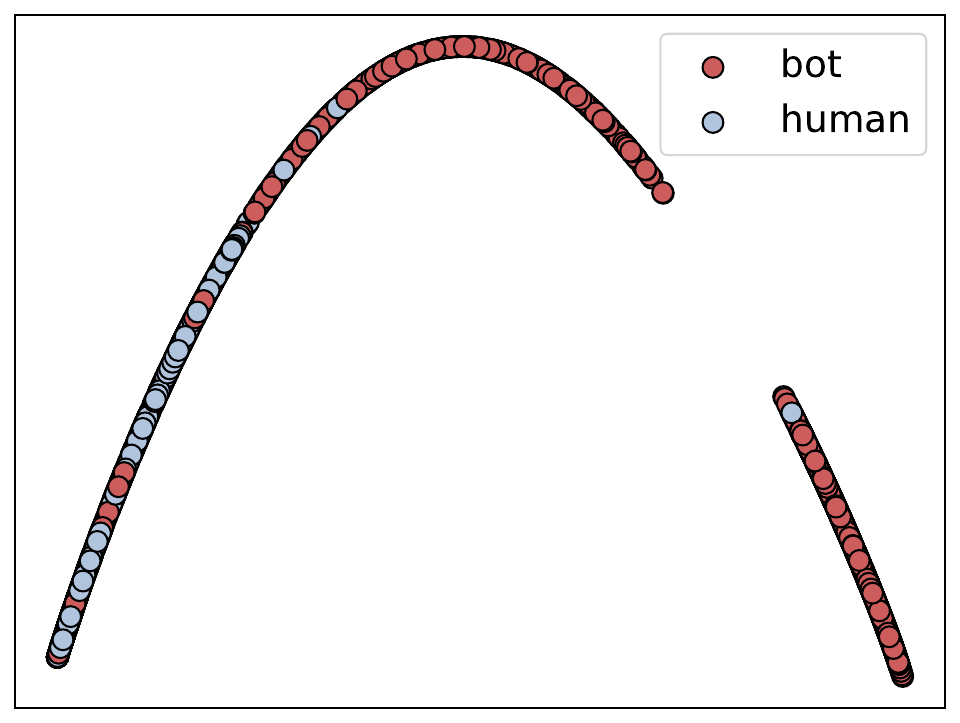}
        \end{minipage}%
    }
    \subfigure[${HOPE}_{MG}$ on Cresci15.]{
        \label{fig: HOPE_cresci15_X}
        \begin{minipage}[t]{0.32\linewidth}
            \centering
            \includegraphics[width=1\linewidth]{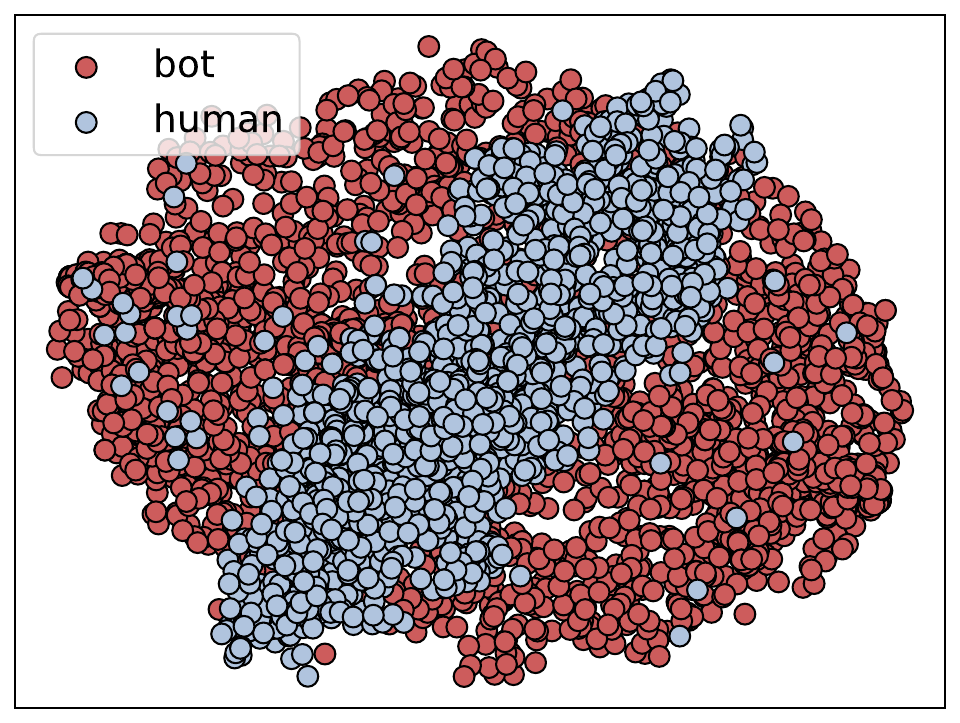}
        \end{minipage}%
    }  

    \subfigure[${DeepWalk}_{MG}$ on Cresci17.]{
        \label{fig: DeepWalk_cresci17_X}
        \begin{minipage}[t]{0.32\linewidth}
            \centering
            \includegraphics[width=1\linewidth]{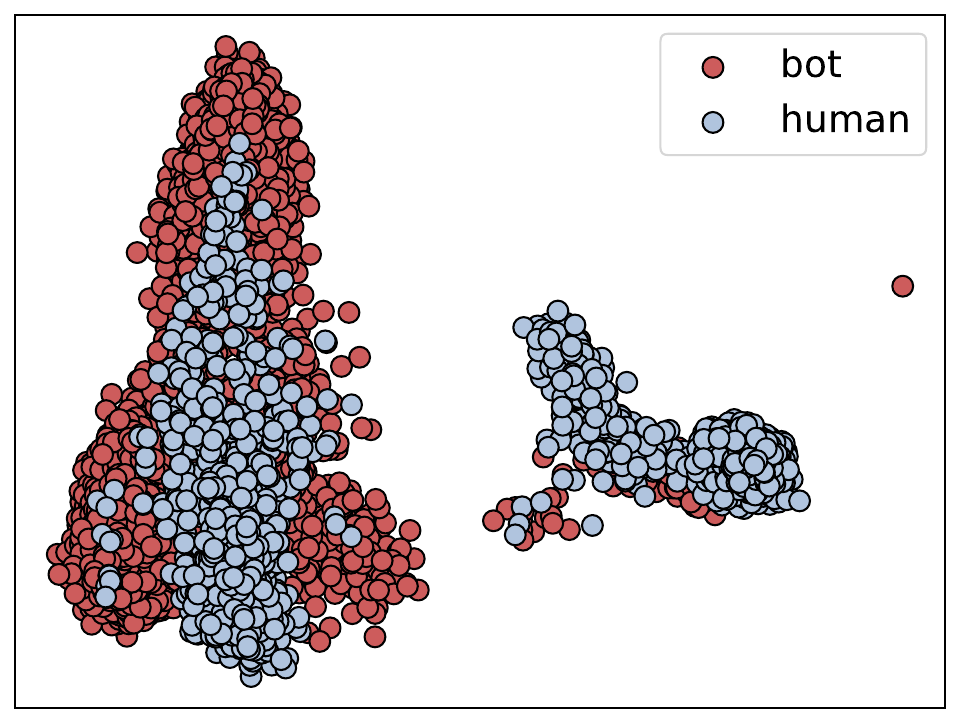}
        \end{minipage}%
    }
    \subfigure[${LINE}_{MG}$ on Cresci17.]{
        \label{fig: LINE_cresci17_X}
        \begin{minipage}[t]{0.32\linewidth}
            \centering
            \includegraphics[width=1\linewidth]{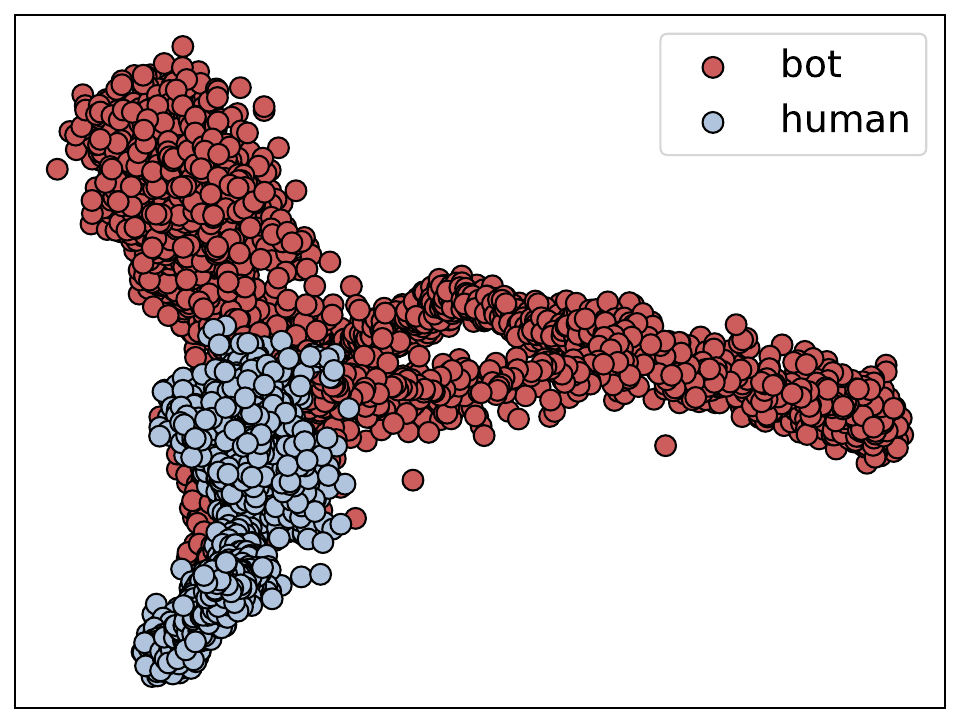}
        \end{minipage}%
    }
    \subfigure[${SDNE}_{MG}$ on Cresci17.]{
        \label{fig: SDNE_cresci17_X}
        \begin{minipage}[t]{0.32\linewidth}
            \centering
            \includegraphics[width=1\linewidth]{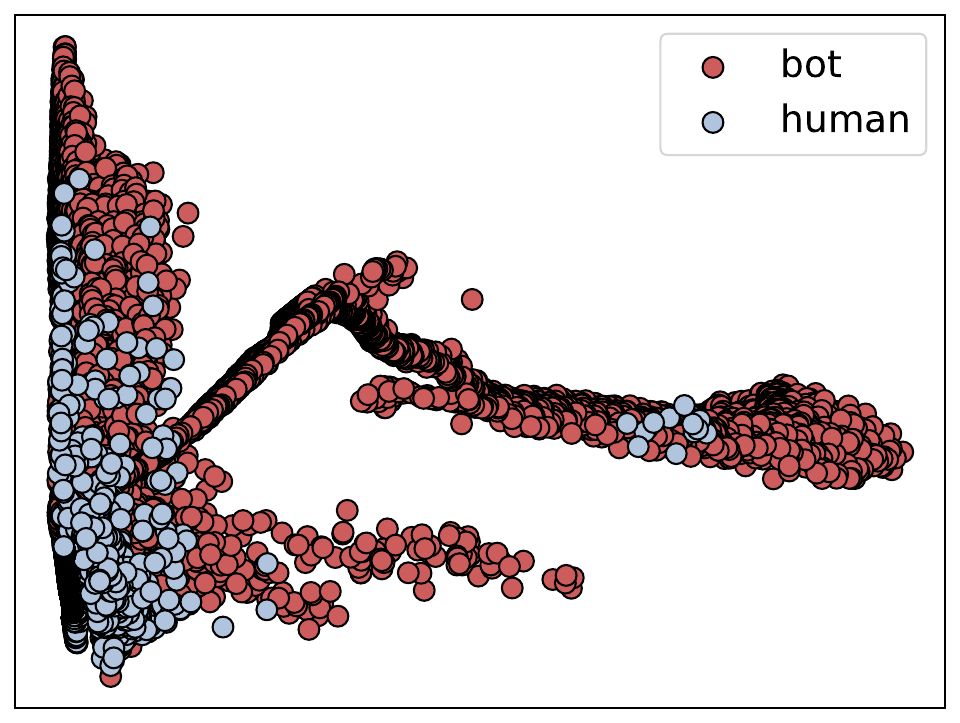}
        \end{minipage}%
    }

    \subfigure[${GraRep}_{MG}$ on Cresci17.]{
        \label{fig: GraRep_cresci17_X}
        \begin{minipage}[t]{0.32\linewidth}
            \centering
            \includegraphics[width=1\linewidth]{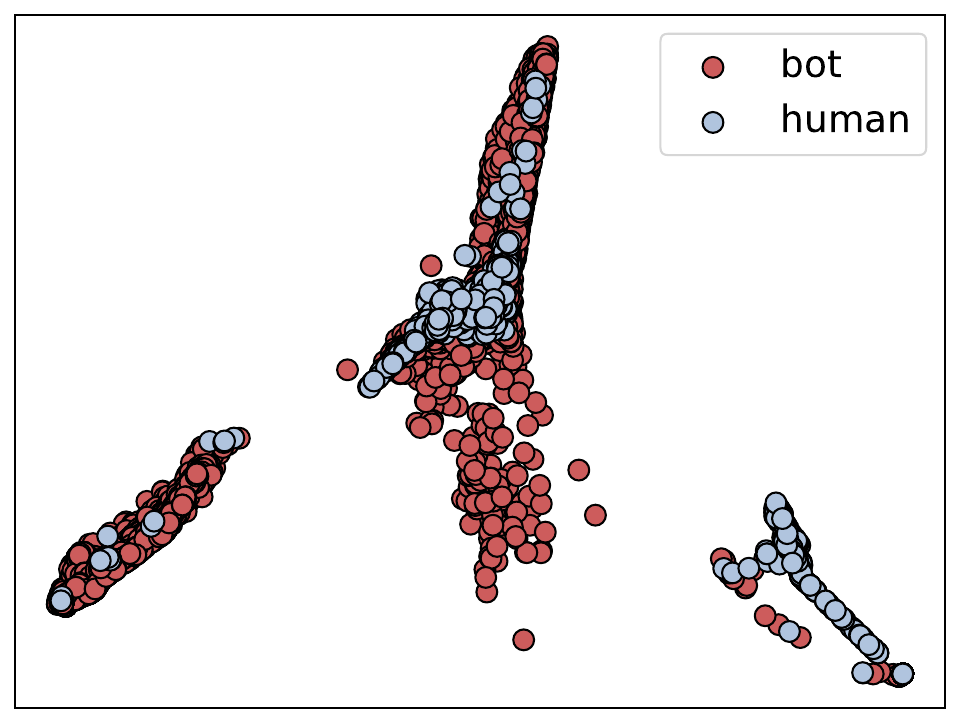}
        \end{minipage}%
    }
    \subfigure[${DNGR}_{MG}$ on Cresci17.]{
        \label{fig: DNGR_cresci17_X}
        \begin{minipage}[t]{0.32\linewidth}
            \centering
            \includegraphics[width=1\linewidth]{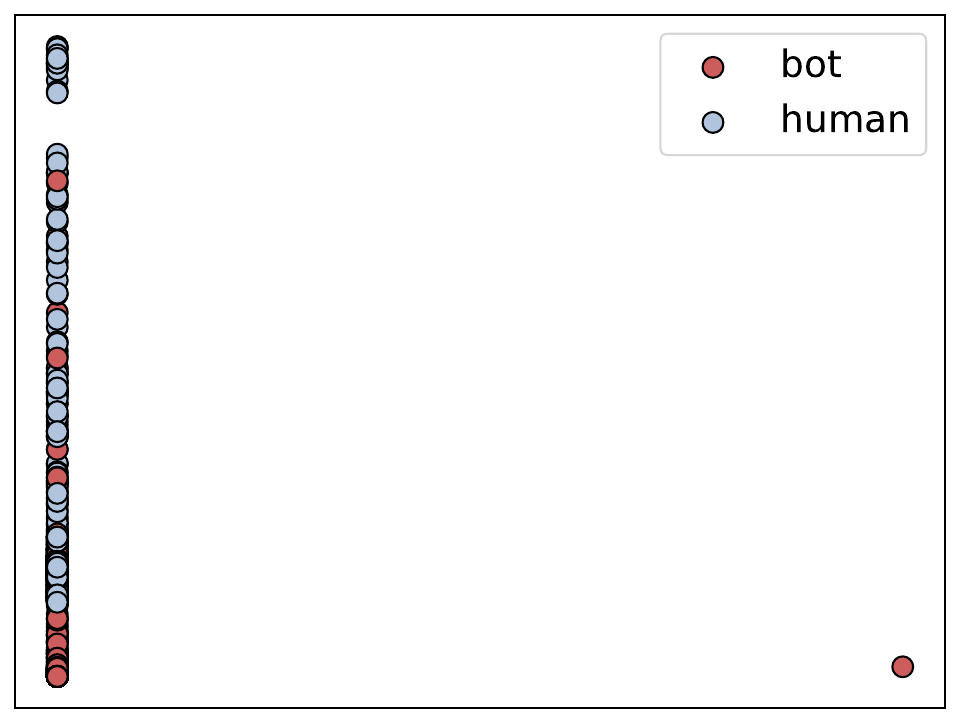}
        \end{minipage}%
    }
    \subfigure[${HOPE}_{MG}$ on Cresci17.]{
        \label{fig: HOPE_cresci17_X}
        \begin{minipage}[t]{0.32\linewidth}
            \centering
            \includegraphics[width=1\linewidth]{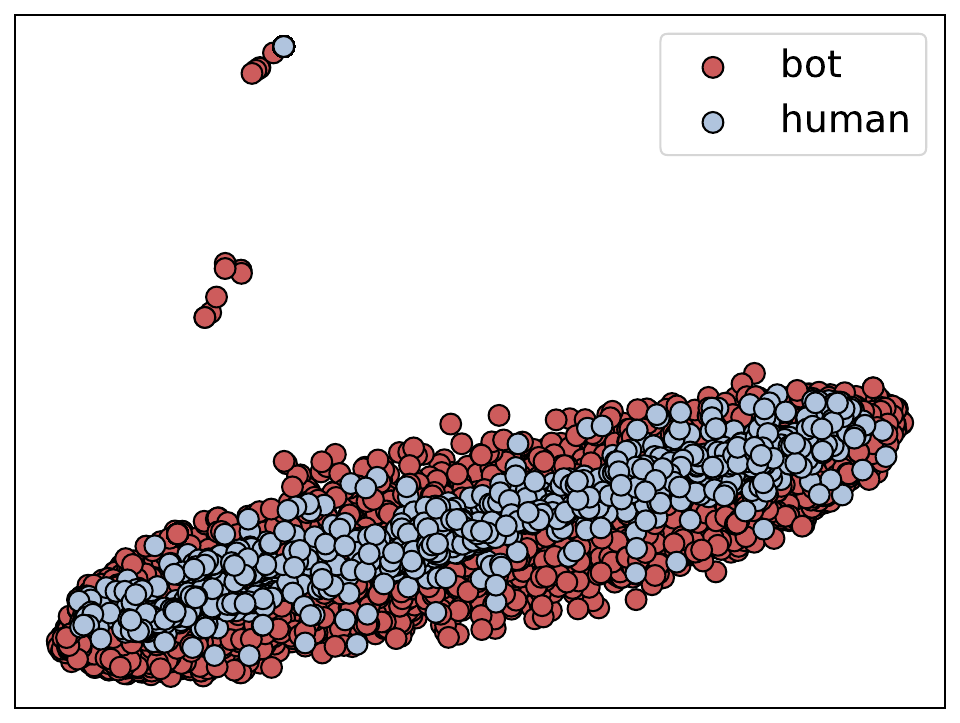}
        \end{minipage}%
    }
    
    \caption{Distribution of user embeddings generated by unsupervised graph learning models on the Cresci-2015 and Cresci-2017 datasets.}
    \label{fig: Cresci-2015 Cresci-2017}
\end{figure}

\begin{figure}[thp]
   \centering
    \subfigure[\revised{${DeepWalk}_{MG}$ on Botwiki19.}]{
        \label{fig: DeepWalk_botwiki19_X}
        \begin{minipage}[t]{0.32\linewidth}
            \centering
            \includegraphics[width=1\linewidth]{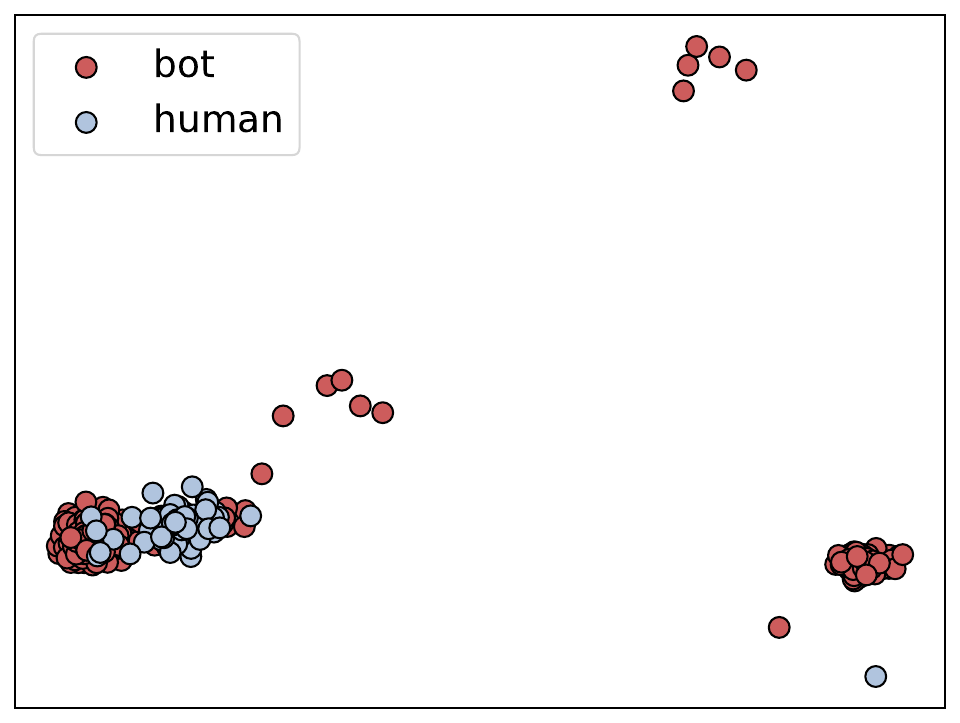}
        \end{minipage}%
    }
    \subfigure[\revised{${LINE}_{MG}$ on Botwiki19.}]{
        \label{fig: LINE_botwiki19_X}
        \begin{minipage}[t]{0.32\linewidth}
            \centering
            \includegraphics[width=1\linewidth]{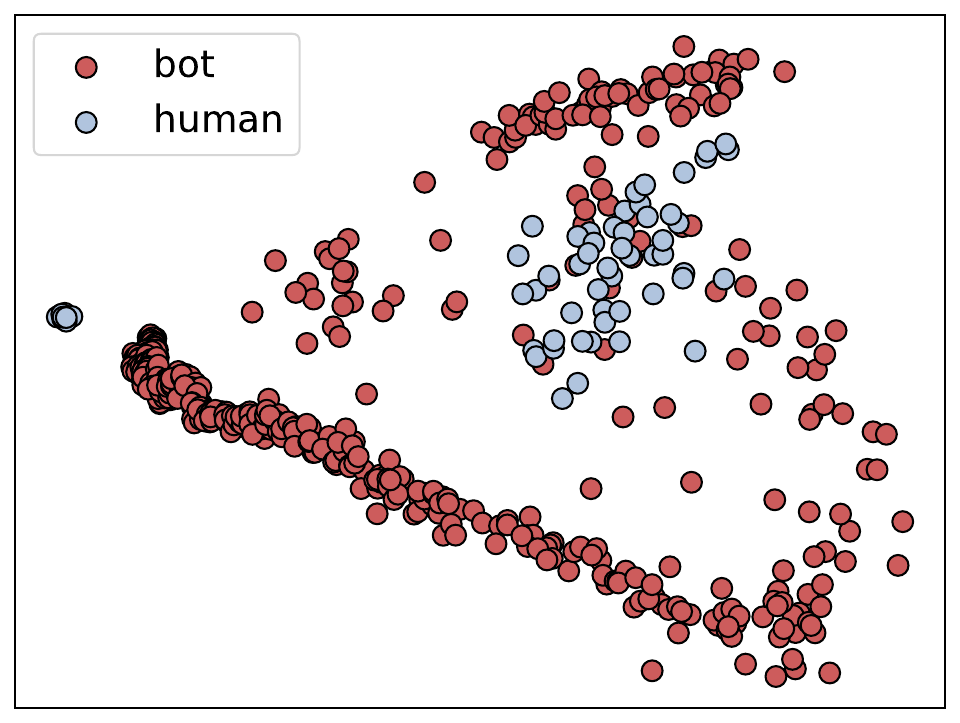}
        \end{minipage}%
    }
    \subfigure[\revised{${SDNE}_{MG}$ on Botwiki19.}]{
        \label{fig: SDNE_botwiki19_X}
        \begin{minipage}[t]{0.32\linewidth}
            \centering
            \includegraphics[width=1\linewidth]{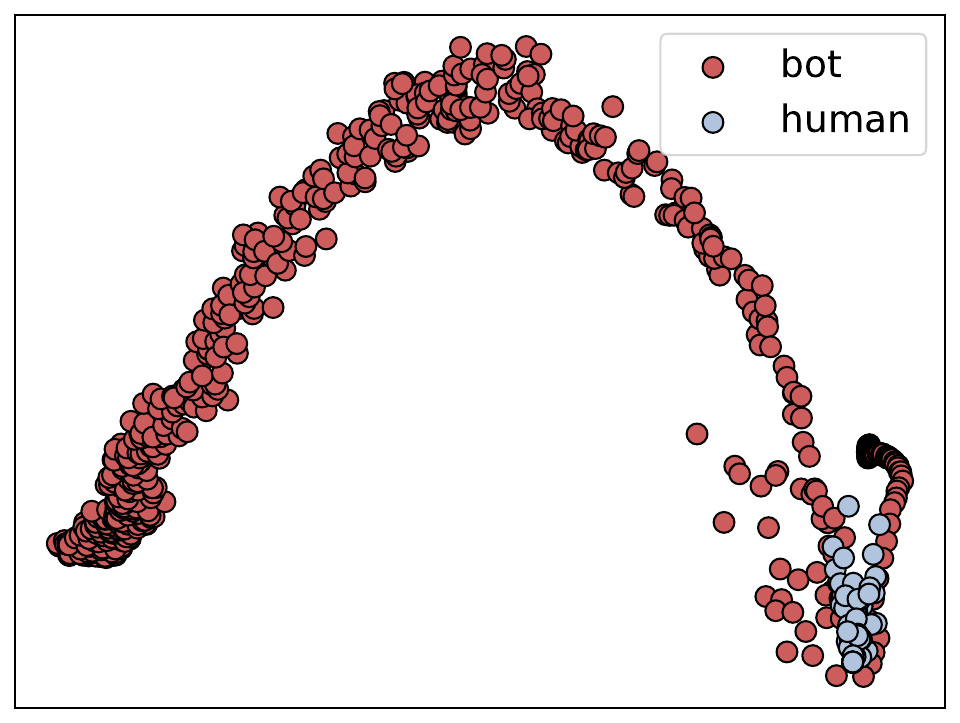}
        \end{minipage}%
    }

    \subfigure[\revised{${GraRep}_{MG}$ on Botwiki19.}]{
        \label{fig: GraRep_botwiki19_X}
        \begin{minipage}[t]{0.32\linewidth}
            \centering
            \includegraphics[width=1\linewidth]{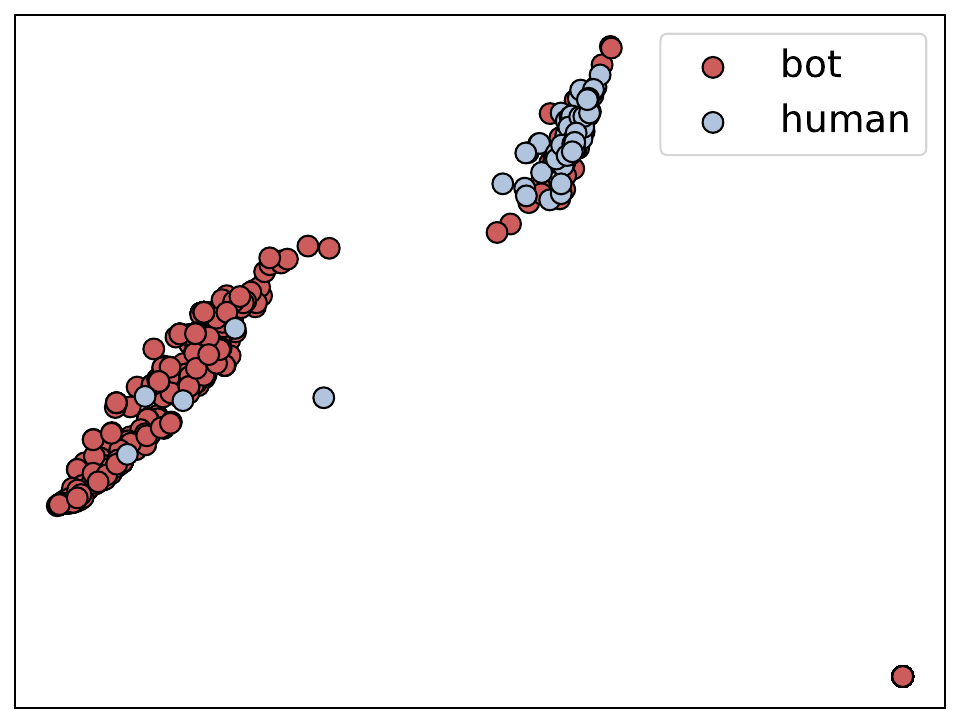}
        \end{minipage}%
    }
    \subfigure[\revised{${DNGR}_{MG}$ on Botwiki19.}]{
        \label{fig: DNGR_botwiki19_X}
        \begin{minipage}[t]{0.32\linewidth}
            \centering
            \includegraphics[width=1\linewidth]{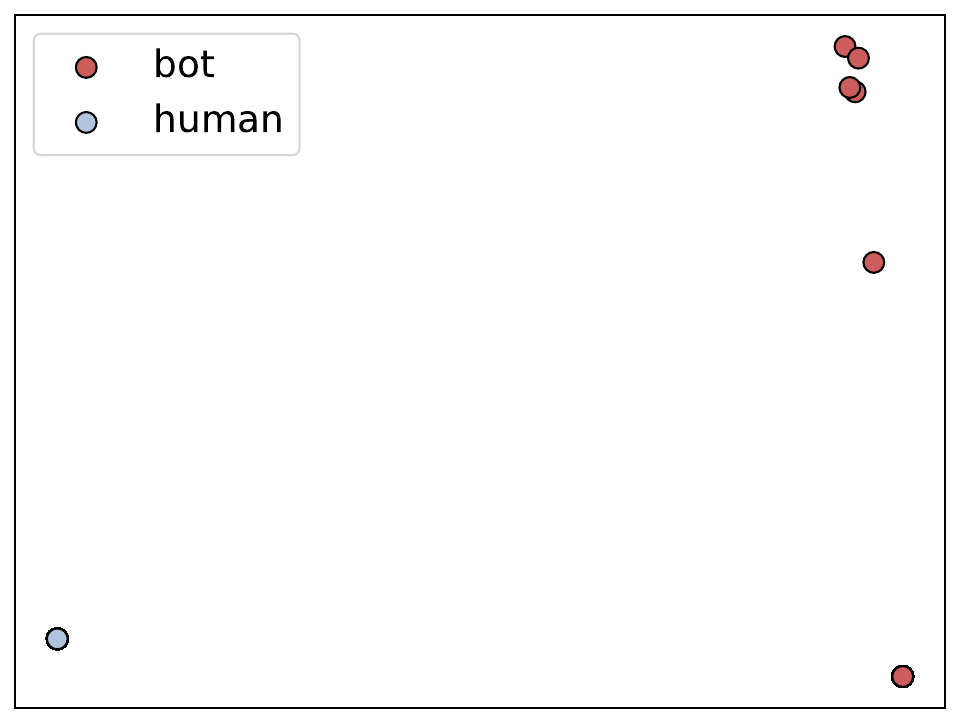}
        \end{minipage}%
    }
    \subfigure[\revised{${HOPE}_{MG}$ on Botwiki19.}]{
        \label{fig: HOPE_botwiki19_X}
        \begin{minipage}[t]{0.32\linewidth}
            \centering
            \includegraphics[width=1\linewidth]{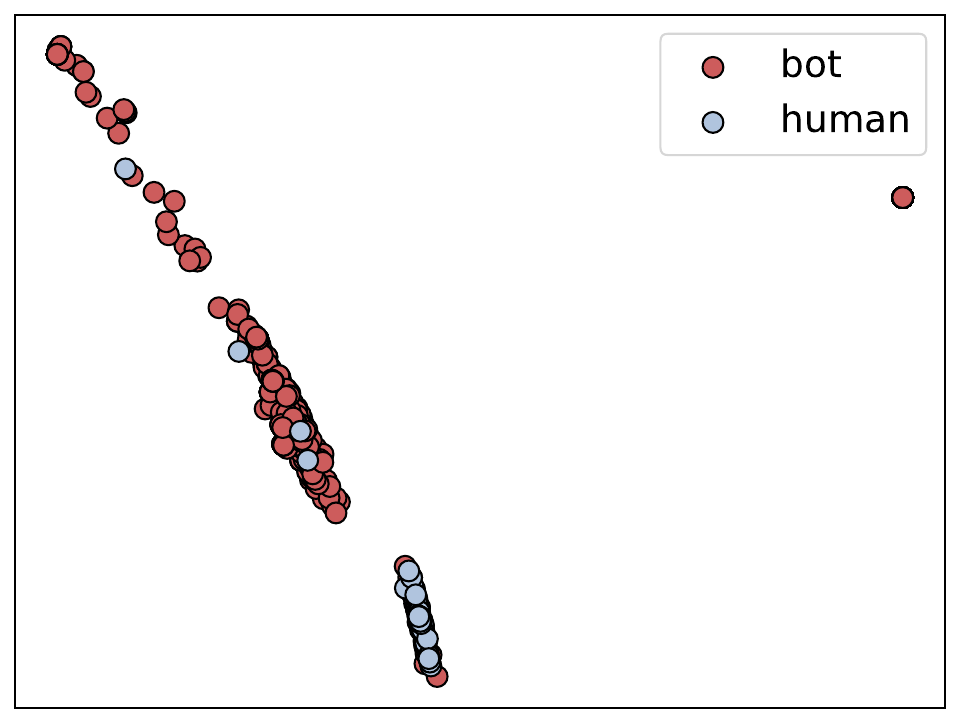}
        \end{minipage}%
    }

    \subfigure[\revised{${DeepWalk}_{MG}$ on Pronbots19.}]{
        \label{fig: DeepWalk_pronbots19_X}
        \begin{minipage}[t]{0.32\linewidth}
            \centering
            \includegraphics[width=1\linewidth]{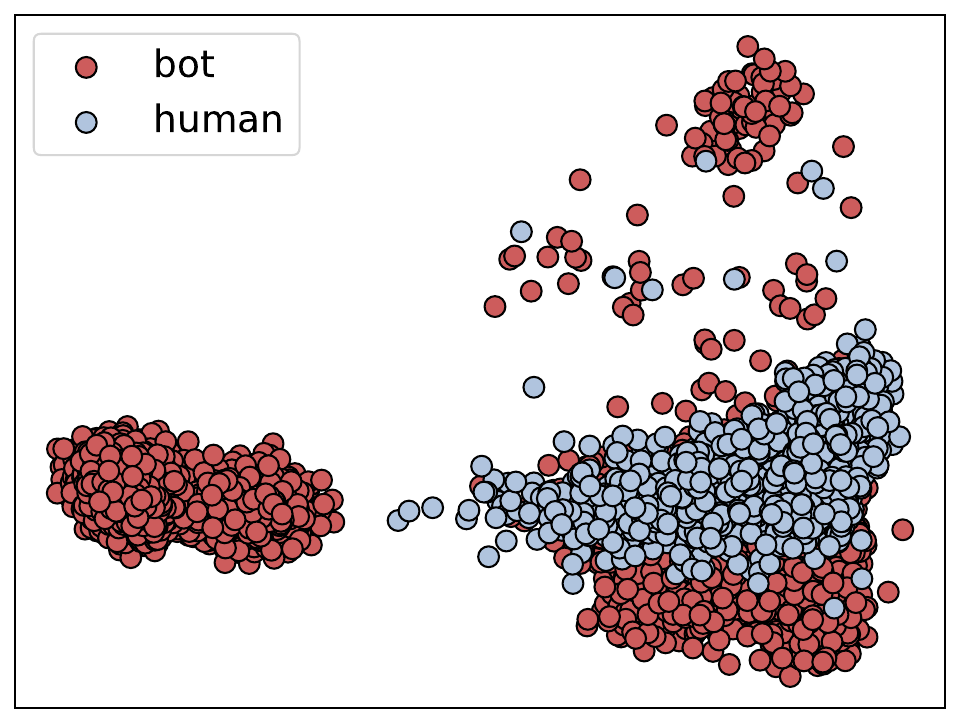}
        \end{minipage}%
    }
    \subfigure[\revised{${LINE}_{MG}$ on Pronbots19.}]{
        \label{fig: LINE_pronbots19_X}
        \begin{minipage}[t]{0.32\linewidth}
            \centering
            \includegraphics[width=1\linewidth]{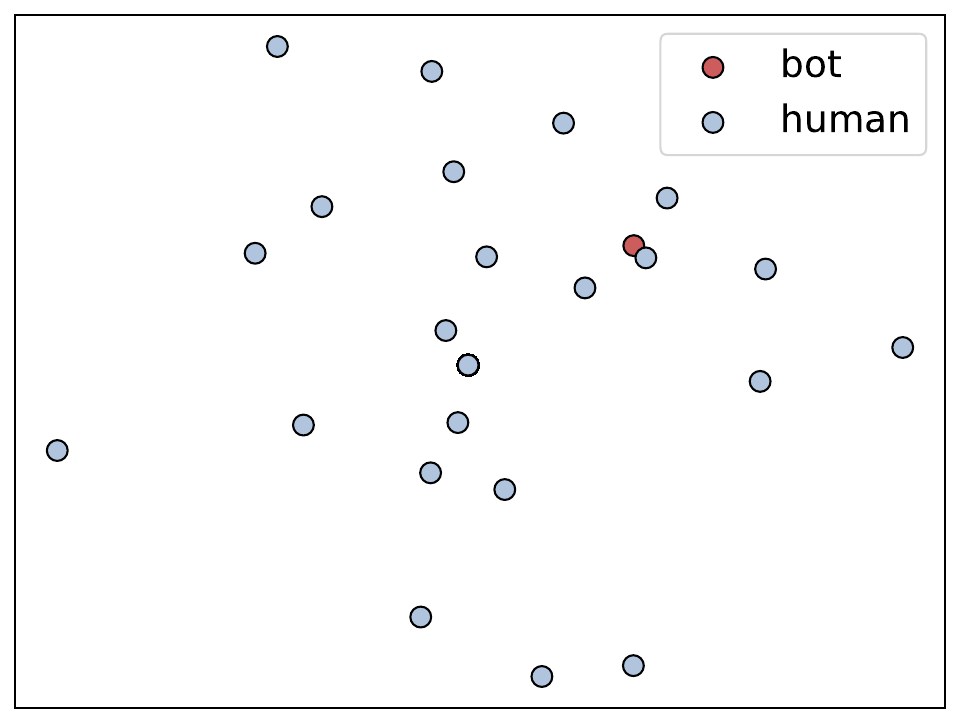}
        \end{minipage}%
    }
    \subfigure[\revised{${SDNE}_{MG}$ on Pronbots19.}]{
        \label{fig: SDNE_pronbots19_X}
        \begin{minipage}[t]{0.32\linewidth}
            \centering
            \includegraphics[width=1\linewidth]{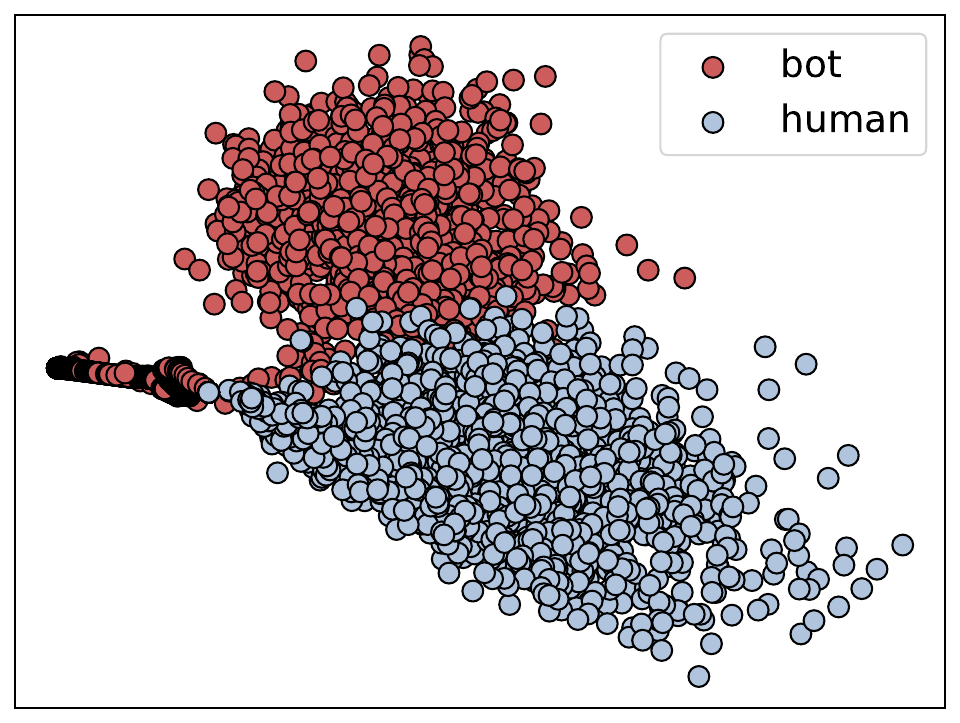}
        \end{minipage}%
    }

    \subfigure[\revised{${GraRep}_{MG}$ on Pronbots19.}]{
        \label{fig: GraRep_pronbots19_X}
        \begin{minipage}[t]{0.32\linewidth}
            \centering
            \includegraphics[width=1\linewidth]{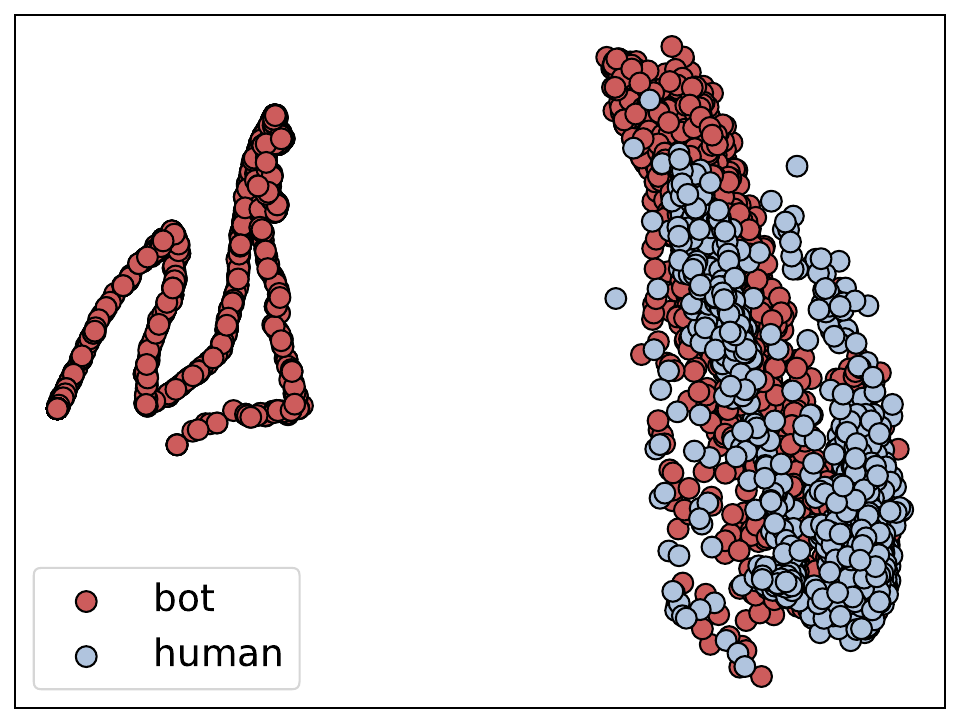}
        \end{minipage}%
    }
    \subfigure[\revised{${DNGR}_{MG}$ on Pronbots19.}]{
        \label{fig: DNGR_pronbots19_X}
        \begin{minipage}[t]{0.32\linewidth}
            \centering
            \includegraphics[width=1\linewidth]{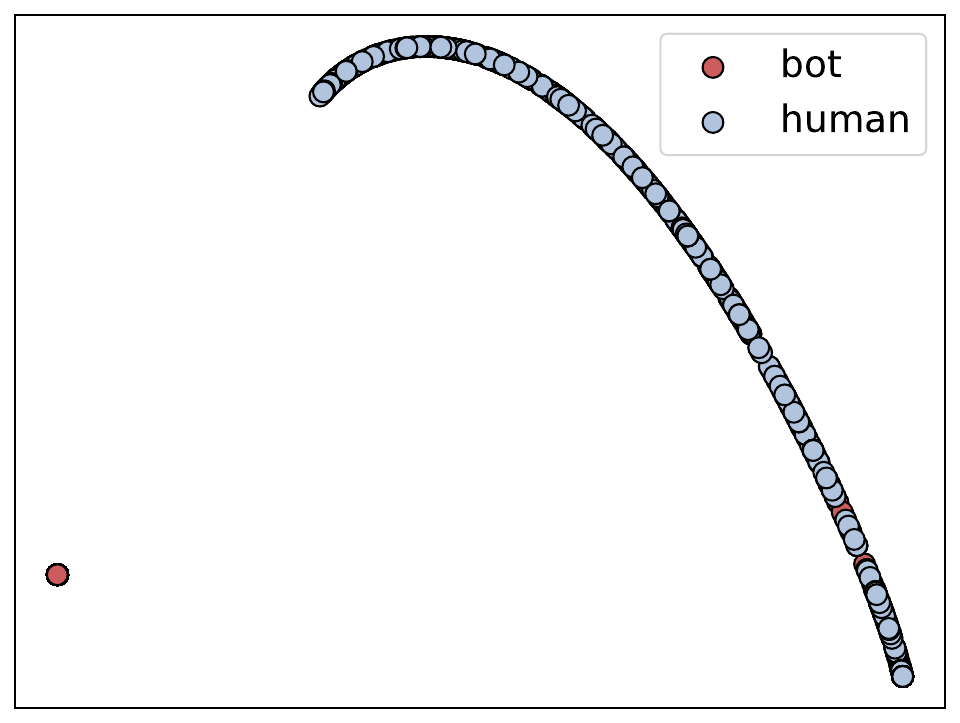}
        \end{minipage}%
    }
    \subfigure[\revised{${HOPE}_{MG}$ on Pronbots19.}]{
        \label{fig: HOPE_pronbots19_X}
        \begin{minipage}[t]{0.32\linewidth}
            \centering
            \includegraphics[width=1\linewidth]{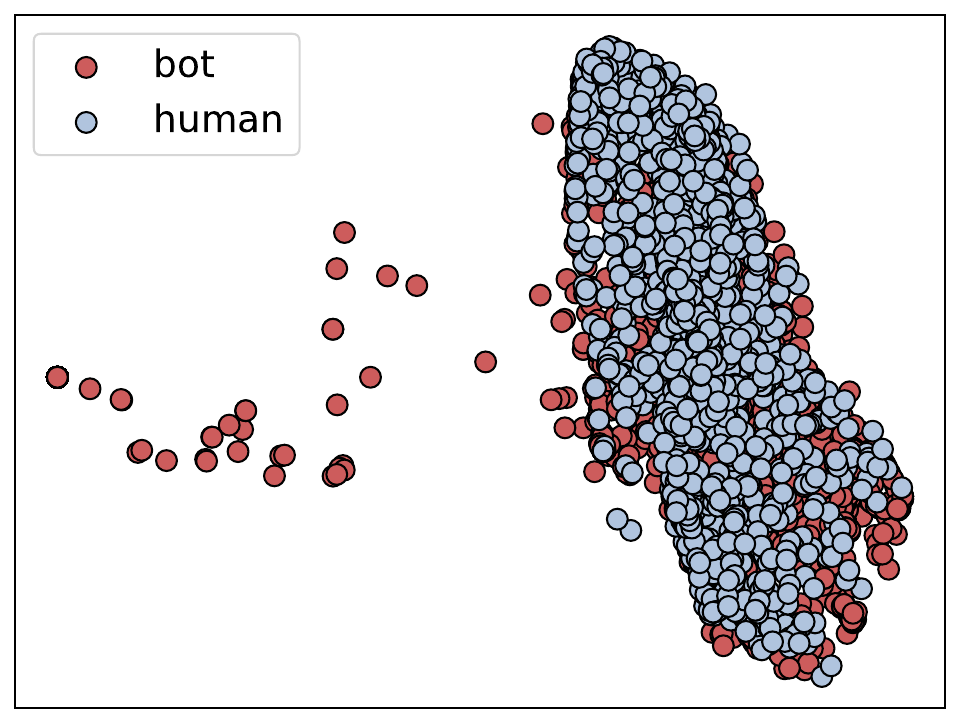}
        \end{minipage}%
    }
    
    \caption{Distribution of user embeddings generated by unsupervised graph learning models on the Botwiki-2019 and Pronbots-2019 datasets.}
    \label{fig: Botwiki-2019 Pronbots-2019}
\end{figure}

Next are seven models based on unsupervised graph representation learning implemented on the multi-relation graph $MG$ constructed in Section~\ref{sec: Multi-relational Graph Construction}.
\emph{DeepWalk} utilizes a random walk approach with depth-first traversal to sample neighboring nodes in a graph and learn node embeddings that capture structural information.
\emph{Node2vec} improves upon \emph{DeepWalk}'s random walk method by incorporating breadth-first traversal. 
\emph{LINE}, on the other hand, employs breadth-first sampling of neighboring nodes and combines first-order and second-order similarities to learn node embeddings. 
\emph{SDNE} extends \emph{LINE} by incorporating autoencoders to optimize the similarities in LINE.
On our multi-relation graph constructed based on behavior similarity, \emph{DeepWalk} achieves higher accuracy by employing depth-first sampling of neighboring nodes. 
However, \emph{Node2vec}, influenced by limited information from breadth-first search, exhibits slightly lower accuracy than \emph{DeepWalk} regarding identification. 
The baselines \emph{LINE} and \emph{SDNE}, which rely on breadth-first sampling, demonstrate generally lower detection performance compared to \emph{Node2vec}. 
The accuracy of \emph{LINE} on the Cresci-2015 dataset is only slightly higher than 50$\%$, and they only learn appropriate node embeddings to distinguish social bots in the experimental setting on the Pronbots-2019 dataset.
\emph{GraRep}, \emph{DNGR}, and \emph{HOPE} are all graph representation learning models based on matrix factorization. 
The difference lies in their approaches. 
\emph{GraRep} can be extended to capture higher-order proximity information and preserve the graph structure in low-dimensional representations, but it overlooks the contextual information of some nodes.
\emph{DNGR} utilizes the PPMI matrix to capture potential complex nonlinear relationships between different vertices. 
\emph{GraRep} and \emph{DNGR} demonstrate similar performance, with overall detection effectiveness superior to \emph{Node2vec} but inferior to \emph{DeepWalk}, especially with low accuracy on the Cresci-2017 dataset.
\emph{HOPE} maintains higher-order similarity relationships between nodes through matrix factorization of higher-order proximity matrices. 
It has limited expressive power and scalability and exhibits the worst performance among unsupervised graph representation learning baselines in social bot detection.
However, it performs well on the Botwiki-2019 dataset.
In comparison, \framework{} has the advantage of utilizing structural information theory to effectively leverage the structural information of the multi-relational graph of social users.
This approach results in hierarchical community partitioning, leading to more accurate classification and identification of social users.\par

To demonstrate the impact of graph construction on the effectiveness of the baselines, we additionally conduct experiments on the seven unsupervised graph representation learning models using the homogeneous graph $G$.
The homogeneous graph $G$ is described in detail in Section~\ref{sec: baselines and variations}.
On the Cresci-2015 dataset, the unsupervised graph representation learning models in the homogeneous graph $G$ exhibit low discriminability for social bots. 
The best-performing model, \emph{DeepWalk}, achieves only 76$\%$ accuracy, while the accuracies of other baseline models are below 70$\%$. 
However, unlike using the multi-relational graph $MG$, these models demonstrate relatively high recall rates, with \emph{SDNE}, \emph{GraRep}, and \emph{HOPE} reaching a Recall of 100$\%$.
For the other three datasets, except for \emph{DNGR}, the unsupervised graph representation learning models on the homogeneous graph $G$ generally exhibit poor performance in detecting social bots. 
We notice that \emph{DNGR} shows relatively high recall rates on the Cresci-2017 dataset and \revised{even outperforms all other baseline models in effectiveness on Pronbots-2019 and Botwiki-2019 datasets.}
However, considering the four datasets together, its overall performance is unsatisfactory.
In comparison, \framework{} achieves high accuracy on all four datasets and maintains a balance between precision and recall.
Each unsupervised graph representation learning baseline achieved significant improvement in detecting social bots on certain datasets after using $MG$.
In the Cresci-2015 dataset, the detection accuracy or precision of the five baselines ($DeepWalk$, $Node2vec$, $SDNE$, $GraRep$, $DNGR$) greatly improved after using $MG$. 
In the Cresci-2017 dataset, the detection accuracy or precision of the five baselines ($DeepWalk$, $LINE$, $Node2vec$, $GraRep$, $HOPE$) greatly improved after using $MG$. 
In the Pronbots-2019 dataset, the detection accuracy or precision of the six baselines ($DeepWalk$, $LINE$, $Node2vec$, $SDNE$, $GraRep$, \revised{$DNGR$}, $HOPE$) greatly improved after using $MG$. 
In the Botwiki-2019 dataset, the detection accuracy or precision of the four baselines ($DeepWalk$, \revised{$LINE$, $SDNE$}, $GraRep$, $HOPE$) greatly improved after using $MG$. 
Overall, the effectiveness of social bot detection using the social user graph $MG$ far surpasses that of using graph $G$ on baselines.
We will delve deeper into the effectiveness of the multi-relational graph $MG$ and analyze the impact of each type of edge within it in Section ~\ref{sec: ablation}.\par

To gain a deeper understanding of the effectiveness of the learned representation vectors, we employed Principal Component Analysis (PCA) to reduce the dimensionality of embeddings learned by unsupervised graph representation learning-based baselines on the multi-relational graph ($MG$) to a two-dimensional vector and visualize them on a plane graph. 
We evaluated six baseline models based on four datasets separately. 
Figure~\ref{fig: Cresci-2015 Cresci-2017} depicts the dimensionality reduction effect of user representation vectors on the Cresci-2015 and Cresci-2017 datasets. 
Figure~\ref{fig: Botwiki-2019 Pronbots-2019} illustrates the dimensionality reduction effect of user representation vectors on the Pronbots-2019 and Botwiki-2019 datasets. 
Blue dots represent humans, while red dots represent social bots.
The unsupervised graph representation learning models embed network structure data into a low-dimensional space reflected in each user's embeddings. 
As illustrated in Figures~\ref{fig: GraRep_cresci15_X}, ~\ref{fig: HOPE_cresci15_X}, ~\ref{fig: SDNE_cresci17_X} and ~\ref{fig: DNGR_cresci17_X}, these embeddings exhibit a limited distinction between social bots and human entities, which leads to the challenge of discerning between social bots and humans.
\revised{Besides, as shown in Figures~\ref{fig: DeepWalk_botwiki19_X}, ~\ref{fig: SDNE_botwiki19_X}, ~\ref{fig: GraRep_botwiki19_X}, ~\ref{fig: DeepWalk_pronbots19_X}, and ~\ref{fig: HOPE_pronbots19_X}, on the Pronbots -2019 and Botwiki-2019 datasets, both of which consist mostly social bots and a small number of humans, the learned embeddings do not exhibit concentration, with high overlap between human and bot embeddings.}
Furthermore, the learned embeddings and the two-dimensional vectors obtained by PCA dimensionality reduction lack interpretability.
In contrast, \framework{} utilizes structural information theory to decode the network structure into a hierarchical encoding tree, resulting in high discriminability for social bots and interpretability as detailed in Section~\ref{sec: interpretability}.

\begin{figure}[thp]
   \centering
    \subfigure[Cresci15.]{
        \label{fig: ROC_Cresci15}
        \begin{minipage}[t]{0.24\linewidth}
            \centering
            \includegraphics[width=1\linewidth]{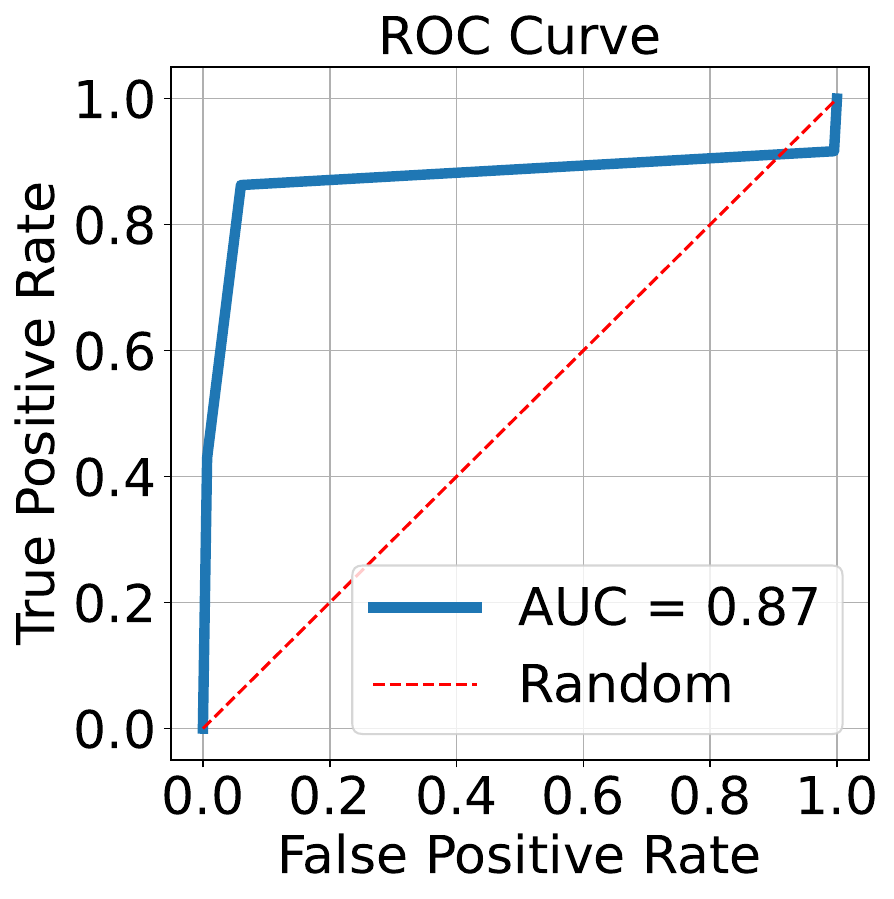}
        \end{minipage}%
    }
    \subfigure[Cresci17.]{
        \label{fig: ROC_Cresci17}
        \begin{minipage}[t]{0.24\linewidth}
            \centering
            \includegraphics[width=1\linewidth]{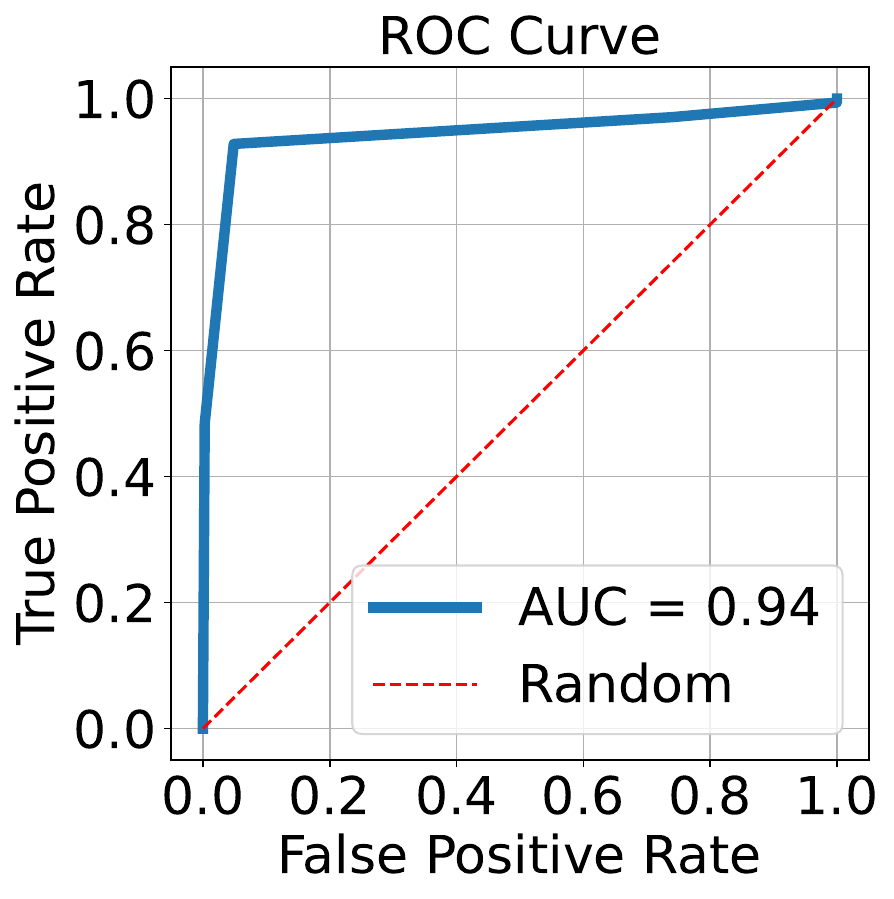}
        \end{minipage}%
    }
    \subfigure[Pronbots19.]{
        \label{fig: ROC_Pronbots19}
        \begin{minipage}[t]{0.24\linewidth}
            \centering
            \includegraphics[width=1\linewidth]{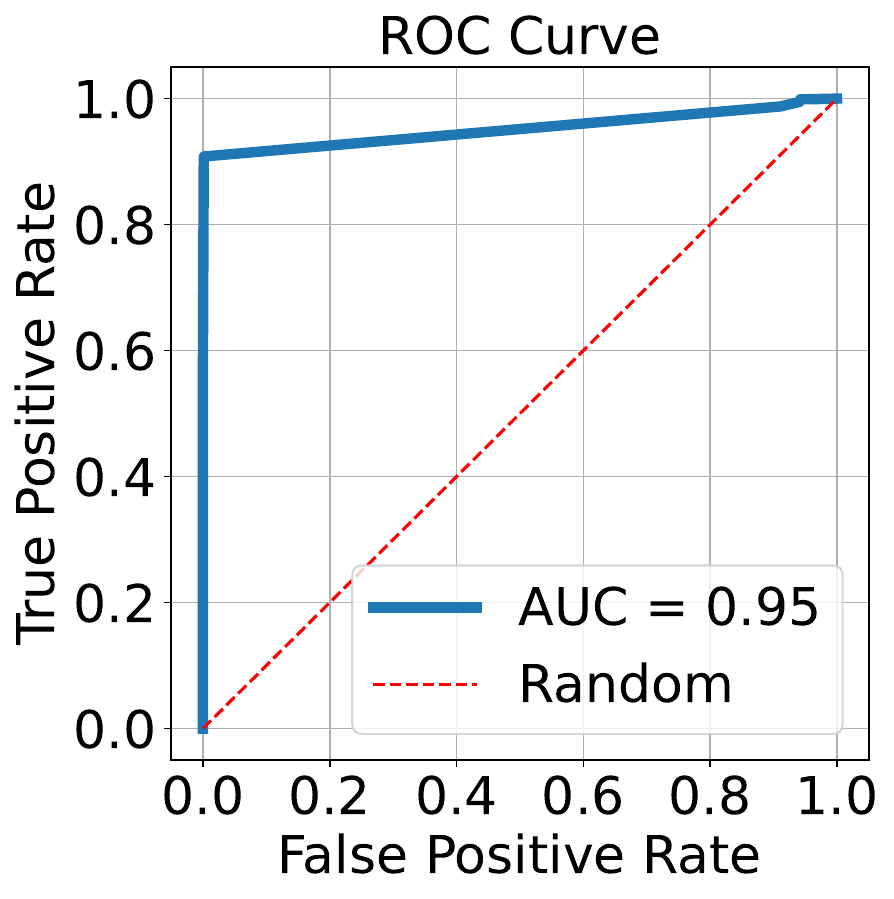}
        \end{minipage}%
    }
    \subfigure[Botwiki19.]{
        \label{fig: ROC_Botwiki19}
        \begin{minipage}[t]{0.24\linewidth}
            \centering
            \includegraphics[width=1\linewidth]{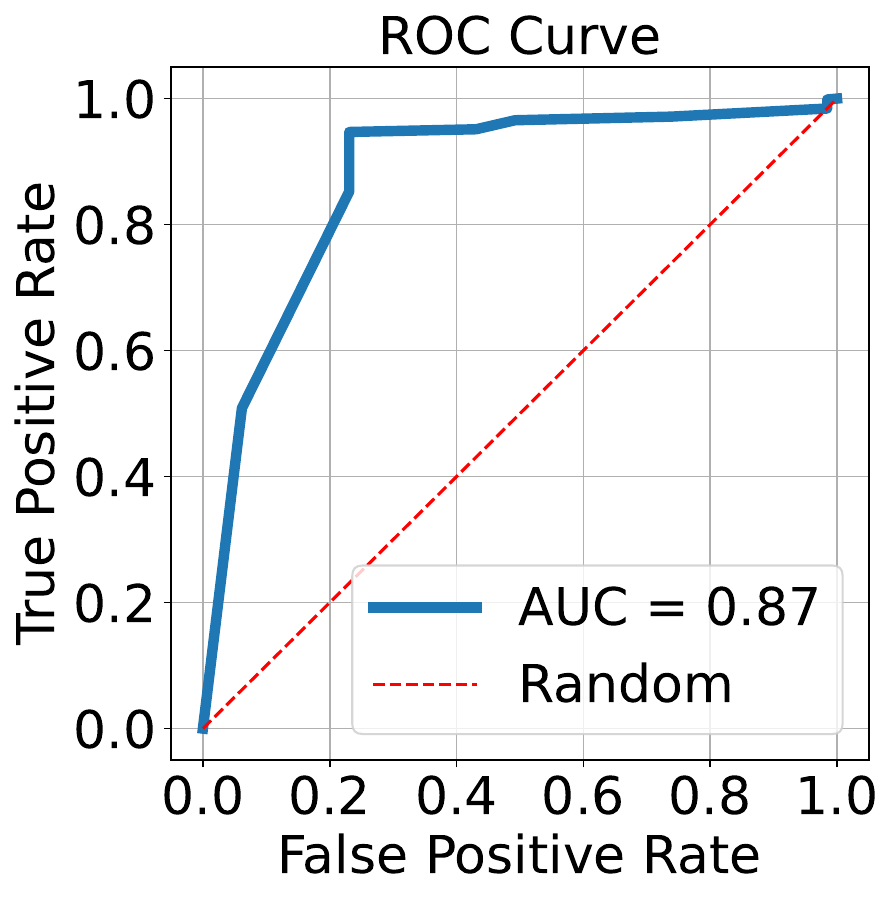}
        \end{minipage}%
    }
    
    \caption{AUC and ROC curves on four datasets.}
    \label{fig: ROC}
\end{figure}
To evaluate the stability of the \framework{} with learned community label index, we treat $E_v(\alpha)$ equally as the predicted value for each user and report the area under the ROC curves.
As shown in Figure~\ref{fig: ROC}, \framework{} achieves good performances in all four datasets, especially for imbalanced datasets like Pronbots19 and Botwiki19.
These results demonstrate that by combining community influence and community cohesion to label social bots, we can effectively deal with unequal distributions of bots and users in the real world.\par

\subsection{Ablation Study}
\label{sec: ablation}
In this section, we conduct ablation experiments on the graph modeling of \framework{} to demonstrate the rationality of social users' multi-relational graph $MG$. 
Additionally, we perform experiments by varying three hyperparameters of \framework{} to analyze the sensitivity of the model to hyperparameters.\par
\begin{table}[thp]
    \caption{\revised{The total number of edges, the average number of edges per user, and the average connection degree of users in Multi-relation Graph, Homogeneity Graph, and Ordinary Multi-relation Graph for different datasets.}}
    \label{tab: edges}
    \centering
    \aboverulesep=0ex
    \belowrulesep=0ex
    \renewcommand\arraystretch{1.3}
    \begin{tabular}{c|ccc|ccc|ccc}      
        \toprule
        \multirow{2}*{Datasets} & \multicolumn{3}{c|}{\multirow{1}*{\textbf{Multi-relation Graph (MG)}}} & \multicolumn{3}{c|}{\multirow{1}*{\textbf{Homogeneity Graph (G)}}} & \multicolumn{3}{c}{\multirow{1}*{\revised{\textbf{Ordinary MG (mg)}}}}\\

        \cline{2-10}

        & \textbf{Total} & \textbf{Avg.} & \multicolumn{1}{c|}{\textbf{Conn.}} & \textbf{Total} & \textbf{Avg.} & \multicolumn{1}{c|}{\textbf{Conn.}} & \revised{\textbf{Total.}} & \revised{\textbf{Avg.}} & \revised{\textbf{Conn.}} \\
        
        \hline
        Cresci-2015 & 4,109,097 & 516.7 & 9.75\% & 4,276 & 1.6 & 0.03\% & \revised{1,966,693} & \revised{247.3} & \revised{4.67\%}\\
        
        Cresci-2017 & 19,958,930 & 1044.6 & 8.20\% & 12,326,304 & 1715.7 & 11.94\% & \revised{9,749,755} & \revised{510.3} & \revised{4.01\%}\\
        
        Pronbots-2019 & \revised{131,935,035} & \revised{4542.5} & \revised{23.46\%} & \revised{40,244,245} & \revised{4156.8} & \revised{21.44\%} & \revised{2,896,693} & \revised{99.7} & \revised{0.52\%} \\
        
        Botwiki-2019 & \revised{181,839} & \revised{158.9} & \revised{20.85\%} & \revised{107,326} & \revised{281.3} & \revised{36.92\%} & \revised{35,294} & \revised{30.8} & \revised{4.05\%} \\
        \bottomrule
    \end{tabular}
\end{table}

\subsubsection{\textbf{Graph Effectiveness}}
To demonstrate the rationality of the social user multi-relational graph $MG$ proposed in Section~\ref{sec: Multi-relational Graph Construction} for social bot detection, we conduct comparative experiments on \framework{} using the homogeneous graph $G$  (denoted as \framework{}-G) \revised{and an ordinary Multi-relation graph $mg$ (denoted as \framework{}-mg)}.
Besides, to evaluate the positive impact of the edges used in the modeling of the social user multi-relation graph in \framework{} on the performance of social bot detection, we also construct three variations by removing one type of edge from the multi-relation graph for ablation experiments (denoted as \framework{}-FT, \framework{}-FI, and \framework{}-TI). 
Except for the graph construction, all other experimental settings remain the same as described in Section~\ref{sec: setting}. 
All experimental results across the four datasets are reported in Table~\ref{tab: variations Cresci-2015}, Table~\ref{tab: variations Cresci-2017}, Table~\ref{tab: variations Pronbots-2019}, and Table~\ref{tab: variations Botwiki-2019}, respectively.
Table~\ref{tab: edges} lists the total edges, the average number of edges under one type per user, and the average connection degree of users in the four datasets under \revised{graph $MG$, graph $G$, and graph $mg$. }
In the case of the Cresci-2015 dataset, the homogeneous graph $G$ is sparse due to the use of the action of following between users. 
For the other datasets, there is no significant difference in the average connectivity level between the homogeneous graph $G$ and the multi-relational graph $MG$. 
\revised{In the case of the Pronbots-2019 dataset, the ordinary Multi-relation graph $mg$ is sparse due to the less similarity between direct user behaviors.}
\revised{The other three datasets have similar degrees of connectivity, but are all sparser than $MG$.}\par

\begin{table}[thp]
    \caption{Comparison with the ACC, Precision, Recall, F1 and the number of communities of different variations on Cresci-2015 dataset (unit:\%).}
    \label{tab: variations Cresci-2015}
    \aboverulesep=0ex
    \belowrulesep=0ex
    \centering
    \renewcommand\arraystretch{1.2}
    \begin{tabular}{m{2.5cm}<{\centering}|m{2cm}<{\centering} m{2cm}<{\centering} m{2cm}<{\centering} m{2cm}<{\centering} m{2cm}<{\centering} }
        \toprule
        
        \multirow{2}*{Variation} & \multicolumn{5}{c}{\multirow{1}*{\textbf{Cresci-2015}}} \\

        \cline{2-6}

        & {\textbf{ACC}} & {\textbf{Precision}} & {\textbf{Recall}} & {\textbf{F1}} & \multicolumn{1}{c}{\textbf{Num\_comm}} \\
        
        \hline

        \textbf{\framework{}-G} & 27.69 & 37.17 & 20.83 & 26.70 & 3731 \\

        \revised{\textbf{\framework{}-mg}} & \revised{63.08} & \revised{63.17} & \revised{\textbf{99.79}} & \revised{77.36} & \revised{9} \\

        \hline
        
        \textbf{~\framework-FT} & 58.06 & 61.76 & 88.36 & 72.71 & 8 \\

        \textbf{~\framework-FI} & 64.97 & 64.48 & 99.28 & 78.18 & 65 \\
        
        \textbf{~\framework-TI} & 59.71 & 61.97 & 93.85 & 74.65 & 10 \\

        \hline
        
        \textbf{~\framework} & \textbf{89.12} & \textbf{96.11} & 86.27 & \textbf{90.93} & \textbf{4}\\
        
        \hline

        \textbf{Gain} & \textbf{24.15-61.43$\uparrow$} & \textbf{31.63-58.94$\uparrow$} & \revised{13.52}$\downarrow$ & \textbf{12.75-64.23$\uparrow$} & \textbf{4-3727$\downarrow$} \\

        \bottomrule
    \end{tabular}
\end{table}

\begin{table}[thp]
    \caption{Comparison with the ACC, Precision, Recall, F1 and the number of communities of different variations on Cresci-2017 dataset (unit:\%).}
    \label{tab: variations Cresci-2017}
    \aboverulesep=0ex
    \belowrulesep=0ex
    \centering
    \renewcommand\arraystretch{1.2}
    \begin{tabular}{m{2.5cm}<{\centering}|m{2cm}<{\centering} m{2cm}<{\centering} m{2cm}<{\centering} m{2cm}<{\centering} m{2cm}<{\centering} }
        \toprule
        
        \multirow{2}*{Variation} & \multicolumn{5}{c}{\multirow{1}*{\textbf{Cresci-2017}}} \\

        \cline{2-6}
       
        & {\textbf{ACC}} & {\textbf{Precision}} & {\textbf{Recall}} & {\textbf{F1}} & \multicolumn{1}{c}{\textbf{Num\_comm}} \\
        
        \hline

        \textbf{\framework{}-G} & 68.79 & 74.59 & 89.24 & 81.26 & 15 \\

        \revised{\textbf{\framework{}-mg}} & \revised{67.30} & \revised{71.25} & \revised{92.28} & \revised{80.41} & \revised{10} \\

        \hline
        
        \textbf{~\framework-FT} & 77.58 & 77.24 & \textbf{98.07} & 86.42 & 19 \\

        \textbf{~\framework-FI} & 86.50 & 87.02 & 95.71 & 91.16 & 2524 \\
        
        \textbf{~\framework-TI} & 75.83 & 76.61 & 96.12 & 85.26 & 13 \\

        \hline
        
        \textbf{\framework{}} & \textbf{93.46} & \textbf{98.07} & 92.83 & \textbf{95.38} & \textbf{6}\\
        
        \hline

        \textbf{Gain} & \textbf{6.96-\revised{26.16}$\uparrow$} & \textbf{11.05-\revised{26.82}$\uparrow$} & 5.24$\downarrow$ & \textbf{4.22-\revised{14.97}$\uparrow$} & \textbf{\revised{4}-2518$\downarrow$} \\

        \bottomrule
    \end{tabular}
    
\end{table}

\begin{table}[thp]
    \caption{Comparison with the ACC, Precision, Recall, F1, and the number of communities of different variations on Pronbots-2019 dataset (unit:\%).}
    \label{tab: variations Pronbots-2019}
    \aboverulesep=0ex
    \belowrulesep=0ex
    \centering
    \renewcommand\arraystretch{1.2}
    \begin{tabular}{m{2.5cm}<{\centering}|m{2cm}<{\centering} m{2cm}<{\centering} m{2cm}<{\centering} m{2cm}<{\centering} m{2cm}<{\centering} }
        \toprule
        
        \multirow{2}*{Variation} & \multicolumn{5}{c}{\multirow{1}*{\textbf{Pronbots-2019}}} \\

       \cline{2-6}

        & {\textbf{ACC}} & {\textbf{Precision}} & {\textbf{Recall}} & {\textbf{F1}} & \multicolumn{1}{c}{\textbf{Num\_comm}} \\
        
        \hline

        \textbf{\framework{}-G} & \revised{93.52} & \revised{95.73} & \revised{97.33} & \revised{96.52} & \revised{\textbf{13}} \\
        
        \revised{\textbf{\framework{}-mg}} & \revised{92.33} & \revised{92.36} & \revised{\textbf{99.96}} & \revised{96.01} & \revised{40} \\

        \hline
        
        \textbf{~\framework-FT} & \revised{\textbf{96.39}} & \revised{\textbf{96.83}} & \revised{99.34} & \revised{\textbf{98.07}} & \revised{928} \\

        \textbf{~\framework-FI} & \revised{16.89} & \revised{69.13} & \revised{18.07} & \revised{28.66} & \revised{14827} \\
        
        \textbf{~\framework-TI} & \revised{93.01} & \revised{93.24} & \revised{99.66} & \revised{96.34} & \revised{101} \\

        \hline
        
        \textbf{~\framework} & \revised{92.68} & \revised{92.75} & \revised{99.88} & \revised{96.18} & \revised{58}\\
        
        \hline

        \textbf{Gain} & \revised{3.71$\downarrow$} & \revised{4.08$\downarrow$} & \revised{0.08$\downarrow$} & \revised{1.89$\downarrow$} & \revised{45$\uparrow$} \\

        \bottomrule
    \end{tabular}
\end{table}

\begin{table}[thp]
    \caption{Comparison with the ACC, Precision, Recall, F1, and the number of communities of different variations on Botwiki-2019 dataset (unit:\%).}
    \label{tab: variations Botwiki-2019}
    \aboverulesep=0ex
    \belowrulesep=0ex
    \centering
    \renewcommand\arraystretch{1.2}
    \begin{tabular}{m{2.5cm}<{\centering}|m{2cm}<{\centering} m{2cm}<{\centering} m{2cm}<{\centering} m{2cm}<{\centering} m{2cm}<{\centering} }
        \toprule
        
        \multirow{2}*{Variation} & \multicolumn{5}{c}{\multirow{1}*{\textbf{Botwiki-2019}}} \\
       
        \cline{2-6}
        
        & {\textbf{ACC}} & {\textbf{Precision}} & {\textbf{Recall}} & {\textbf{F1}} & \multicolumn{1}{c}{\textbf{Num\_comm}} \\
        
        \hline

        \textbf{\framework{}-G} & \revised{86.37} & \revised{92.79} & \revised{92.26} & \revised{92.53} & \revised{14} \\

        \revised{\textbf{\framework{}-mg}} & \revised{87.42} & \revised{91.46} & \revised{\textbf{95.13}} & \revised{93.26} & \revised{18} \\

        \hline
        
        \textbf{~\framework-FT} & \revised{90.43} & \revised{94.45} & \revised{\textbf{95.13}} & \revised{94.79} & \revised{29} \\

        \textbf{~\framework-FI} & \revised{75.23} & \revised{90.33} & \revised{81.66} & \revised{85.78} & \revised{153} \\
        
        \textbf{~\framework-TI} & \revised{84.80} & \revised{\textbf{97.86}} & \revised{85.24} & \revised{91.11} & \revised{\textbf{11}} \\

        \hline
        
        \textbf{~\framework} & \revised{\textbf{93.18}} & \revised{97.78} & \revised{94.70} & \revised{\textbf{96.22}} & \revised{\textbf{11}}\\
        
        \hline

        \textbf{Gain} & \revised{\textbf{2.75-17.95$\uparrow$}} & \revised{0.08$\downarrow$} & \revised{0.43$\downarrow$} & \revised{\textbf{1.43-10.44$\uparrow$}} & \revised{\textbf{0-142$\downarrow$}} \\

        \bottomrule
    \end{tabular}
    
\end{table}

Table~\ref{tab: variations Cresci-2015} shows that \framework{}-$G$ with sparse edges leads to many communities and low accuracy after clustering based on structural entropy.
Besides, as shown in Table~\ref{tab: variations Cresci-2017} and Table~\ref{tab: variations Botwiki-2019}, the accuracy of using graph $G$ is much lower compared to the $MG$.
This is because social bots make their attributes indistinguishable from humans through camouflage.
Therefore, utilizing the proposed social user multi-relational graph $MG$ for \framework{} is more rational. 
To further analyze $MG$ on \framework{}, we also investigate the efficacy of three types of edges within the multi-relational graph $MG$ \revised{as well as the results of using other types of edges}.
For $MG$ in the variation \framework{}-FT, only the edges related to the \emph{following-follower ratio} and \emph{posting type distribution} are retained.
When the influence of \emph{posting influence} is excluded in the \revised{Cresci-2015, Cresci-2017, and Botwiki-2019} datasets, the detection accuracy significantly decreases, but the recall rate increases to 98.07\% on Cresci-2017 \revised{and 95.13\% on Botwiki-2019}. 
\revised{However, in the Pronbots-2019 dataset, the accuracy, precision, and F1 score of the variation \framework{}-FT increases by 3.71\%, 4.08\%, and 1.89\% compared to the original \framework{}.}
For $MG$ in the variation \framework{}-FI, only the edges related to the \emph{following-follower ratio} and \emph{posting influence} are retained. 
This significantly decreases the detection accuracy and precision in all four datasets. 
Especially on the Pronbots-2019 dataset, the accuracy rate is even less than 20\%.
Additionally, due to the lack of rich connectivity structure related to \emph{posting type distribution}, the variation \framework{}-FI forms a larger number of communities with a smaller average size, resulting in a more dispersed social bot community.
It can be observed that the number of communities increases most compared with \framework{}.
For $MG$ in the variation \framework{}-TI, only the edges related to \emph{posting type distribution} and \emph{posting influence} are retained. 
In the \revised{Pronbots-2019} datasets, the accuracy, precision, recall, F1 score, and the number of communities are all comparable to \revised{or slightly higher than} the original \framework{} model. 
However, \revised{in the Cresci-2015, Cresci-2017, and Botwiki-2019 datasets}, the accuracy and precision drop significantly even with less than 60\% accuracy on Cresci-2015.
\revised{For ordinary Multi-relation graph $mg$ in the variation \framework{}-mg, three new relations (\emph{U-twe-U}, \emph{U-fri-U}, \emph{U-fav-U}) is constructed based on direct user behaviors to prove the validity of the three relationships selected in $MG$}.
\revised{\framework{}-mg only reflects a higher recall rate. In addition, the accuracy, precision, and F1 on the four datasets are far inferior to \framework{}.}
Overall, integrating three types of edges can significantly enhance the detection accuracy of \framework{}. 
Therefore, it is necessary to jointly model the multi-relational social user graph of the three relationships to unleash the full potential of \framework{}'s performance optimization.\par

\subsubsection{\textbf{Hyperparameter Sensitivity}}
\begin{figure}[thp]
    \centering
    \includegraphics[width=1\textwidth]{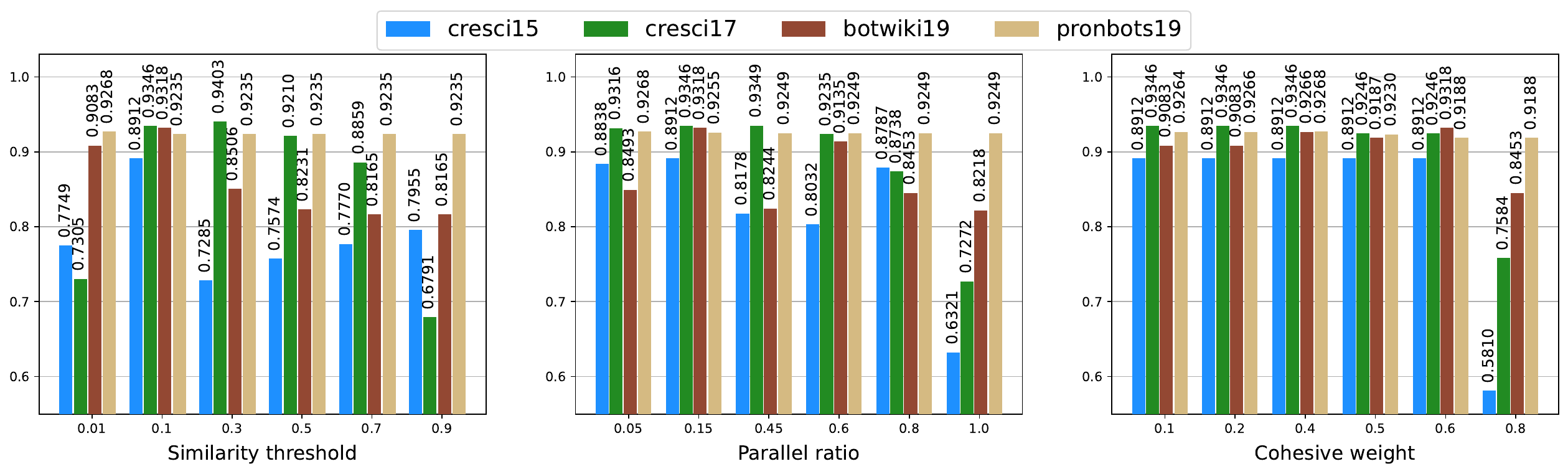}
    \caption{\revised{Hyperparameter sensitivity.}}
    \label{fig: hyperparameter}
\end{figure}
For the three hyperparameters involved in \framework{} \revised{(similarity threshold $\xi$, parallel ratio $p$, cohesive weight $\pi$)}, we conduct a series of hyperparameter sensitivity experiments by varying individual parameters at a time and repeating the experiments on \framework{}. 
Specifically, \revised{similarity threshold} $\xi$ represents the threshold for user behavior feature bias during the graph construction process. 
\revised{parallel ratio} $p$ represents the maximum scale ratio for parallel merge operators, indicating the proportion of community numbers that can be fused compared to the maximum fusion limit.
Lastly, \revised{cohesive weight} $\pi$ represents the community cohesion proportion in the evaluation metric.
The experimental results, as shown in Figure~\ref{fig: hyperparameter}, indicate that the accuracy of social bot detection is significantly influenced by the \revised{similarity threshold} $\xi$ parameter on \revised{the first three} datasets. 
In particular, the detection performance is better when \revised{similarity threshold} $\xi$ is around 0.1.
This is because a very small $\xi$ value leads to a lack of important connections in the constructed graph, while a very large $\xi$ value introduces excessive noise into the graph.
\revised{However, this has little impact on Pronbots-2019, on the contrary, the minimum $\xi$ of 0.01 achieves the best performance. This may be because the users themselves in this dataset have high similarity and the smaller $\xi$ filters out redundant information.}
In the context of parallel operations for fusion operators, apart from the Pronbots-2019 dataset, selecting a relatively small value for \revised{parallel ratio} $p$ can improve the detection performance. 
This is because having an excessive number of parallel operands can potentially lead to the merging of suboptimal communities, resulting in a suboptimal encoding tree and a less stable partitioning of communities.
However, setting \revised{parallel ratio} $p$ to a lower value will also increase the number of rounds for fusion operator execution, thereby reducing the model's efficiency. 
Hence, we uniformly set $p$ to \revised{0.05 for Pronbots-2019 and 0.15 for other datasets}, considering both performance and efficiency factors.
Besides, we also find that the \revised{cohesive weight $\pi$ parameter significantly impacts the Botwiki-2019 dataset, while its value does not affect the detection accuracy of Cresci-2015, Cresci-2017, and Pronbots-2019 datasets under non-extreme conditions.}

\subsection{Interpretability}
\label{sec: interpretability}

\begin{figure}[thp]
    \centering    
    \subfigure[\revised{Network with true label.}]{
        \label{fig: network_true_botwiki19}
        \begin{minipage}[t]{0.49\linewidth}
            \centering
            \includegraphics[width=1.0\linewidth]{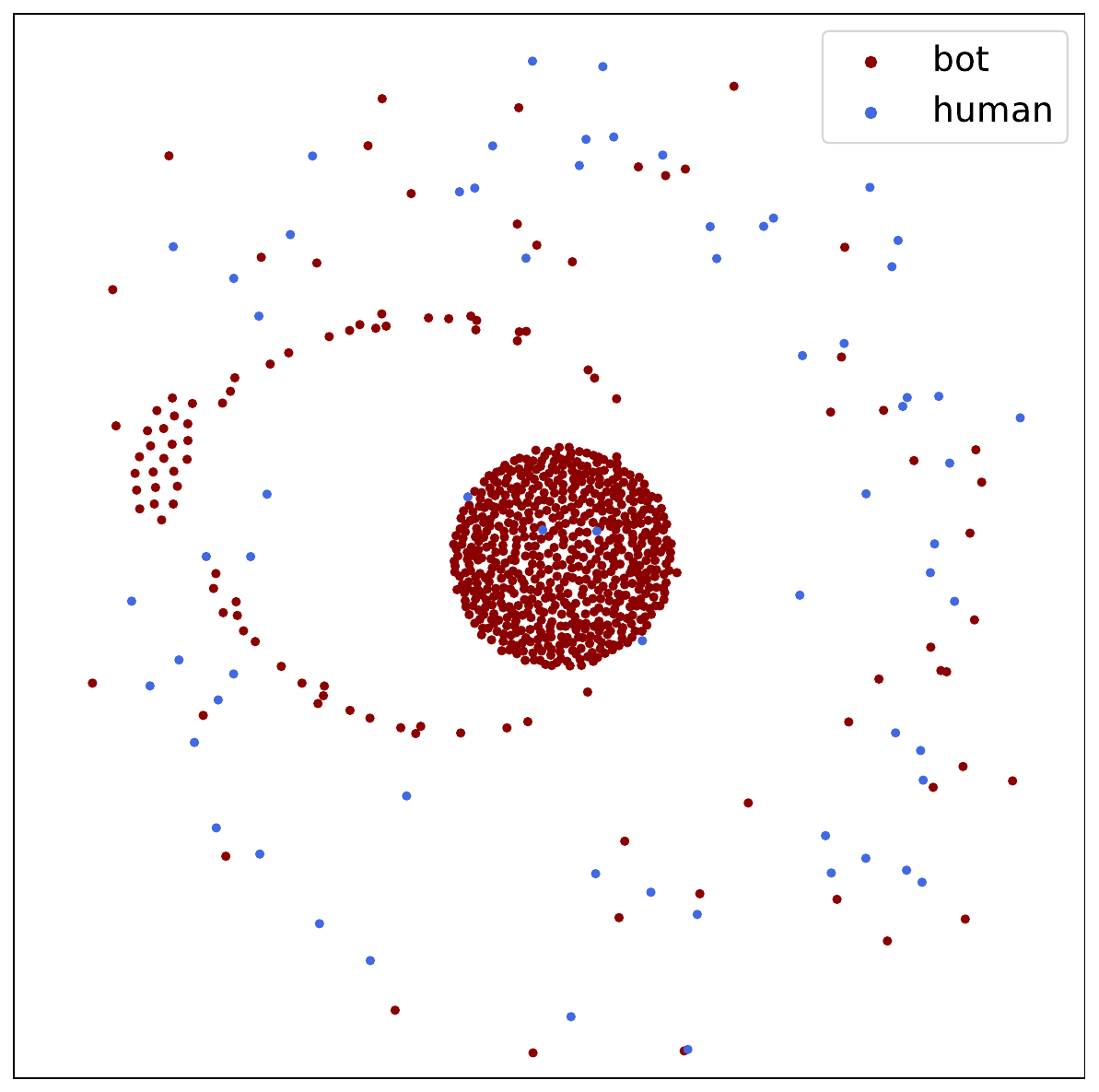}
        \end{minipage}%
    }
    \subfigure[\revised{Network with pred label.}]{
        \label{fig: network_pred_botwiki19}
        \begin{minipage}[t]{0.49\linewidth}
            \centering
            \includegraphics[width=1.0\linewidth]{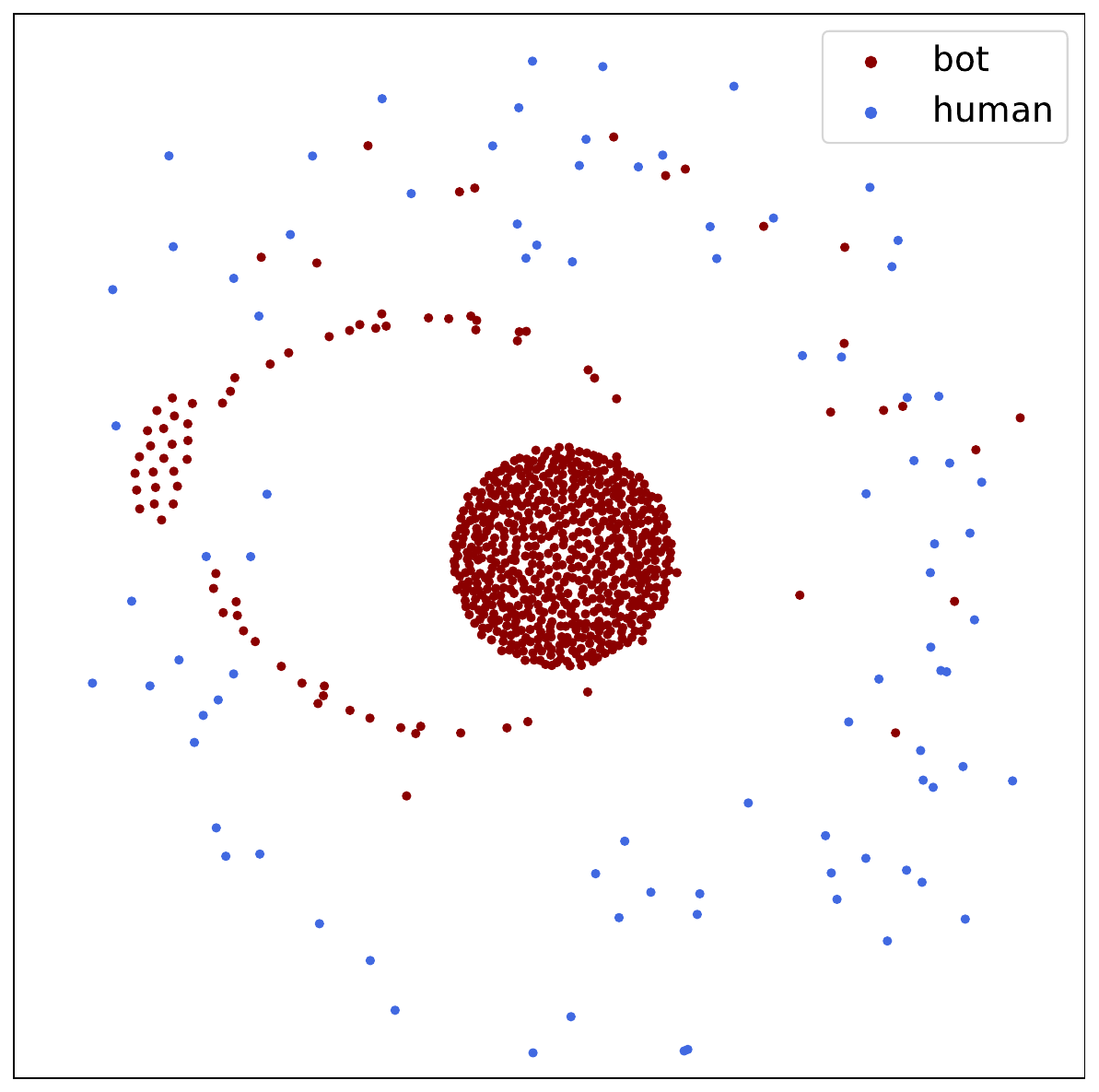}
        \end{minipage}%
    }

    \subfigure[\revised{The optimal encoding tree of Botwiki-2019 dataset in \framework{} (square tree nodes are communities, circular leaf nodes are social users, and the size of the nodes indicates the number of social users. For visualization, each large leaf node represents 20 social users of Botwiki-2019). }]{
        \label{fig: tree_botwiki19}
        \begin{minipage}[t]{1\linewidth}
            \centering
            \includegraphics[width=1\linewidth]{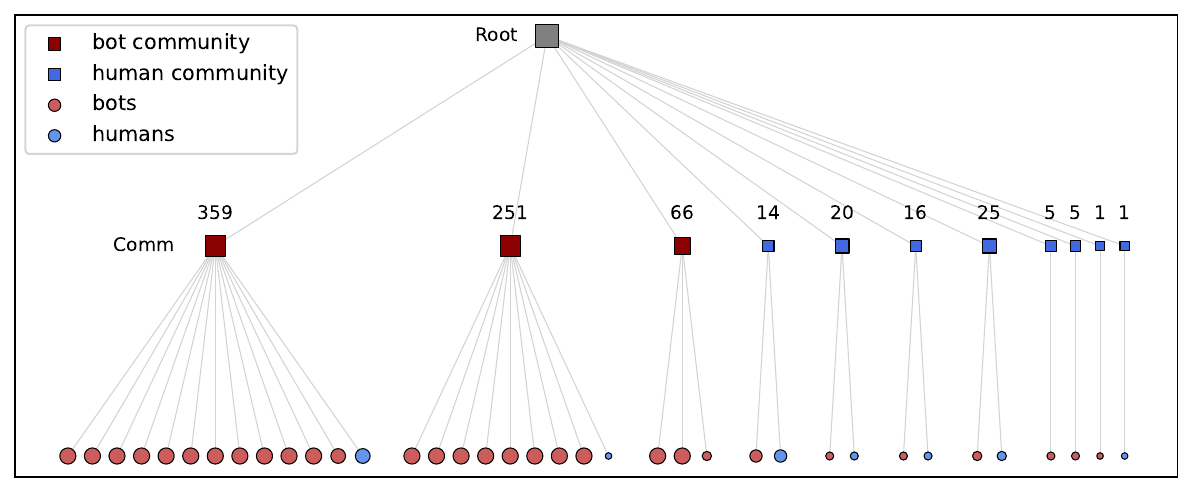}
        \end{minipage}%
    }
    
    \caption{\revised{Renderings of Botwiki-2019.}}
    \label{fig: rendering_botwiki}
\end{figure}

\begin{figure}[thp]
    \centering    
    \subfigure[\revised{Network with true label.}]{
        \label{fig: network_true_pronbots19}
        \begin{minipage}[t]{0.49\linewidth}
            \centering
            \includegraphics[width=1.0\linewidth]{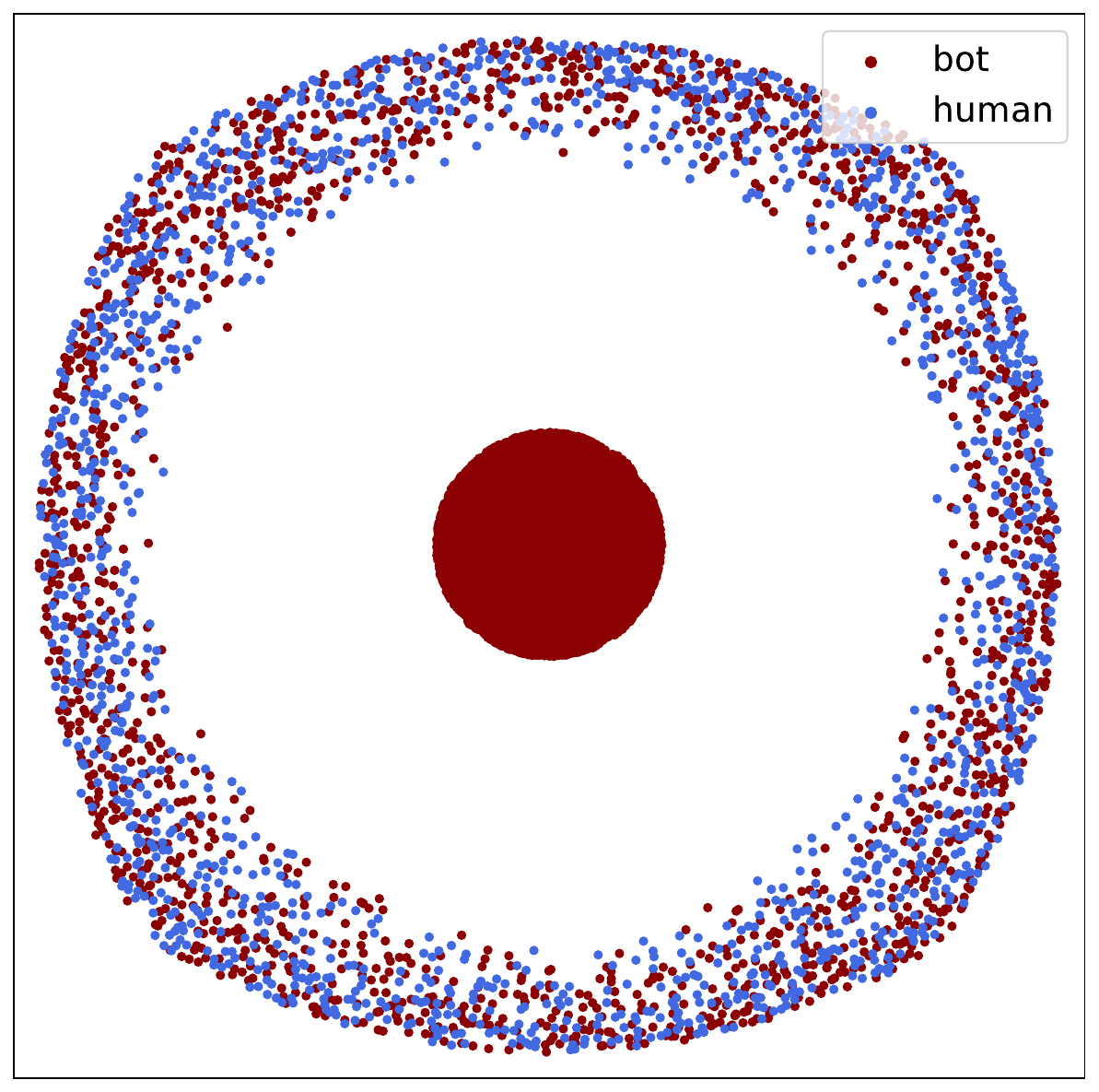}
        \end{minipage}%
    }
    \subfigure[\revised{Network with pred label.}]{
        \label{fig: network_pred_pronbots19}
        \begin{minipage}[t]{0.49\linewidth}
            \centering
            \includegraphics[width=1.0\linewidth]{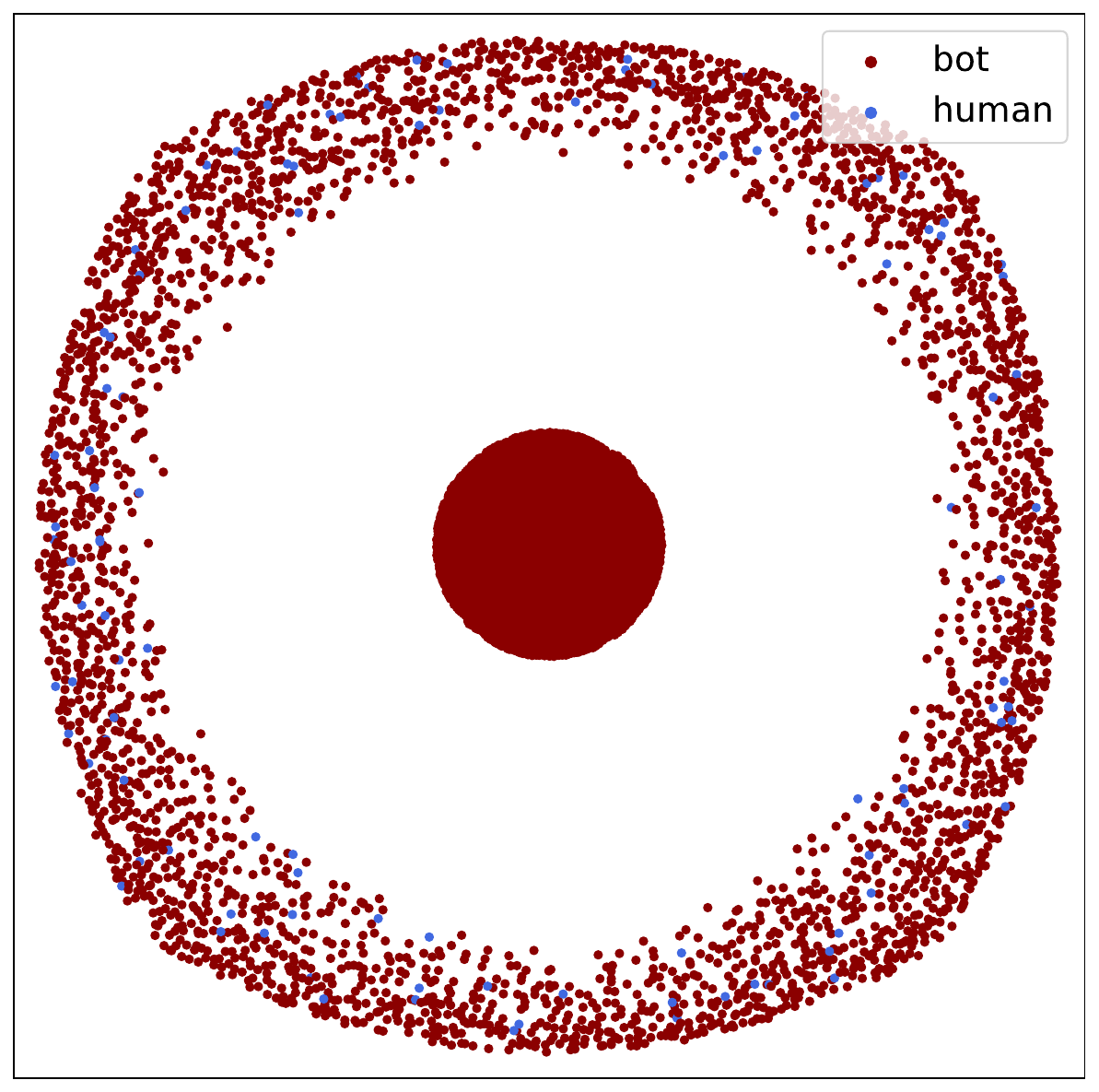}
        \end{minipage}%
    }

    \subfigure[\revised{The optimal encoding tree of Pronbots-2019 dataset in \framework{} (square tree nodes are communities, circular leaf nodes are social users, and the size of the nodes indicates the number of social users. For visualization, each large leaf node represents 200 social users of Pronbots-2019).}]{
        \label{fig: tree_pronbots19}
        \begin{minipage}[t]{1\linewidth}
            \centering
            \includegraphics[width=1\linewidth]{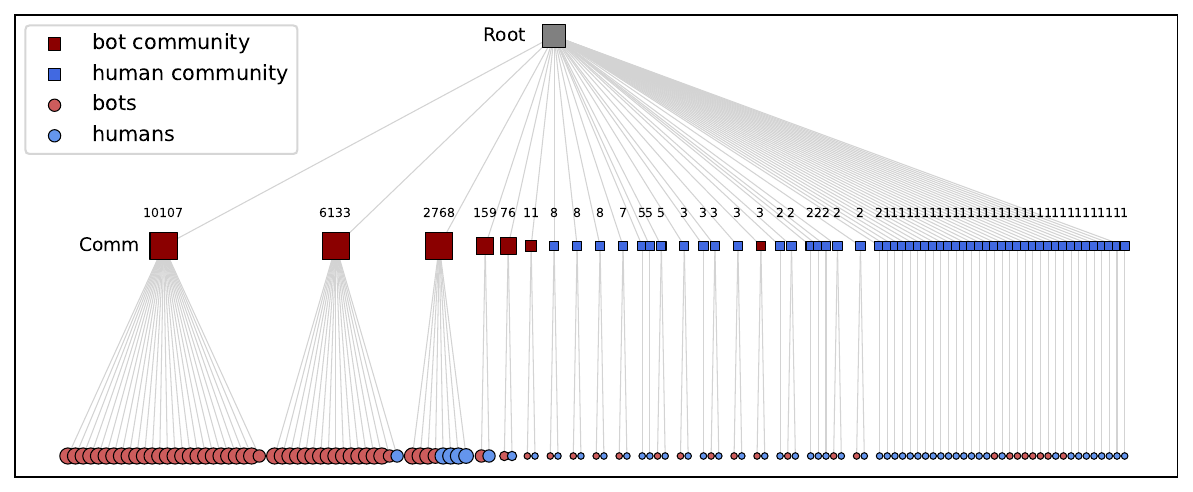}
        \end{minipage}%
    }
    
    \caption{Renderings of Pronbots-2019.}
    \label{fig: rendering_pronbots19}
\end{figure}

\begin{figure}[thp]
    \centering    
    \subfigure[Network with true label.]{
        \label{fig: network_true_cresci15}
        \begin{minipage}[t]{0.49\linewidth}
            \centering
            \includegraphics[width=1.0\linewidth]{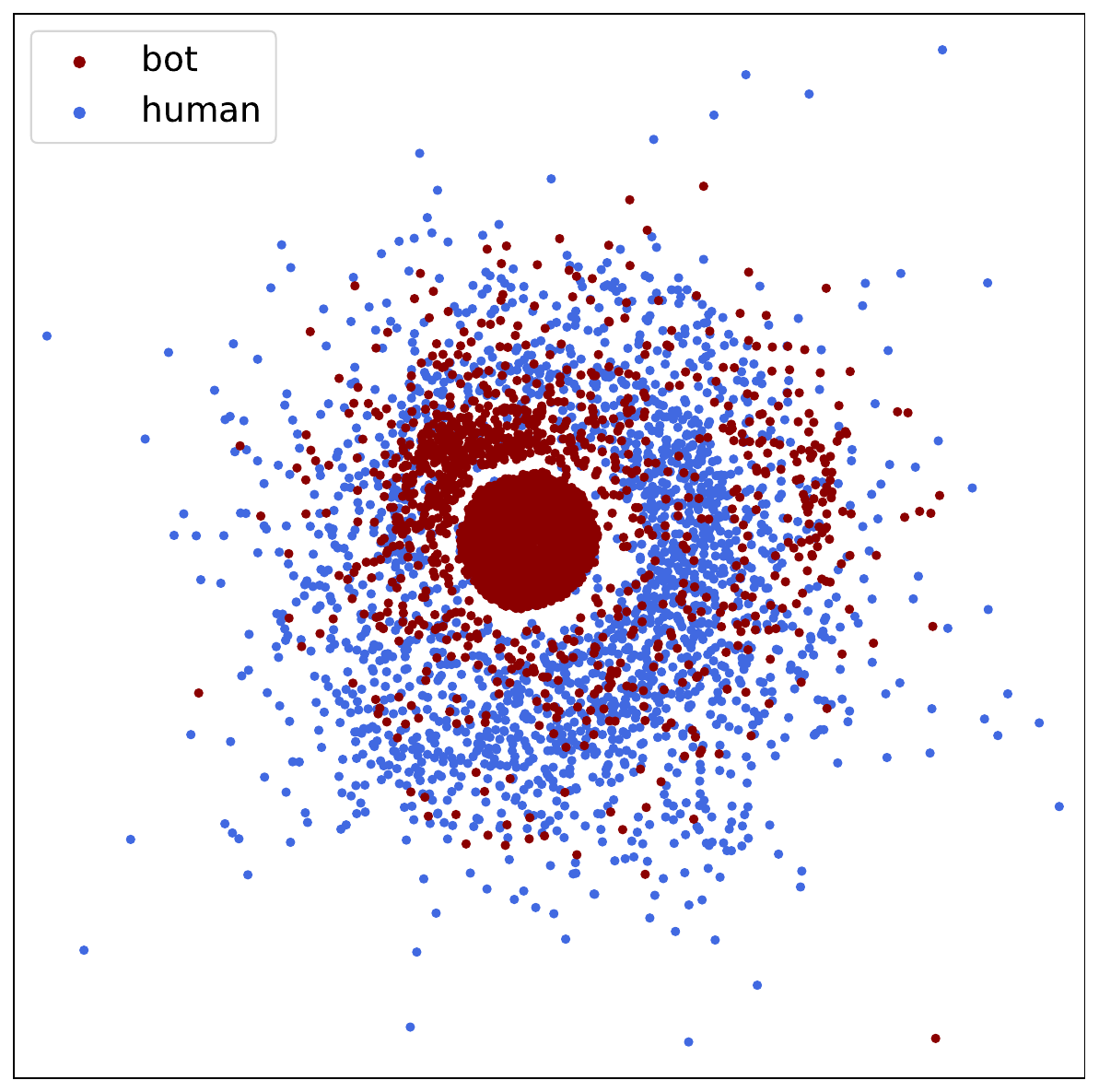}
        \end{minipage}%
    }
    \subfigure[Network with pred label.]{
        \label{fig: network_pred_cresci15}
        \begin{minipage}[t]{0.49\linewidth}
            \centering
            \includegraphics[width=1.0\linewidth]{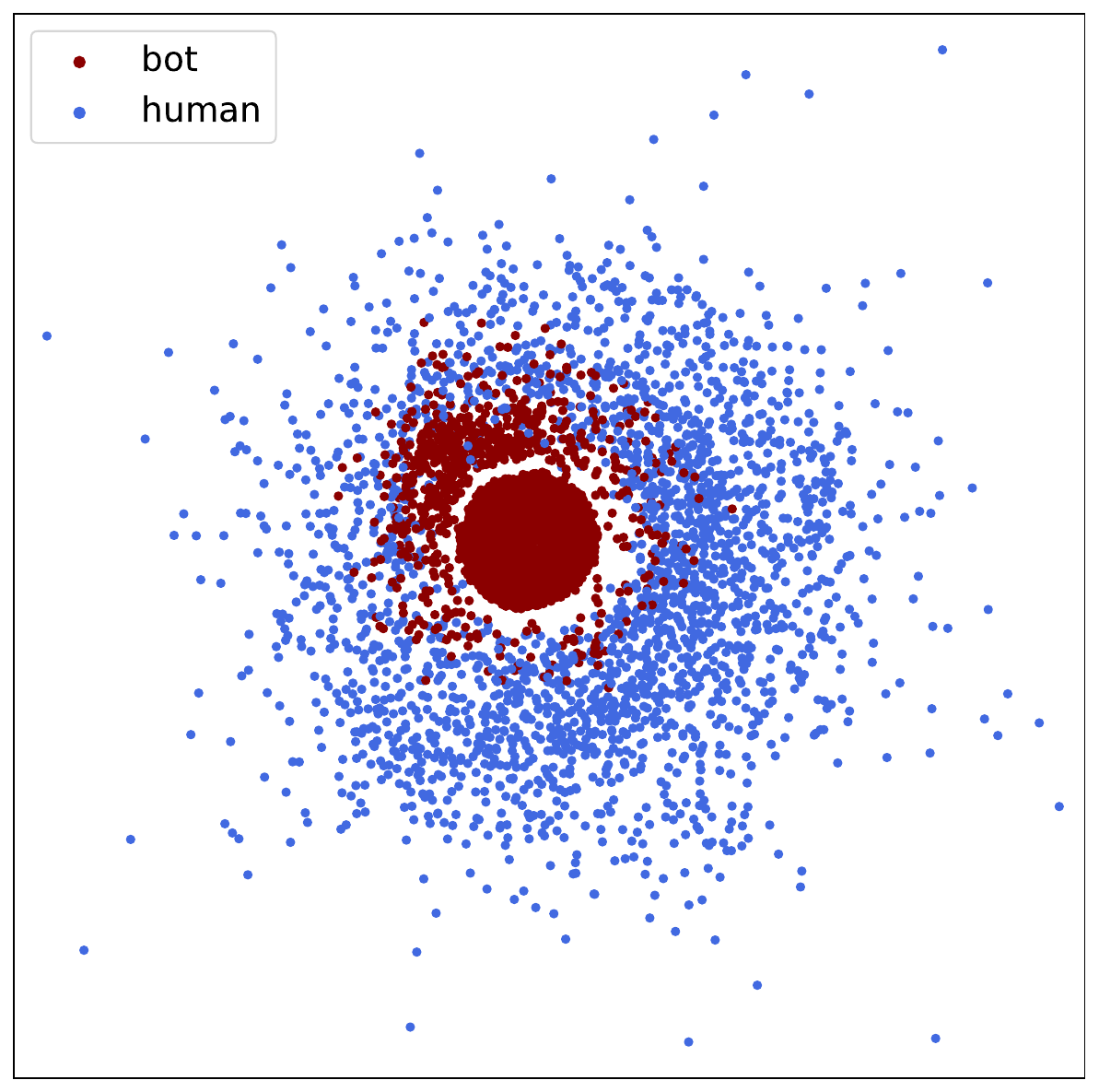}
        \end{minipage}%
    }

    \subfigure[\revised{The optimal encoding tree of Cresci-2015 dataset in \framework{} (square tree nodes are communities, circular leaf nodes are social users, and the size of the nodes indicates the number of social users. For visualization, each large leaf node represents 100 social users of Cresci-2015).}]{
        \label{fig: tree_cresci15}
        \begin{minipage}[t]{1\linewidth}
            \centering
            \includegraphics[width=1\linewidth]{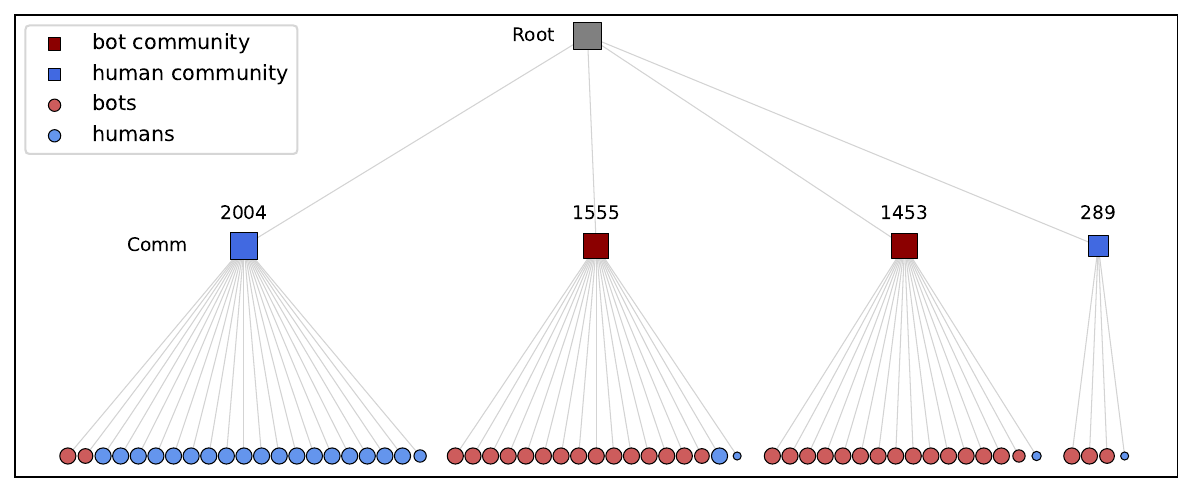}
        \end{minipage}%
    }
    
    \caption{Renderings of Cresci-2015.}
    \label{fig: rendering_cresci15}
\end{figure}

\begin{figure}[thp]
    \centering    
    \subfigure[Network with true label.]{
        \label{fig: network_true_cresci17}
        \begin{minipage}[t]{0.49\linewidth}
            \centering
            \includegraphics[width=1.0\linewidth]{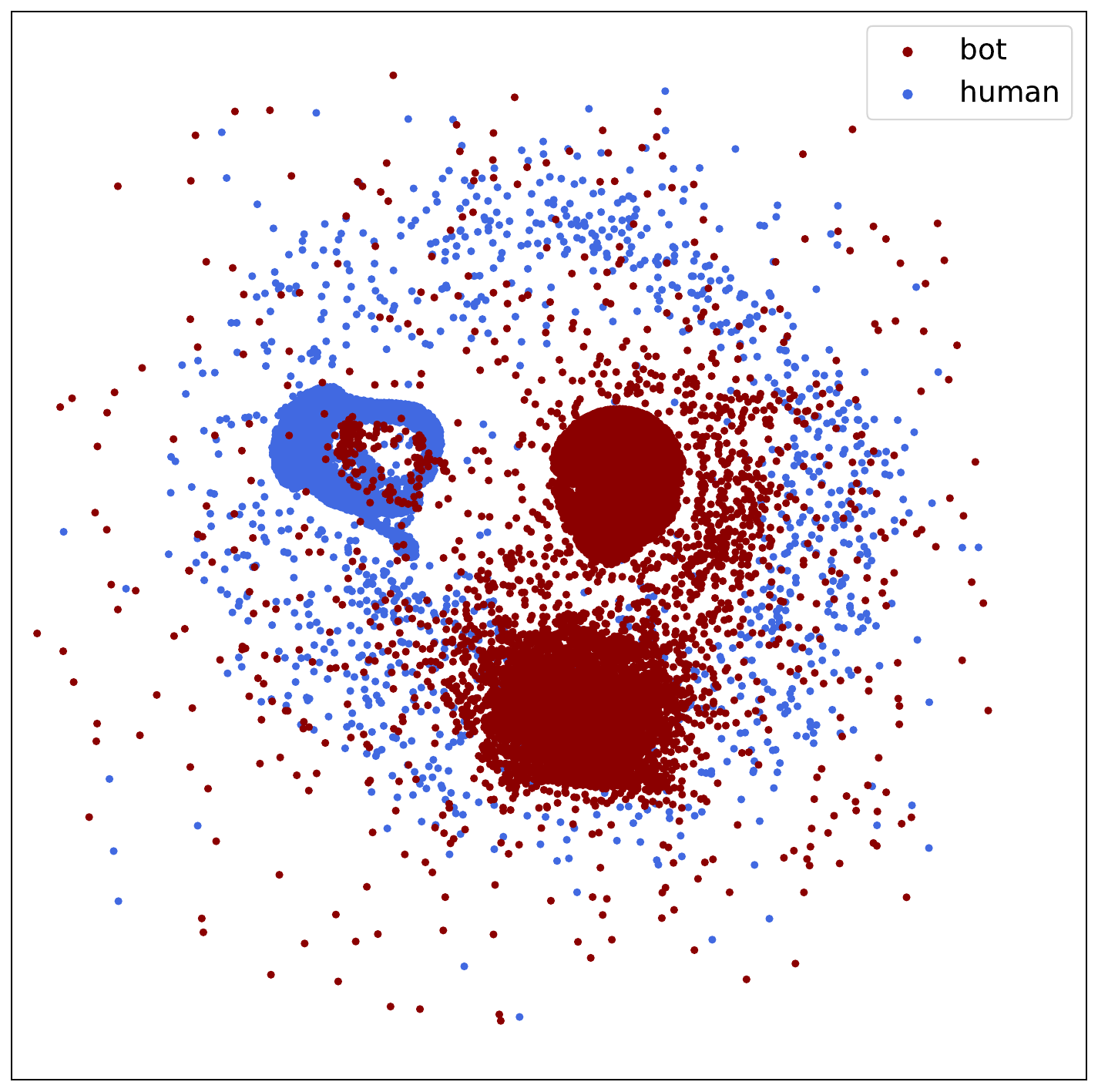}
        \end{minipage}%
    }
    \subfigure[Network with pred label.]{
        \label{fig: network_pred_cresci17}
        \begin{minipage}[t]{0.49\linewidth}
            \centering
            \includegraphics[width=1.0\linewidth]{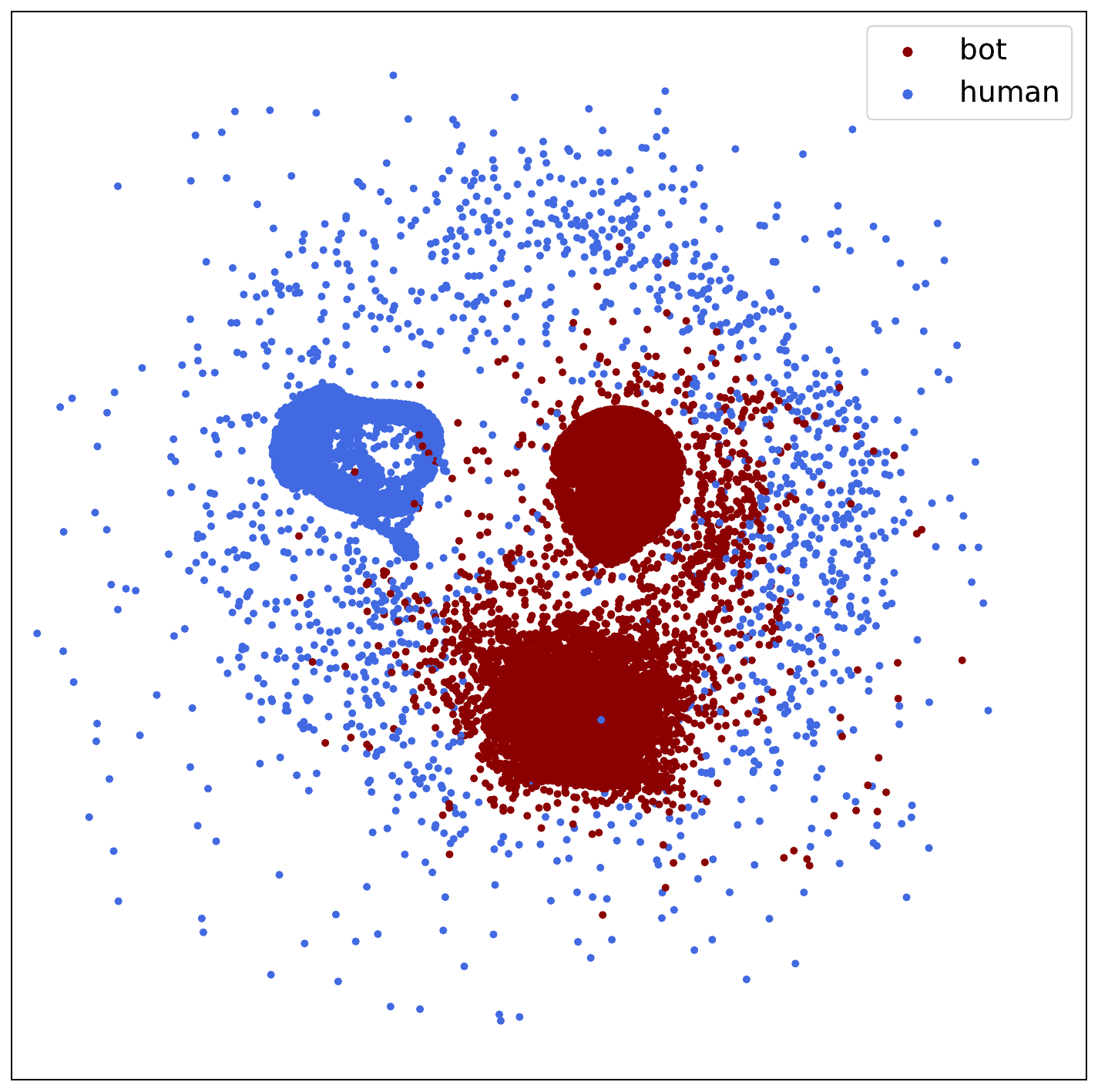}
        \end{minipage}%
    }

    \subfigure[\revised{The optimal encoding tree of Cresci-2017 dataset in \framework{} (square tree nodes are communities, circular leaf nodes are social users, and the size of the nodes indicates the number of social users. For visualization, each large leaf node represents 200 social users of Cresci-2017).}]{
        \label{fig: tree_cresci17}
        \begin{minipage}[t]{1\linewidth}
            \centering
            \includegraphics[width=1\linewidth]{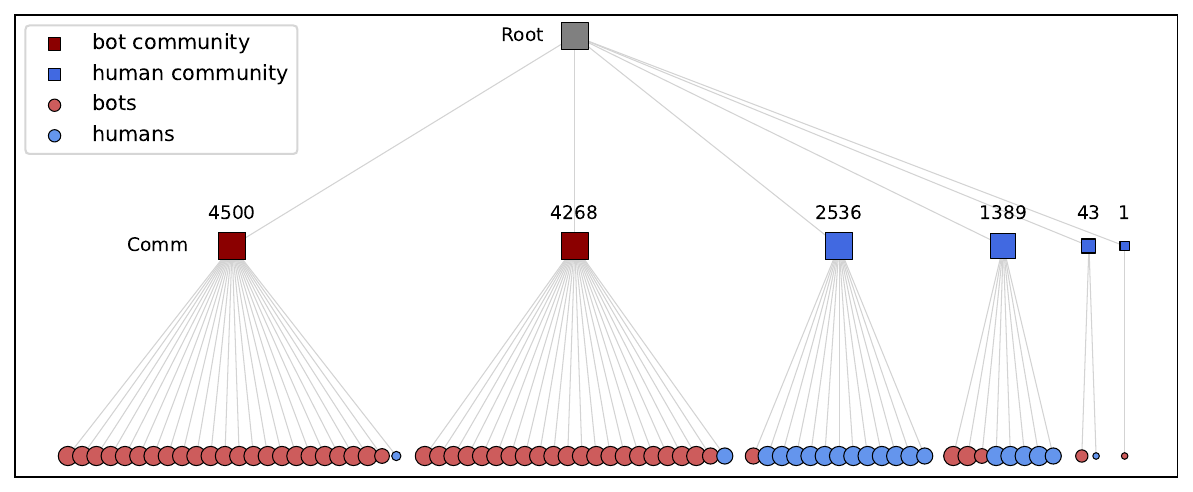}
        \end{minipage}%
    }
    
    \caption{Renderings of Cresci-2017.}
    \label{fig: rendering_cresci17}
\end{figure}

Interpretability refers to the extraction of knowledge about the relationships contained in the data from a model, providing insights to the audience regarding the detection of social bots~\cite{murdoch2019definitions}.
To gain a deeper understanding of the interpretability of the \framework{}, we visualize social user network graphs and structural entropy encoding trees on four datasets.
The social user networks in Figure~\ref{fig: rendering_botwiki}, Figure~\ref{fig: rendering_pronbots19}, Figure~\ref{fig: rendering_cresci15}, and Figure~\ref{fig: rendering_cresci17} utilize the spring\_layout algorithm~\cite{kobourov2012spring} to calculate the positional parameters of each user node, taking into account both the nodes and edges. 
This algorithm uses Fruchterman-Reingold~\cite{gajdovs2016parallel} to arrange the nodes, introducing a physical model into the nodes so that the overall layout achieves a dynamic balance by minimizing the system's total energy, resulting in nodes with connected edges being closer in position.
In social user networks, red dots represent social bots, and blue dots represent genuine users. 
Figure~\ref{fig: network_true_botwiki19}, Figure~\ref{fig: network_true_pronbots19}, Figure~\ref{fig: network_true_cresci15}, and Figure~\ref{fig: network_true_cresci17} show the network with true labels of users, while Figure~\ref{fig: network_pred_botwiki19}, Figure~\ref{fig: network_pred_pronbots19}, Figure~\ref{fig: network_pred_cresci15} and Figure~\ref{fig: network_pred_cresci17} depict the network with labels obtained through the detection of \framework{}.
In the structure entropy encoding tree, square tree nodes represent social user communities, and circular leaf nodes represent social users. 
Blue nodes represent human communities (users), while red nodes represent social bot communities (users). 
The numbers above the communities indicate the number of users in each community, representing the community size.\par

\revised{Figure~\ref{fig: network_true_botwiki19}, Figure~\ref{fig: network_true_pronbots19}, Figure~\ref{fig: network_true_cresci15}, and Figure~\ref{fig: network_true_cresci17} illustrate that social bots are concentrated in the distribution of the multi-relational graph ($MG$) constructed based on social behavior similarity while humans are distributed more widely on the periphery.}
Social bots are typically controlled by machines or programs, serving specific purposes such as advertising, information theft, or inciting public opinion. 
So their behavior tends to be more similar compared to humans, resulting in a higher density of connections among social bots in the $MG$, whereas connections between humans or between humans and social bots are relatively sparse.
\framework{} intends to identify communities with a higher average number of connections (greater influence) and a higher number of internal connections (stronger cohesion) as social bot communities. 
Figure~\ref{fig: network_pred_botwiki19}, Figure~\ref{fig: network_pred_pronbots19},Figure~\ref{fig: network_pred_cresci15}, and Figure~\ref{fig: network_pred_cresci17} illustrate the identification results of \framework{} for social users. 
Users with dense distributions are classified as social bots, while users with relatively dispersed distributions are identified as humans.\par

Figure~\ref{fig: tree_botwiki19}, Figure~\ref{fig: tree_pronbots19}, Figure~\ref{fig: tree_cresci15}, and Figure~\ref{fig: tree_cresci17} depict the community division on the encoding tree, respectively. 
\revised{To provide a clearer representation, each large circular leaf node represents 30 users in Botwiki-2019, 300 users in Pronbots-2019, 100 users in Cresci-2015, and 200 users in Cresci-2017.}
The smaller the diameter of the node, the fewer users it represents. 
The encoding tree partitions social bots and humans into different communities.
\revised{In Figure~\ref{fig: tree_botwiki19} and Figure~\ref{fig: tree_pronbots19}}, the communities exhibit a long-tail distribution, with fewer large-scale communities and more small-scale communities.
The communities marked as social bot communities in Botwiki-2019 and Pronbots-2019 are predominantly large-scale communities.
On the other hand, in the encoding trees of Cresci-2015 and Cresci-2017, the distribution of community sizes is relatively uniform, and there is a high proportion of single-type user components within each community.\par

\framework{} employs structural information theory to decode the essential structure of the social bot network from the multi-relationship graph. 
It extracts knowledge of user community division from the encoding tree.
From the perspective of the network modeled by behavioral similarity, the distinction between social bots and humans lies in the fact that social bots exhibit dense distributions and strong intra-community cohesion. 
This provides interpretable insights for the detection of social bots by \framework{}.

\subsection{Time Analysis}
\label{sec: efficient}
\begin{table}[thp]
    \caption{\revised{Average time per run for \framework{} and all baselines (Unit: Second).}}
    \label{tab: efficient}
    \aboverulesep=0ex
    \belowrulesep=0ex
    \centering
    \renewcommand\arraystretch{1.3}
    \begin{tabular}{m{2.0cm}<{\centering}|m{2.5cm}<{\centering} m{2.5cm}<{\centering} m{2.5cm}<{\centering} m{2.5cm}<{\centering}}
        \toprule
        Method & \textbf{Cresci-2015} & \textbf{Cresci-2017} & \textbf{Pronbots-2019} & \textbf{Botwiki-2019} \\
        \hline
        K-means & 4.39 & 8.76 & \revised{5.78} & \revised{4.05} \\
        DNA & 63052.45 & 224525.64 & \revised{43832.82} & \revised{419.31} \\
        GAE & 2.36 & 1.63 & \revised{2.07} & \revised{0.39} \\
        DeepWalk & 60.78 & 308.22 & \revised{1841.80} & \revised{2.78} \\
        LINE & 859.57 & 4542.45 & \revised{49379.05} & \revised{214.55} \\
        Node2vec & 39.38 & 165.18 & \revised{1308.81} & \revised{8.47} \\
        SDNE & 22.12 & 127.65 & \revised{792.07} & \revised{1.48} \\
        GraRep & 179.92 & 925.80 & \revised{3394.07} & \revised{4.85} \\
        DNGR & 47.44 & 209.86 & \revised{1346.57} & \revised{1.64} \\
        HOPE & 24.42 & 141.69 & \revised{1167.06} & \revised{1.15} \\
        \hline
        \textbf{\framework{}} & 7.37 & 55.98 & \revised{541.86} & \revised{1.12}\\
        \bottomrule
    \end{tabular}
\end{table}
This section evaluates the efficiency of \framework{} and various baselines. 
We primarily assess efficiency by measuring the runtime of the models. 
To facilitate a fairer comparison for unsupervised graph learning-based models, we conduct temporal analysis on the multi-relational social user graph $MG$. 
Table~\ref{tab: efficient} presents the average runtime of each model over 10 runs. 
Overall, \framework{} achieves a balance between runtime and effectiveness. 
Although the \emph{K-means} model is too simplistic and the $GAE$ package allows for parallel computation to accelerate the training process, resulting in less required running time, the accuracy and precision achieved by both of them are significantly lower than \framework{} across all datasets, particularly on Pronbots-2019 and Botwiki-2019 where the data types are relatively homogeneous.
It is worth noting that, compared to other baselines, \framework{}'s runtime is generally reduced by over threefold.
The runtime of the $DNA$ model is the longest, as it involves considering all possible combinations of actions for tweeting.

\begin{figure}[thp]
    \centering
    \subfigure[\emph{Follow-To-Follower} Statistics.]{
        \label{fig: follow-to-follower}
        \begin{minipage}[t]{0.49\linewidth}
            \centering
            \includegraphics[width=1.0\linewidth]{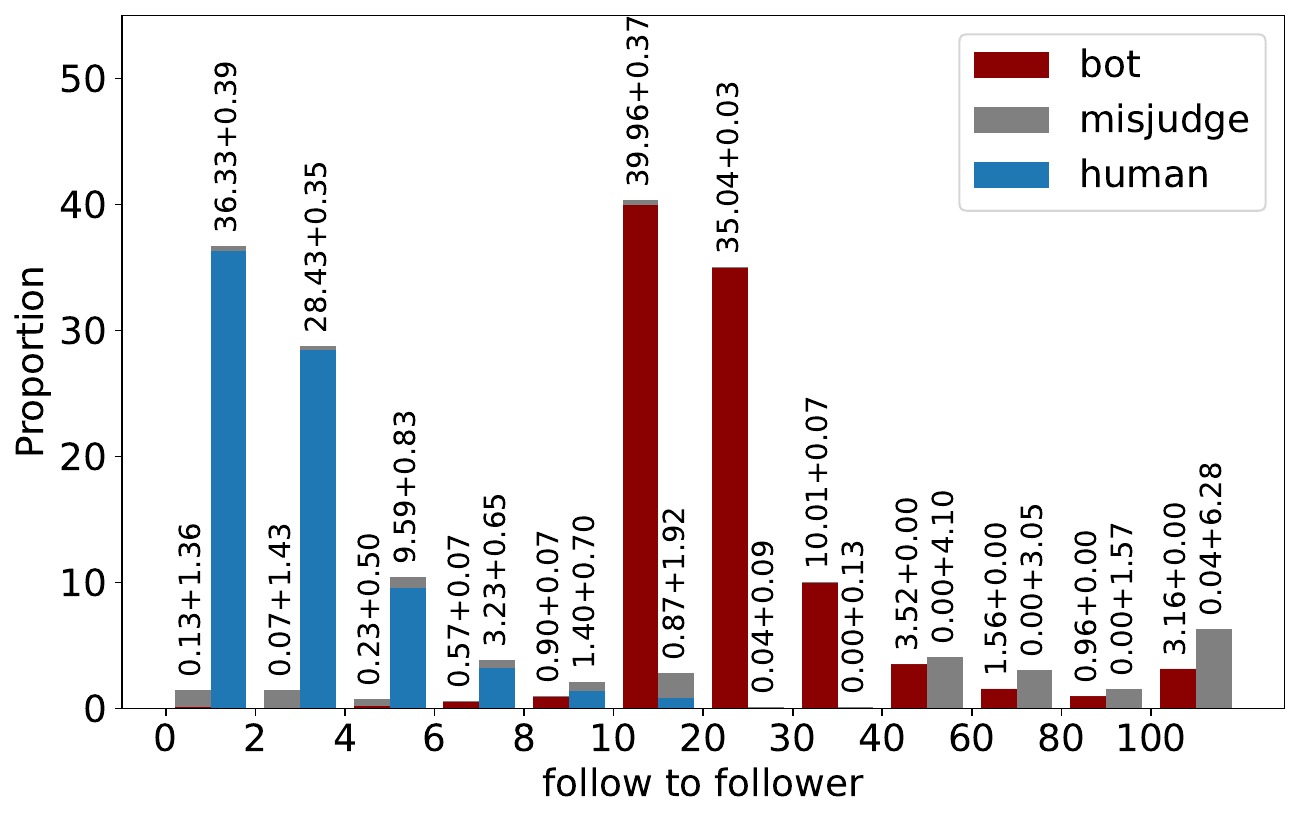}
        \end{minipage}%
    }
    \subfigure[\emph{Posting Influence} Statistics.]{
        \label{fig: influence}
        \begin{minipage}[t]{0.49\linewidth}
            \centering
            \includegraphics[width=1.0\linewidth]{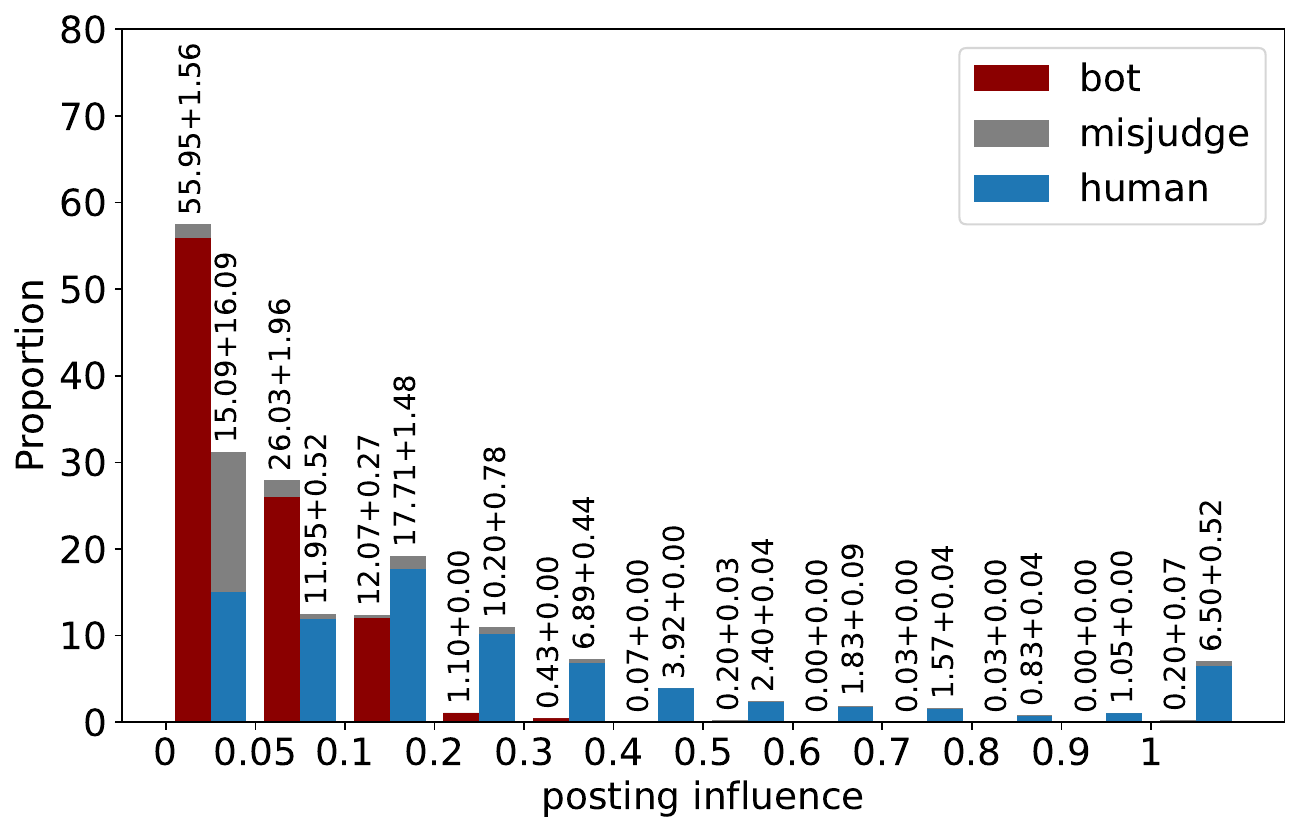}
        \end{minipage}%
    }

    \subfigure[\emph{Posting Type Distribution} Statistics on Social Bot.]{
        \label{fig: post type bot}
        \begin{minipage}[t]{0.49\linewidth}
            \centering
            \includegraphics[width=1.0\linewidth]{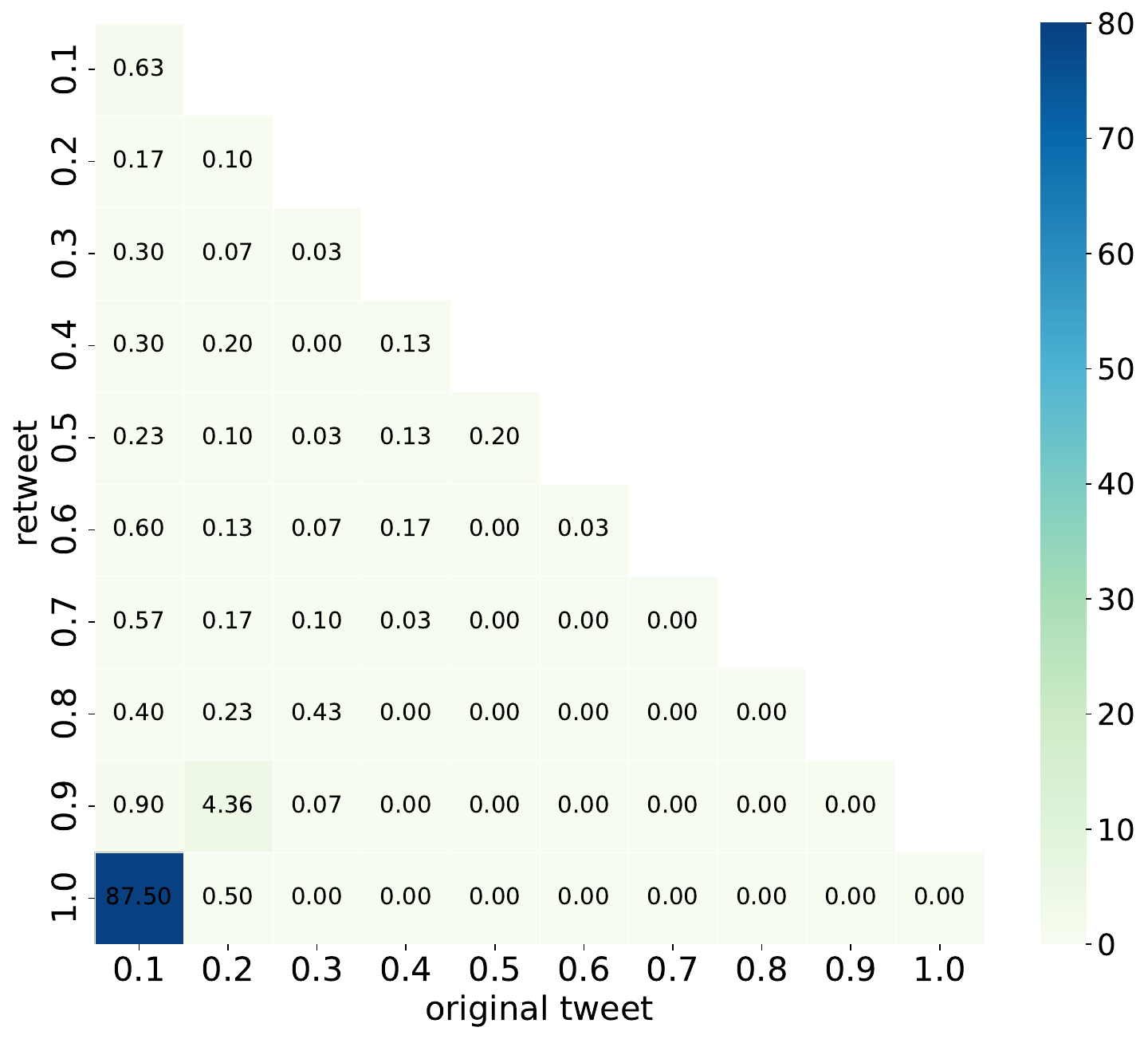}
        \end{minipage}%
    }
    \subfigure[\emph{Posting Type Distribution} Statistics on Social Human.]{
        \label{fig: post type human}
        \begin{minipage}[t]{0.49\linewidth}
            \centering
            \includegraphics[width=1.0\linewidth]{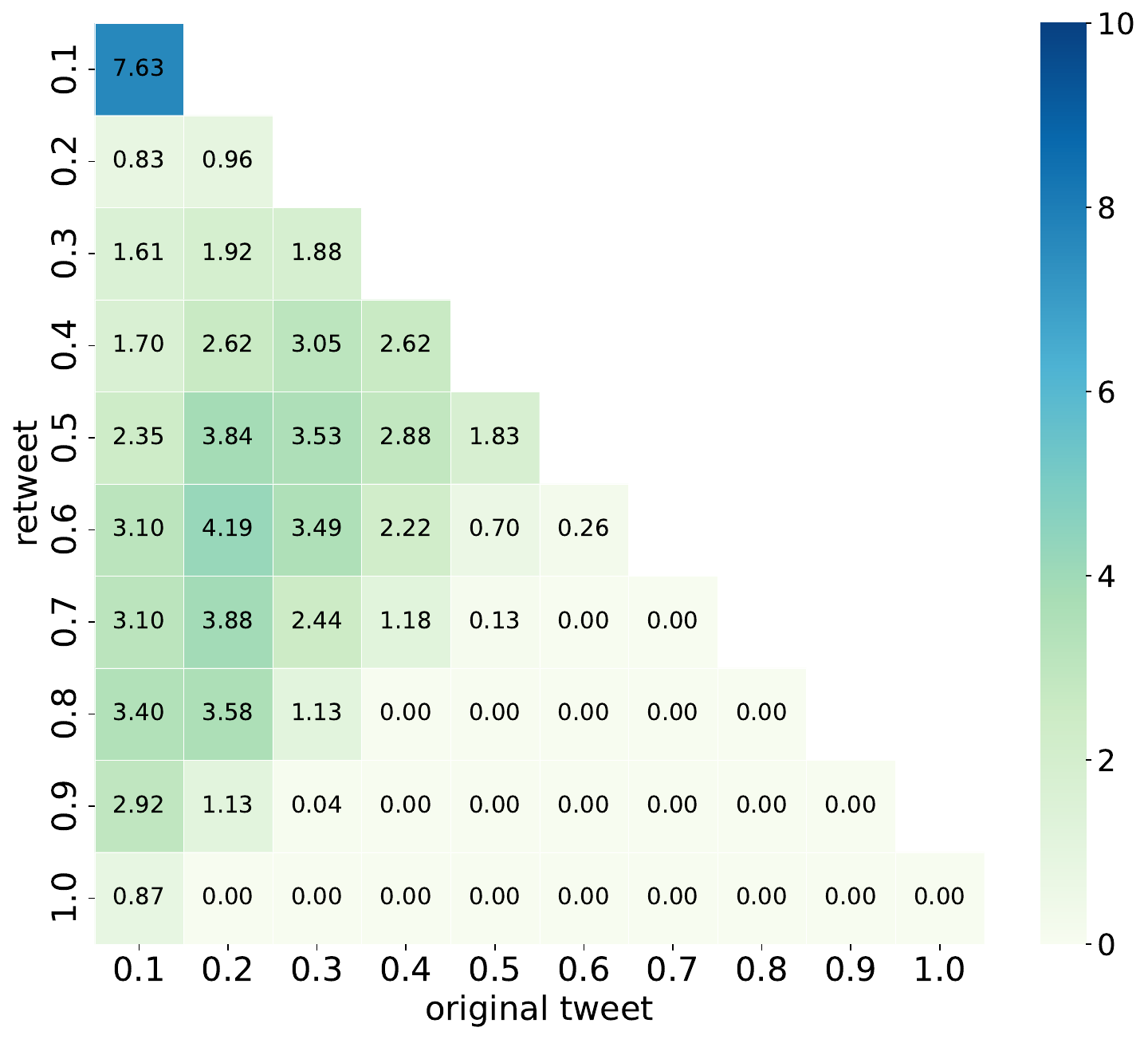}
        \end{minipage}%
    }

    \caption{Feature Statistics}
    \label{fig: Feature Statistics}
\end{figure}
\subsection{Case Study}
\label{sec: case study}
To further understand the distinctions between social bots and humans, we compare various characteristics under \framework{}, such as the follower-to-following ratio, posting type distribution, and posting influence. 
We collate the statistical distribution of identified social bots and humans for each feature value separately. 
This analysis aims to determine whether these social users align with human judgment.
Figure~\ref{fig: Feature Statistics} displays the results of the statistical analysis of Cresci-2015 based on the identification performance of \framework{}. 
For one-dimensional features such as \emph{follow-to-follower ratio} and \emph{posting influence}, bar charts Figures~\ref{fig: follow-to-follower} and ~\ref{fig: influence} illustrate the proportion of users falling into each interval, with the horizontal axis representing the value intervals of the feature values and the vertical axis representing the proportion of users. 
Red represents the social bots identified by \framework{}, blue represents humans, and gray represents misjudged users. 
Each bar consists of two components: the proportion of social bots (humans) and the misjudged ratio.
Heat maps Figures~\ref{fig: post type bot} and ~\ref{fig: post type human} depict the proportions of social bots and humans falling into each interval of the \emph{posting type distribution}. 
The horizontal axis represents the proportion of original tweets, while the vertical axis represents the proportion of retweets.
Darker colors indicate a higher proportion of users.
It can be found from the results that the social bots identified by \framework{} generally have a high follow-to-following ratio, very low posting influence, and the type of posting is single (such as retweeting), which is consistent with the human judgment of social bots.
Furthermore, the main reason for misjudgments is that the characteristics of some social bots have not been identified. 
This is because they are relatively similar to humans in certain characteristics or have a smaller scale. 
This also provides us with inspiration for future research directions. 
When the behavior of social bots under AI control is almost indistinguishable from humans or when the behavior similarity between social bots is not high, how do we explore more effective features to identify them?
\section{Related Work}
\label{sec: Related Work}
In this section, we \revised{summarize the related literature, baselines, and approaches related to the proposed framework.}
It is primarily divided into \revised{three} parts: \revised{social bot detection, network embedding, and structural information theory based applications}.

\subsection{Social Bot Detection}
\revised{Early social bot detection methods were mainly based on the user's inherent characteristics and text content, such as the length of the username, number of tweets, number of fans, number of friends, number of likes in the username information, part-of-speech features, frequency features in the tweet information, emotional characteristics, etc.}
\revised{The earliest method~\cite{miller2014twitter} treated spam identification as anomaly detection to handle the streaming nature of tweets effectively.}
\revised{Subsequently, a method of analyzing digital DNA sequences in user behavior to model online user behavior~\cite{cresci2016dna} provided a new idea for behavioral representation.}
\revised{And fully analyzing users' basic information and social activities~\cite{efthimion2018supervised} has been proven to improve the accuracy.}
\revised{A later proposed multi-model “toolbox” approach~\cite{beskow2019its} focused on random string detection in usernames, while language-independent different feature groups based on specific characteristics~\cite{knauth2019language} demonstrated the effectiveness of a small number of expressive features was demonstrated.}
\revised{However, the performance of traditional machine learning methods is often limited due to its low-dimensional and low-order features.}
\revised{With the development of graph neural networks, deep learning technologies are applied to social bot detection tasks.}
\revised{BiLSTM~\cite{wei2019twitter} proposed the use of bidirectional long short-term memory recurrent neural network to effectively capture the characteristics of tweets.}
\revised{The recent BotRGCN~\cite{feng2021botrgcn} and Bot-AHGCN~\cite{zhao2020multi} used graph convolutional network training data to enhance the ability to capture social bots with multiple disguises.}
\revised{In addition, large language models have also been applied in bot detection. }
\revised{BGSRD~\cite{guo2021social} combined large-scale pre-training and transduction learning to symmetrically combine BERT and the graph convolution network GCN for joint training. }
\revised{However, the above data-driven models require a large number of labeled social user samples and even run the risk of weak generalization ability due to the lack of interpretability.}
\revised{Here the proposed work addresses this issue by remodeling the multi-relational social user graph and designing user community division and classification indicators based on structural entropy.}

\subsection{Network Embedding}
\label{subsec: network embedding}
The term "network embedding" mentioned here mainly refers to embedding methods that preserve the structure of a network (excluding node attributes and side information). 
It is an important tool for studying complex structural systems and processing network data.
Common models for network embedding include methods based on random walks, matrix factorization, and deep neural networks~\cite{cui2018survey}. 
Like Word2vec, random walk-based methods focus on learning the neighborhood structure by maximizing the probability of neighboring nodes during the random walk process.
The latest Bot2vec~\cite{pham2022bot2vec} model designs a new network representation learning method specifically for social bot detection, which automatically preserves the neighbor relationship and community structure of users.
Matrix factorization methods represent the network by learning the low-rank space of the adjacency matrix. 
Deep neural network methods, such as SDAE~\cite{cao2016deep}, SINE~\cite{wang2017signed}, and LIME~\cite{peng2021lime} introduce nonlinear functions to learn highly complex and nonlinear networks. 
Network embedding provides important support for downstream tasks such as node classification, clustering, anomaly detection~\cite{hu2016embedding}, and link prediction. 
In node classification~\cite{perozzi2014deepwalk,tang2015line,peng2020dynamic}, network embedding provides effective feature representations, thereby improving the effectiveness of node classification. 
It has been widely applied in fields such as social networks~\cite{huang2017label}, citation networks~\cite{pan2016tri}, and biological networks~\cite{grover2016node2vec}. 
In anomaly detection, the advantage of network embedding lies in its ability to encode the structural information of the network into vector representations, thereby better capturing the relationships between nodes. 
Compared to traditional anomaly detection methods, network embedding can more accurately identify abnormal nodes and can handle large-scale network data. In addition, network embedding also has another important function, which is network visualization. 
Network embedding can generate meaningful node layouts by providing low-dimensional representations of nodes and displaying complex networks in two-dimensional space for further research and analysis.

\subsection{\revised{Structural Information Theory based Applications}}
The Structural Information Theory decoding network's ability to capture the structure's essence has been validated in many applications. 
Introducing structural entropy in neural networks captures the underlying connectivity graph and reduces random interference~\cite{wang2023user}. 
\revised{The hierarchical nature of the structure entropy encoding tree provides new methods for hierarchical structure pooling in graph neural network~\cite{wu2022structural}, unsupervised image segmentation~\cite{zeng2023unsupervised}, dimension estimation~\cite{yang2023minimum}, state abstraction~\cite{zeng2023hierarchical} in reinforcement learning, ensemble of constraints in semi-supervised clustering~\cite{zeng2024semi}, and unsupervised social event detection~\cite{cao2024hierarchical}. }
Additionally, reconstructing the graph structure on the hierarchical encoding tree suppresses edge noise and enhances the learning ability of the graph structure~\cite{zou2023se, zou2024multispans}. 
Furthermore, modifying the network structure based on minimizing structural entropy achieves maximum deception of community structure~\cite{liu2019rem}. 
Similarly, the anchor view, guided by the principle of minimizing structural entropy, improves the performance of graph contrastive learning~\cite{wu2023sega}.
\section{Conclusion}
\label{sec: conclusion} 
This paper investigates a framework for supporting the unsupervised detection of social bots. 
The proposed \framework{} framework adaptively performs hierarchical community partitioning and identifies social bot communities during social bot detection. 
By modeling social user networks based on social behavior similarity, we effectively enhance the connections between social bots. 
The parallel execution of fusion operators during community partitioning maintains a stable community structure while ensuring high operational efficiency. 
Unsupervised detection is accomplished by employing a community binary classification that differentiates social bots from humans based on influence and cohesion.
Experimental results demonstrate that \framework{} outperforms all unsupervised models in terms of accuracy, achieving a balance between operational efficiency and performance while exhibiting strong interpretability.
Our work demonstrates the potential of unsupervised social bot detection and may open up new directions for research in social bot detection. 
In the future, our goal is to investigate how graph structure modeling with more behavioral and semantic features can improve the detection effectiveness of the model and expand \framework{} to other detection domains.

\section*{Acknowledgments}
This work is supported by the National Key R\&D Program of China through grant 2022YFB3104700, NSFC through grants 62322202, 61932002, U21B2027, U23A20388, and 62266028, Beijing Natural Science Foundation through grant 4222030, Guangdong Basic and Applied Basic Research Foundation through grant 2023B1515120020, Shijiazhuang Science and Technology Plan Project through grant 231130459A, Foundation of State Key Laboratory of Public Big Data through grant PBD2022-04, Research Fund of Guangxi Key Lab of Multi-source Information Mining \& Security (MIMS23-M-01), Yunnan Provincial Major Science and Technology Special Plan Projects through grants 202302AD080003, 202202AD080003 and 202303AP140008, General Projects of Basic Research in Yunnan Province through grant 202301AS070047, 202301AT070471, and the Fundamental Research Funds for the Central Universities.
Philip S. Yu was partly supported by NSF under grant III-2106758, and POSE-2346158. 
\bibliography{reference}


\begin{thebibliography}{104}


\ifx \showCODEN    \undefined \def \showCODEN     #1{\unskip}     \fi
\ifx \showDOI      \undefined \def \showDOI       #1{#1}\fi
\ifx \showISBNx    \undefined \def \showISBNx     #1{\unskip}     \fi
\ifx \showISBNxiii \undefined \def \showISBNxiii  #1{\unskip}     \fi
\ifx \showISSN     \undefined \def \showISSN      #1{\unskip}     \fi
\ifx \showLCCN     \undefined \def \showLCCN      #1{\unskip}     \fi
\ifx \shownote     \undefined \def \shownote      #1{#1}          \fi
\ifx \showarticletitle \undefined \def \showarticletitle #1{#1}   \fi
\ifx \showURL      \undefined \def \showURL       {\relax}        \fi
\providecommand\bibfield[2]{#2}
\providecommand\bibinfo[2]{#2}
\providecommand\natexlab[1]{#1}
\providecommand\showeprint[2][]{arXiv:#2}

\bibitem[\protect\citeauthoryear{Ali~Alhosseini, Bin~Tareaf, Najafi, and
  Meinel}{Ali~Alhosseini et~al\mbox{.}}{2019}]%
        {ali2019detect}
\bibfield{author}{\bibinfo{person}{Seyed Ali~Alhosseini}, \bibinfo{person}{Raad
  Bin~Tareaf}, \bibinfo{person}{Pejman Najafi}, {and}
  \bibinfo{person}{Christoph Meinel}.} \bibinfo{year}{2019}\natexlab{}.
\newblock \showarticletitle{Detect me if you can: Spam bot detection using
  inductive representation learning}. In \bibinfo{booktitle}{\emph{Companion
  Proceedings of The 2019 World Wide Web Conference}}.
  \bibinfo{pages}{148--153}.
\newblock


\bibitem[\protect\citeauthoryear{Allem and Ferrara}{Allem and Ferrara}{2018}]%
        {Allem2018could}
\bibfield{author}{\bibinfo{person}{Jon-Patrick Allem} {and}
  \bibinfo{person}{Emilio Ferrara}.} \bibinfo{year}{2018}\natexlab{}.
\newblock \showarticletitle{Could Social Bots Pose a Threat to Public Health?}
\newblock \bibinfo{journal}{\emph{American journal of public health}}
  \bibinfo{volume}{108} (\bibinfo{date}{08} \bibinfo{year}{2018}),
  \bibinfo{pages}{1005--1006}.
\newblock


\bibitem[\protect\citeauthoryear{Anwar and Yaqub}{Anwar and Yaqub}{2020}]%
        {anwar2020bot}
\bibfield{author}{\bibinfo{person}{Ahmed Anwar} {and} \bibinfo{person}{Ussama
  Yaqub}.} \bibinfo{year}{2020}\natexlab{}.
\newblock \showarticletitle{Bot detection in twitter landscape using
  unsupervised learning}. In \bibinfo{booktitle}{\emph{Proceedings of the 21st
  Annual International Conference on Digital Government Research}}.
  \bibinfo{pages}{329--330}.
\newblock


\bibitem[\protect\citeauthoryear{Arin and Kutlu}{Arin and Kutlu}{2023}]%
        {arin2023deep}
\bibfield{author}{\bibinfo{person}{Efe Arin} {and} \bibinfo{person}{Mucahid
  Kutlu}.} \bibinfo{year}{2023}\natexlab{}.
\newblock \showarticletitle{Deep learning based social bot detection on
  twitter}.
\newblock \bibinfo{journal}{\emph{IEEE Transactions on Information Forensics
  and Security}}  \bibinfo{volume}{18} (\bibinfo{year}{2023}),
  \bibinfo{pages}{1763--1772}.
\newblock


\bibitem[\protect\citeauthoryear{Beskow and Carley}{Beskow and Carley}{2019}]%
        {beskow2019its}
\bibfield{author}{\bibinfo{person}{David~M Beskow} {and}
  \bibinfo{person}{Kathleen~M Carley}.} \bibinfo{year}{2019}\natexlab{}.
\newblock \showarticletitle{Its all in a name: detecting and labeling bots by
  their name}.
\newblock \bibinfo{journal}{\emph{Computational and mathematical organization
  theory}} \bibinfo{volume}{25}, \bibinfo{number}{1} (\bibinfo{year}{2019}),
  \bibinfo{pages}{24--35}.
\newblock


\bibitem[\protect\citeauthoryear{Breuer, Eilat, and Weinsberg}{Breuer
  et~al\mbox{.}}{2020}]%
        {breuer2020friend}
\bibfield{author}{\bibinfo{person}{Adam Breuer}, \bibinfo{person}{Roee Eilat},
  {and} \bibinfo{person}{Udi Weinsberg}.} \bibinfo{year}{2020}\natexlab{}.
\newblock \showarticletitle{Friend or faux: Graph-based early detection of fake
  accounts on social networks}. In \bibinfo{booktitle}{\emph{Proceedings of the
  Web conference}}. \bibinfo{pages}{1287--1297}.
\newblock


\bibitem[\protect\citeauthoryear{Cao, Lu, and Xu}{Cao et~al\mbox{.}}{2015}]%
        {cao2015grarep}
\bibfield{author}{\bibinfo{person}{Shaosheng Cao}, \bibinfo{person}{Wei Lu},
  {and} \bibinfo{person}{Qiongkai Xu}.} \bibinfo{year}{2015}\natexlab{}.
\newblock \showarticletitle{Grarep: Learning graph representations with global
  structural information}. In \bibinfo{booktitle}{\emph{Proceedings of the 24th
  ACM international on conference on information and knowledge management}}.
  \bibinfo{pages}{891--900}.
\newblock


\bibitem[\protect\citeauthoryear{Cao, Lu, and Xu}{Cao et~al\mbox{.}}{2016}]%
        {cao2016deep}
\bibfield{author}{\bibinfo{person}{Shaosheng Cao}, \bibinfo{person}{Wei Lu},
  {and} \bibinfo{person}{Qiongkai Xu}.} \bibinfo{year}{2016}\natexlab{}.
\newblock \showarticletitle{Deep neural networks for learning graph
  representations}. In \bibinfo{booktitle}{\emph{Proceedings of the AAAI
  conference on artificial intelligence}}. \bibinfo{pages}{1145--1152}.
\newblock


\bibitem[\protect\citeauthoryear{Cao, Peng, Yu, and Yu}{Cao
  et~al\mbox{.}}{2024}]%
        {cao2024hierarchical}
\bibfield{author}{\bibinfo{person}{Yuwei Cao}, \bibinfo{person}{Hao Peng},
  \bibinfo{person}{Zhengtao Yu}, {and} \bibinfo{person}{Philip~S. Yu}.}
  \bibinfo{year}{2024}\natexlab{}.
\newblock \showarticletitle{Hierarchical and Incremental Structural Entropy
  Minimization for Unsupervised Social Event Detection}. In
  \bibinfo{booktitle}{\emph{Proceedings of the AAAI Conference on Artificial
  Intelligence}}, Vol.~\bibinfo{volume}{38}. \bibinfo{pages}{8255--8264}.
\newblock


\bibitem[\protect\citeauthoryear{Chavoshi, Hamooni, and Mueen}{Chavoshi
  et~al\mbox{.}}{2016}]%
        {chavoshi2016debot}
\bibfield{author}{\bibinfo{person}{Nikan Chavoshi}, \bibinfo{person}{Hossein
  Hamooni}, {and} \bibinfo{person}{Abdullah Mueen}.}
  \bibinfo{year}{2016}\natexlab{}.
\newblock \showarticletitle{Debot: Twitter bot detection via warped
  correlation.}. In \bibinfo{booktitle}{\emph{Proceedings of the IEEE ICDM}},
  Vol.~\bibinfo{volume}{18}. \bibinfo{pages}{28--65}.
\newblock


\bibitem[\protect\citeauthoryear{Chen, Yin, Li, Wang, Chen, and Chen}{Chen
  et~al\mbox{.}}{2017b}]%
        {chen2017people}
\bibfield{author}{\bibinfo{person}{Hongxu Chen}, \bibinfo{person}{Hongzhi Yin},
  \bibinfo{person}{Xue Li}, \bibinfo{person}{Meng Wang},
  \bibinfo{person}{Weitong Chen}, {and} \bibinfo{person}{Tong Chen}.}
  \bibinfo{year}{2017}\natexlab{b}.
\newblock \showarticletitle{People opinion topic model: opinion based user
  clustering in social networks}. In \bibinfo{booktitle}{\emph{Proceedings of
  the 26th International Conference on World Wide Web Companion}}.
  \bibinfo{pages}{1353--1359}.
\newblock


\bibitem[\protect\citeauthoryear{Chen, Pacheco, Yang, and Menczer}{Chen
  et~al\mbox{.}}{2021}]%
        {chen2021neutral}
\bibfield{author}{\bibinfo{person}{Wen Chen}, \bibinfo{person}{Diogo Pacheco},
  \bibinfo{person}{Kai-Cheng Yang}, {and} \bibinfo{person}{Filippo Menczer}.}
  \bibinfo{year}{2021}\natexlab{}.
\newblock \showarticletitle{Neutral bots probe political bias on social media}.
\newblock \bibinfo{journal}{\emph{Nature Communications}} \bibinfo{volume}{12},
  \bibinfo{number}{1} (\bibinfo{year}{2021}), \bibinfo{pages}{1--10}.
\newblock


\bibitem[\protect\citeauthoryear{Chen, Tanash, Stoll, and Subramanian}{Chen
  et~al\mbox{.}}{2017a}]%
        {chen2017hunting}
\bibfield{author}{\bibinfo{person}{Zhouhan Chen}, \bibinfo{person}{Rima~S
  Tanash}, \bibinfo{person}{Richard Stoll}, {and} \bibinfo{person}{Devika
  Subramanian}.} \bibinfo{year}{2017}\natexlab{a}.
\newblock \showarticletitle{Hunting malicious bots on twitter: An unsupervised
  approach}. In \bibinfo{booktitle}{\emph{Proceedings of the Social
  Informatics: 9th International Conference}}. Springer,
  \bibinfo{pages}{501--510}.
\newblock


\bibitem[\protect\citeauthoryear{Cheng, Luo, and Yu}{Cheng
  et~al\mbox{.}}{2020}]%
        {cheng2020dynamic}
\bibfield{author}{\bibinfo{person}{Chun Cheng}, \bibinfo{person}{Yun Luo},
  {and} \bibinfo{person}{Changbin Yu}.} \bibinfo{year}{2020}\natexlab{}.
\newblock \showarticletitle{Dynamic mechanism of social bots interfering with
  public opinion in network}.
\newblock \bibinfo{journal}{\emph{Physica A: statistical mechanics and its
  applications}}  \bibinfo{volume}{551} (\bibinfo{year}{2020}),
  \bibinfo{pages}{1--8}.
\newblock


\bibitem[\protect\citeauthoryear{Cresci}{Cresci}{2019}]%
        {cresci2019detecting}
\bibfield{author}{\bibinfo{person}{Stefano Cresci}.}
  \bibinfo{year}{2019}\natexlab{}.
\newblock \showarticletitle{Detecting Malicious Social Bots: Story of a
  Never-Ending Clash}. In \bibinfo{booktitle}{\emph{Proceedings of the
  Disinformation in Open Online Media: First Multidisciplinary International
  Symposium, MISDOOM 2019}}. \bibinfo{publisher}{Springer-Verlag},
  \bibinfo{pages}{77–88}.
\newblock


\bibitem[\protect\citeauthoryear{Cresci}{Cresci}{2020}]%
        {cresci2020decade}
\bibfield{author}{\bibinfo{person}{Stefano Cresci}.}
  \bibinfo{year}{2020}\natexlab{}.
\newblock \showarticletitle{A decade of social bot detection}.
\newblock \bibinfo{journal}{\emph{Commun. ACM}} \bibinfo{volume}{63},
  \bibinfo{number}{10} (\bibinfo{year}{2020}), \bibinfo{pages}{72--83}.
\newblock


\bibitem[\protect\citeauthoryear{Cresci, Di~Pietro, Petrocchi, Spognardi, and
  Tesconi}{Cresci et~al\mbox{.}}{2015}]%
        {cresci2015fame}
\bibfield{author}{\bibinfo{person}{Stefano Cresci}, \bibinfo{person}{Roberto
  Di~Pietro}, \bibinfo{person}{Marinella Petrocchi}, \bibinfo{person}{Angelo
  Spognardi}, {and} \bibinfo{person}{Maurizio Tesconi}.}
  \bibinfo{year}{2015}\natexlab{}.
\newblock \showarticletitle{Fame for sale: Efficient detection of fake Twitter
  followers}.
\newblock \bibinfo{journal}{\emph{Decision Support Systems}}
  \bibinfo{volume}{80} (\bibinfo{year}{2015}), \bibinfo{pages}{56--71}.
\newblock


\bibitem[\protect\citeauthoryear{Cresci, Di~Pietro, Petrocchi, Spognardi, and
  Tesconi}{Cresci et~al\mbox{.}}{2016}]%
        {cresci2016dna}
\bibfield{author}{\bibinfo{person}{Stefano Cresci}, \bibinfo{person}{Roberto
  Di~Pietro}, \bibinfo{person}{Marinella Petrocchi}, \bibinfo{person}{Angelo
  Spognardi}, {and} \bibinfo{person}{Maurizio Tesconi}.}
  \bibinfo{year}{2016}\natexlab{}.
\newblock \showarticletitle{DNA-inspired online behavioral modeling and its
  application to spambot detection}.
\newblock \bibinfo{journal}{\emph{IEEE Intelligent Systems}}
  \bibinfo{volume}{31}, \bibinfo{number}{5} (\bibinfo{year}{2016}),
  \bibinfo{pages}{58--64}.
\newblock


\bibitem[\protect\citeauthoryear{Cresci, Di~Pietro, Petrocchi, Spognardi, and
  Tesconi}{Cresci et~al\mbox{.}}{2017a}]%
        {cresci2017paradigm}
\bibfield{author}{\bibinfo{person}{Stefano Cresci}, \bibinfo{person}{Roberto
  Di~Pietro}, \bibinfo{person}{Marinella Petrocchi}, \bibinfo{person}{Angelo
  Spognardi}, {and} \bibinfo{person}{Maurizio Tesconi}.}
  \bibinfo{year}{2017}\natexlab{a}.
\newblock \showarticletitle{The paradigm-shift of social spambots: Evidence,
  theories, and tools for the arms race}. In
  \bibinfo{booktitle}{\emph{Proceedings of the 26th international conference on
  world wide web companion}}. \bibinfo{pages}{963--972}.
\newblock


\bibitem[\protect\citeauthoryear{Cresci, Di~Pietro, Petrocchi, Spognardi, and
  Tesconi}{Cresci et~al\mbox{.}}{2017b}]%
        {cresci2017social}
\bibfield{author}{\bibinfo{person}{Stefano Cresci}, \bibinfo{person}{Roberto
  Di~Pietro}, \bibinfo{person}{Marinella Petrocchi}, \bibinfo{person}{Angelo
  Spognardi}, {and} \bibinfo{person}{Maurizio Tesconi}.}
  \bibinfo{year}{2017}\natexlab{b}.
\newblock \showarticletitle{Social fingerprinting: detection of spambot groups
  through DNA-inspired behavioral modeling}.
\newblock \bibinfo{journal}{\emph{IEEE Transactions on Dependable and Secure
  Computing}} \bibinfo{volume}{15}, \bibinfo{number}{4} (\bibinfo{year}{2017}),
  \bibinfo{pages}{561--576}.
\newblock


\bibitem[\protect\citeauthoryear{Cui, Wang, Pei, and Zhu}{Cui
  et~al\mbox{.}}{2018}]%
        {cui2018survey}
\bibfield{author}{\bibinfo{person}{Peng Cui}, \bibinfo{person}{Xiao Wang},
  \bibinfo{person}{Jian Pei}, {and} \bibinfo{person}{Wenwu Zhu}.}
  \bibinfo{year}{2018}\natexlab{}.
\newblock \showarticletitle{A survey on network embedding}.
\newblock \bibinfo{journal}{\emph{IEEE Transactions on Knowledge and Data
  Engineering}} \bibinfo{volume}{31}, \bibinfo{number}{5}
  (\bibinfo{year}{2018}), \bibinfo{pages}{833--852}.
\newblock


\bibitem[\protect\citeauthoryear{Dehghan, Siuta, Skorupka, Dubey, Betlen,
  Miller, Xu, Kami{\'n}ski, and Pra{\l}at}{Dehghan et~al\mbox{.}}{2023}]%
        {dehghan2023detecting}
\bibfield{author}{\bibinfo{person}{Ashkan Dehghan}, \bibinfo{person}{Kinga
  Siuta}, \bibinfo{person}{Agata Skorupka}, \bibinfo{person}{Akshat Dubey},
  \bibinfo{person}{Andrei Betlen}, \bibinfo{person}{David Miller},
  \bibinfo{person}{Wei Xu}, \bibinfo{person}{Bogumi{\l} Kami{\'n}ski}, {and}
  \bibinfo{person}{Pawe{\l} Pra{\l}at}.} \bibinfo{year}{2023}\natexlab{}.
\newblock \showarticletitle{Detecting bots in social-networks using node and
  structural embeddings}.
\newblock \bibinfo{journal}{\emph{Journal of Big Data}} \bibinfo{volume}{10},
  \bibinfo{number}{1} (\bibinfo{year}{2023}), \bibinfo{pages}{119}.
\newblock


\bibitem[\protect\citeauthoryear{Duan, Li, Lukito, Yang, Chen, Shah, and
  Yang}{Duan et~al\mbox{.}}{2022}]%
        {duan2022algorithmic}
\bibfield{author}{\bibinfo{person}{Zening Duan}, \bibinfo{person}{Jianing Li},
  \bibinfo{person}{Josephine Lukito}, \bibinfo{person}{Kai-Cheng Yang},
  \bibinfo{person}{Fan Chen}, \bibinfo{person}{Dhavan~V Shah}, {and}
  \bibinfo{person}{Sijia Yang}.} \bibinfo{year}{2022}\natexlab{}.
\newblock \showarticletitle{Algorithmic Agents in the Hybrid Media System:
  Social Bots, Selective Amplification, and Partisan News about COVID-19}.
\newblock \bibinfo{journal}{\emph{Human Communication Research}}
  (\bibinfo{year}{2022}).
\newblock


\bibitem[\protect\citeauthoryear{Efthimion, Payne, and Proferes}{Efthimion
  et~al\mbox{.}}{2018}]%
        {efthimion2018supervised}
\bibfield{author}{\bibinfo{person}{Phillip~George Efthimion},
  \bibinfo{person}{Scott Payne}, {and} \bibinfo{person}{Nicholas Proferes}.}
  \bibinfo{year}{2018}\natexlab{}.
\newblock \showarticletitle{Supervised machine learning bot detection
  techniques to identify social twitter bots}.
\newblock \bibinfo{journal}{\emph{SMU Data Science Review}}
  \bibinfo{volume}{1}, \bibinfo{number}{2} (\bibinfo{year}{2018}),
  \bibinfo{pages}{1--70}.
\newblock


\bibitem[\protect\citeauthoryear{Fazil and Abulaish}{Fazil and
  Abulaish}{2017}]%
        {fazil2017identifying}
\bibfield{author}{\bibinfo{person}{Mohd Fazil} {and} \bibinfo{person}{Muhammad
  Abulaish}.} \bibinfo{year}{2017}\natexlab{}.
\newblock \showarticletitle{Identifying active, reactive, and inactive targets
  of socialbots in twitter}. In \bibinfo{booktitle}{\emph{Proceedings of the
  International Conference on Web Intelligence}}. \bibinfo{pages}{573--580}.
\newblock


\bibitem[\protect\citeauthoryear{Feng, Tan, Li, and Luo}{Feng
  et~al\mbox{.}}{2022}]%
        {feng2022heterogeneity}
\bibfield{author}{\bibinfo{person}{Shangbin Feng}, \bibinfo{person}{Zhaoxuan
  Tan}, \bibinfo{person}{Rui Li}, {and} \bibinfo{person}{Minnan Luo}.}
  \bibinfo{year}{2022}\natexlab{}.
\newblock \showarticletitle{Heterogeneity-aware twitter bot detection with
  relational graph transformers}. In \bibinfo{booktitle}{\emph{Proceedings of
  the AAAI Conference on Artificial Intelligence}}, Vol.~\bibinfo{volume}{36}.
  \bibinfo{pages}{3977--3985}.
\newblock


\bibitem[\protect\citeauthoryear{Feng, Wan, Wang, Li, and Luo}{Feng
  et~al\mbox{.}}{2021b}]%
        {feng2021satar}
\bibfield{author}{\bibinfo{person}{Shangbin Feng}, \bibinfo{person}{Herun Wan},
  \bibinfo{person}{Ningnan Wang}, \bibinfo{person}{Jundong Li}, {and}
  \bibinfo{person}{Minnan Luo}.} \bibinfo{year}{2021}\natexlab{b}.
\newblock \showarticletitle{Satar: A self-supervised approach to twitter
  account representation learning and its application in bot detection}. In
  \bibinfo{booktitle}{\emph{Proceedings of the 30th ACM International
  Conference on Information \& Knowledge Management}}.
  \bibinfo{pages}{3808--3817}.
\newblock


\bibitem[\protect\citeauthoryear{Feng, Wan, Wang, and Luo}{Feng
  et~al\mbox{.}}{2021a}]%
        {feng2021botrgcn}
\bibfield{author}{\bibinfo{person}{Shangbin Feng}, \bibinfo{person}{Herun Wan},
  \bibinfo{person}{Ningnan Wang}, {and} \bibinfo{person}{Minnan Luo}.}
  \bibinfo{year}{2021}\natexlab{a}.
\newblock \showarticletitle{BotRGCN: Twitter bot detection with relational
  graph convolutional networks}. In \bibinfo{booktitle}{\emph{Proceedings of
  the 2021 IEEE/ACM International Conference on Advances in Social Networks
  Analysis and Mining}}. \bibinfo{pages}{236--239}.
\newblock


\bibitem[\protect\citeauthoryear{Ferrara, Varol, Davis, Menczer, and
  Flammini}{Ferrara et~al\mbox{.}}{2016}]%
        {ferrara2016rise}
\bibfield{author}{\bibinfo{person}{Emilio Ferrara}, \bibinfo{person}{Onur
  Varol}, \bibinfo{person}{Clayton Davis}, \bibinfo{person}{Filippo Menczer},
  {and} \bibinfo{person}{Alessandro Flammini}.}
  \bibinfo{year}{2016}\natexlab{}.
\newblock \showarticletitle{The rise of social bots}.
\newblock \bibinfo{journal}{\emph{Commun. ACM}} \bibinfo{volume}{59},
  \bibinfo{number}{7} (\bibinfo{year}{2016}), \bibinfo{pages}{96--104}.
\newblock


\bibitem[\protect\citeauthoryear{Gajdo{\v{s}}, Je{\v{z}}owicz, Uher, and
  Dohn{\'a}lek}{Gajdo{\v{s}} et~al\mbox{.}}{2016}]%
        {gajdovs2016parallel}
\bibfield{author}{\bibinfo{person}{Petr Gajdo{\v{s}}},
  \bibinfo{person}{Tom{\'a}{\v{s}} Je{\v{z}}owicz},
  \bibinfo{person}{Vojt{\v{e}}ch Uher}, {and} \bibinfo{person}{Pavel
  Dohn{\'a}lek}.} \bibinfo{year}{2016}\natexlab{}.
\newblock \showarticletitle{A parallel Fruchterman--Reingold algorithm
  optimized for fast visualization of large graphs and swarms of data}.
\newblock \bibinfo{journal}{\emph{Swarm and evolutionary computation}}
  \bibinfo{volume}{26} (\bibinfo{year}{2016}), \bibinfo{pages}{56--63}.
\newblock


\bibitem[\protect\citeauthoryear{Grover and Leskovec}{Grover and
  Leskovec}{2016}]%
        {grover2016node2vec}
\bibfield{author}{\bibinfo{person}{Aditya Grover} {and} \bibinfo{person}{Jure
  Leskovec}.} \bibinfo{year}{2016}\natexlab{}.
\newblock \showarticletitle{node2vec: Scalable feature learning for networks}.
  In \bibinfo{booktitle}{\emph{Proceedings of the 22nd ACM SIGKDD international
  conference on Knowledge discovery and data mining}}.
  \bibinfo{pages}{855--864}.
\newblock


\bibitem[\protect\citeauthoryear{Guo, Xie, Li, Ma, and Zhang}{Guo
  et~al\mbox{.}}{2021}]%
        {guo2021social}
\bibfield{author}{\bibinfo{person}{Qinglang Guo}, \bibinfo{person}{Haiyong
  Xie}, \bibinfo{person}{Yangyang Li}, \bibinfo{person}{Wen Ma}, {and}
  \bibinfo{person}{Chao Zhang}.} \bibinfo{year}{2021}\natexlab{}.
\newblock \showarticletitle{Social Bots Detection via Fusing BERT and Graph
  Convolutional Networks}.
\newblock \bibinfo{journal}{\emph{Symmetry}} \bibinfo{volume}{14},
  \bibinfo{number}{1} (\bibinfo{year}{2021}), \bibinfo{pages}{1--14}.
\newblock


\bibitem[\protect\citeauthoryear{Heidari and Jones}{Heidari and Jones}{2020}]%
        {heidari2020using}
\bibfield{author}{\bibinfo{person}{Maryam Heidari} {and}
  \bibinfo{person}{James~H Jones}.} \bibinfo{year}{2020}\natexlab{}.
\newblock \showarticletitle{Using bert to extract topic-independent sentiment
  features for social media bot detection}. In
  \bibinfo{booktitle}{\emph{Proceedings of the 2020 11th IEEE Annual Ubiquitous
  Computing, Electronics \& Mobile Communication Conference (UEMCON)}}. IEEE,
  \bibinfo{pages}{0542--0547}.
\newblock


\bibitem[\protect\citeauthoryear{Himelein-Wachowiak, Giorgi, Devoto, Rahman,
  Ungar, Schwartz, Epstein, Leggio, Curtis, et~al\mbox{.}}{Himelein-Wachowiak
  et~al\mbox{.}}{2021}]%
        {himelein2021bots}
\bibfield{author}{\bibinfo{person}{McKenzie Himelein-Wachowiak},
  \bibinfo{person}{Salvatore Giorgi}, \bibinfo{person}{Amanda Devoto},
  \bibinfo{person}{Muhammad Rahman}, \bibinfo{person}{Lyle Ungar},
  \bibinfo{person}{H~Andrew Schwartz}, \bibinfo{person}{David~H Epstein},
  \bibinfo{person}{Lorenzo Leggio}, \bibinfo{person}{Brenda Curtis},
  {et~al\mbox{.}}} \bibinfo{year}{2021}\natexlab{}.
\newblock \showarticletitle{Bots and misinformation spread on social media:
  Implications for COVID-19}.
\newblock \bibinfo{journal}{\emph{Journal of Medical Internet Research}}
  \bibinfo{volume}{23}, \bibinfo{number}{5} (\bibinfo{year}{2021}),
  \bibinfo{pages}{e26933}.
\newblock


\bibitem[\protect\citeauthoryear{Hu, Aggarwal, Ma, and Huai}{Hu
  et~al\mbox{.}}{2016}]%
        {hu2016embedding}
\bibfield{author}{\bibinfo{person}{Renjun Hu}, \bibinfo{person}{Charu~C
  Aggarwal}, \bibinfo{person}{Shuai Ma}, {and} \bibinfo{person}{Jinpeng Huai}.}
  \bibinfo{year}{2016}\natexlab{}.
\newblock \showarticletitle{An embedding approach to anomaly detection}. In
  \bibinfo{booktitle}{\emph{Proceedings of the 2016 IEEE 32nd International
  Conference on Data Engineering (ICDE)}}. IEEE, \bibinfo{pages}{385--396}.
\newblock


\bibitem[\protect\citeauthoryear{Huang, Li, and Hu}{Huang
  et~al\mbox{.}}{2017}]%
        {huang2017label}
\bibfield{author}{\bibinfo{person}{Xiao Huang}, \bibinfo{person}{Jundong Li},
  {and} \bibinfo{person}{Xia Hu}.} \bibinfo{year}{2017}\natexlab{}.
\newblock \showarticletitle{Label informed attributed network embedding}. In
  \bibinfo{booktitle}{\emph{Proceedings of the tenth ACM international
  conference on web search and data mining}}. \bibinfo{pages}{731--739}.
\newblock


\bibitem[\protect\citeauthoryear{Jia, Wang, and Gong}{Jia
  et~al\mbox{.}}{2017}]%
        {jia2017random}
\bibfield{author}{\bibinfo{person}{Jinyuan Jia}, \bibinfo{person}{Binghui
  Wang}, {and} \bibinfo{person}{Neil~Zhenqiang Gong}.}
  \bibinfo{year}{2017}\natexlab{}.
\newblock \showarticletitle{Random walk based fake account detection in online
  social networks}. In \bibinfo{booktitle}{\emph{2017 47th annual IEEE/IFIP
  international conference on dependable systems and networks (DSN)}}. IEEE,
  \bibinfo{pages}{273--284}.
\newblock


\bibitem[\protect\citeauthoryear{Kantepe and Ganiz}{Kantepe and Ganiz}{2017}]%
        {kantepe2017preprocessing}
\bibfield{author}{\bibinfo{person}{M{\"u}cahit Kantepe} {and}
  \bibinfo{person}{Murat~Can Ganiz}.} \bibinfo{year}{2017}\natexlab{}.
\newblock \showarticletitle{Preprocessing framework for Twitter bot detection}.
  In \bibinfo{booktitle}{\emph{Proceedings of the 2017 International conference
  on computer science and engineering (ubmk)}}. IEEE,
  \bibinfo{pages}{630--634}.
\newblock


\bibitem[\protect\citeauthoryear{Keller and Klinger}{Keller and
  Klinger}{2019}]%
        {keller2019social}
\bibfield{author}{\bibinfo{person}{Tobias~R Keller} {and}
  \bibinfo{person}{Ulrike Klinger}.} \bibinfo{year}{2019}\natexlab{}.
\newblock \showarticletitle{Social bots in election campaigns: Theoretical,
  empirical, and methodological implications}.
\newblock \bibinfo{journal}{\emph{Political Communication}}
  \bibinfo{volume}{36}, \bibinfo{number}{1} (\bibinfo{year}{2019}),
  \bibinfo{pages}{171--189}.
\newblock


\bibitem[\protect\citeauthoryear{Knauth}{Knauth}{2019}]%
        {knauth2019language}
\bibfield{author}{\bibinfo{person}{J{\"u}rgen Knauth}.}
  \bibinfo{year}{2019}\natexlab{}.
\newblock \showarticletitle{Language-agnostic twitter-bot detection}. In
  \bibinfo{booktitle}{\emph{Proceedings of the International Conference on
  Recent Advances in Natural Language Processing (RANLP 2019)}}.
  \bibinfo{pages}{550--558}.
\newblock


\bibitem[\protect\citeauthoryear{K{\"o}bis, Bonnefon, and Rahwan}{K{\"o}bis
  et~al\mbox{.}}{2021}]%
        {kobis2021bad}
\bibfield{author}{\bibinfo{person}{Nils K{\"o}bis},
  \bibinfo{person}{Jean-Fran{\c{c}}ois Bonnefon}, {and} \bibinfo{person}{Iyad
  Rahwan}.} \bibinfo{year}{2021}\natexlab{}.
\newblock \showarticletitle{Bad machines corrupt good morals}.
\newblock \bibinfo{journal}{\emph{Nature Human Behaviour}} \bibinfo{volume}{5},
  \bibinfo{number}{6} (\bibinfo{year}{2021}), \bibinfo{pages}{679--685}.
\newblock


\bibitem[\protect\citeauthoryear{Kobourov}{Kobourov}{2012}]%
        {kobourov2012spring}
\bibfield{author}{\bibinfo{person}{Stephen~G Kobourov}.}
  \bibinfo{year}{2012}\natexlab{}.
\newblock \showarticletitle{Spring embedders and force directed graph drawing
  algorithms}.
\newblock \bibinfo{journal}{\emph{arXiv preprint arXiv:1201.3011}}
  (\bibinfo{year}{2012}).
\newblock


\bibitem[\protect\citeauthoryear{Le, Tran-Thanh, and Lee}{Le
  et~al\mbox{.}}{2022}]%
        {le2022socialbots}
\bibfield{author}{\bibinfo{person}{Thai Le}, \bibinfo{person}{Long Tran-Thanh},
  {and} \bibinfo{person}{Dongwon Lee}.} \bibinfo{year}{2022}\natexlab{}.
\newblock \showarticletitle{Socialbots on Fire: Modeling Adversarial Behaviors
  of Socialbots via Multi-Agent Hierarchical Reinforcement Learning}. In
  \bibinfo{booktitle}{\emph{Proceedings of the ACM Web Conference}}.
  \bibinfo{pages}{545--554}.
\newblock


\bibitem[\protect\citeauthoryear{Li, Li, and Pan}{Li et~al\mbox{.}}{2015}]%
        {li2015discovering}
\bibfield{author}{\bibinfo{person}{Angsheng Li}, \bibinfo{person}{Jiankou Li},
  {and} \bibinfo{person}{Yicheng Pan}.} \bibinfo{year}{2015}\natexlab{}.
\newblock \showarticletitle{Discovering natural communities in networks}.
\newblock \bibinfo{journal}{\emph{Physica A: Statistical Mechanics and its
  Applications}}  \bibinfo{volume}{436} (\bibinfo{year}{2015}),
  \bibinfo{pages}{878--896}.
\newblock


\bibitem[\protect\citeauthoryear{Li and Pan}{Li and Pan}{2016}]%
        {li2016structural}
\bibfield{author}{\bibinfo{person}{Angsheng Li} {and} \bibinfo{person}{Yicheng
  Pan}.} \bibinfo{year}{2016}\natexlab{}.
\newblock \showarticletitle{Structural information and dynamical complexity of
  networks}.
\newblock \bibinfo{journal}{\emph{IEEE Transactions on Information Theory}}
  \bibinfo{volume}{62}, \bibinfo{number}{6} (\bibinfo{year}{2016}),
  \bibinfo{pages}{3290--3339}.
\newblock


\bibitem[\protect\citeauthoryear{Li, Wang, Huang, Jin, Xu, Zhang, and Gao}{Li
  et~al\mbox{.}}{2024}]%
        {li2024graph}
\bibfield{author}{\bibinfo{person}{Zhao Li}, \bibinfo{person}{Biao Wang},
  \bibinfo{person}{Jiaming Huang}, \bibinfo{person}{Yilun Jin},
  \bibinfo{person}{Zenghui Xu}, \bibinfo{person}{Ji Zhang}, {and}
  \bibinfo{person}{Jianliang Gao}.} \bibinfo{year}{2024}\natexlab{}.
\newblock \showarticletitle{A graph-powered large-scale fraud detection
  system}.
\newblock \bibinfo{journal}{\emph{International Journal of Machine Learning and
  Cybernetics}} \bibinfo{volume}{15}, \bibinfo{number}{1}
  (\bibinfo{year}{2024}), \bibinfo{pages}{115--128}.
\newblock


\bibitem[\protect\citeauthoryear{Liang, Li, Yan, Li, and Jiang}{Liang
  et~al\mbox{.}}{2021}]%
        {liang2021explaining}
\bibfield{author}{\bibinfo{person}{Yu Liang}, \bibinfo{person}{Siguang Li},
  \bibinfo{person}{Chungang Yan}, \bibinfo{person}{Maozhen Li}, {and}
  \bibinfo{person}{Changjun Jiang}.} \bibinfo{year}{2021}\natexlab{}.
\newblock \showarticletitle{Explaining the black-box model: A survey of local
  interpretation methods for deep neural networks}.
\newblock \bibinfo{journal}{\emph{Neurocomputing}}  \bibinfo{volume}{419}
  (\bibinfo{year}{2021}), \bibinfo{pages}{168--182}.
\newblock


\bibitem[\protect\citeauthoryear{Liu, Liu, Zhang, Zhu, and Li}{Liu
  et~al\mbox{.}}{2019}]%
        {liu2019rem}
\bibfield{author}{\bibinfo{person}{Yiwei Liu}, \bibinfo{person}{Jiamou Liu},
  \bibinfo{person}{Zijian Zhang}, \bibinfo{person}{Liehuang Zhu}, {and}
  \bibinfo{person}{Angsheng Li}.} \bibinfo{year}{2019}\natexlab{}.
\newblock \showarticletitle{REM: From structural entropy to community structure
  deception}.
\newblock \bibinfo{journal}{\emph{Proceedings of the Advances in Neural
  Information Processing Systems}}  \bibinfo{volume}{32}
  (\bibinfo{year}{2019}), \bibinfo{pages}{1--11}.
\newblock


\bibitem[\protect\citeauthoryear{Liu, Chen, Yang, Zhou, Li, and Song}{Liu
  et~al\mbox{.}}{2018}]%
        {liu2018heterogeneous}
\bibfield{author}{\bibinfo{person}{Ziqi Liu}, \bibinfo{person}{Chaochao Chen},
  \bibinfo{person}{Xinxing Yang}, \bibinfo{person}{Jun Zhou},
  \bibinfo{person}{Xiaolong Li}, {and} \bibinfo{person}{Le Song}.}
  \bibinfo{year}{2018}\natexlab{}.
\newblock \showarticletitle{Heterogeneous graph neural networks for malicious
  account detection}. In \bibinfo{booktitle}{\emph{Proceedings of the 27th ACM
  international conference on information and knowledge management}}.
  \bibinfo{pages}{2077--2085}.
\newblock


\bibitem[\protect\citeauthoryear{Lo, Kulatilleke, Sarhan, Layeghy, and
  Portmann}{Lo et~al\mbox{.}}{2023}]%
        {lo2023xg}
\bibfield{author}{\bibinfo{person}{Wai~Weng Lo}, \bibinfo{person}{Gayan
  Kulatilleke}, \bibinfo{person}{Mohanad Sarhan}, \bibinfo{person}{Siamak
  Layeghy}, {and} \bibinfo{person}{Marius Portmann}.}
  \bibinfo{year}{2023}\natexlab{}.
\newblock \showarticletitle{XG-BoT: An explainable deep graph neural network
  for botnet detection and forensics}.
\newblock \bibinfo{journal}{\emph{Internet of Things}}  \bibinfo{volume}{22}
  (\bibinfo{year}{2023}), \bibinfo{pages}{1--10}.
\newblock


\bibitem[\protect\citeauthoryear{Makovi, Sargsyan, Li, Bonnefon, and
  Rahwan}{Makovi et~al\mbox{.}}{2023}]%
        {Makovi2023trust}
\bibfield{author}{\bibinfo{person}{Kinga Makovi}, \bibinfo{person}{Anahit
  Sargsyan}, \bibinfo{person}{Wendi Li}, \bibinfo{person}{Jean-Fran{\c{c}}ois
  Bonnefon}, {and} \bibinfo{person}{Talal Rahwan}.}
  \bibinfo{year}{2023}\natexlab{}.
\newblock \showarticletitle{Trust within human-machine collectives depends on
  the perceived consensus about cooperative norms}.
\newblock \bibinfo{journal}{\emph{Nature Communications}} \bibinfo{volume}{14},
  \bibinfo{number}{1} (\bibinfo{year}{2023}), \bibinfo{pages}{1--12}.
\newblock


\bibitem[\protect\citeauthoryear{Mannocci, Cresci, Monreale, Vakali, and
  Tesconi}{Mannocci et~al\mbox{.}}{2022}]%
        {mannocci2022mulbot}
\bibfield{author}{\bibinfo{person}{Lorenzo Mannocci}, \bibinfo{person}{Stefano
  Cresci}, \bibinfo{person}{Anna Monreale}, \bibinfo{person}{Athina Vakali},
  {and} \bibinfo{person}{Maurizio Tesconi}.} \bibinfo{year}{2022}\natexlab{}.
\newblock \showarticletitle{MulBot: Unsupervised Bot Detection Based on
  Multivariate Time Series}. In \bibinfo{booktitle}{\emph{2022 IEEE
  International Conference on Big Data (Big Data)}}. IEEE,
  \bibinfo{pages}{1485--1494}.
\newblock


\bibitem[\protect\citeauthoryear{Mart{\'\i}n-Guti{\'e}rrez,
  Hern{\'a}ndez-Pe{\~n}aloza, Hern{\'a}ndez, Lozano-Diez, and
  {\'A}lvarez}{Mart{\'\i}n-Guti{\'e}rrez et~al\mbox{.}}{2021}]%
        {martin2021deep}
\bibfield{author}{\bibinfo{person}{David Mart{\'\i}n-Guti{\'e}rrez},
  \bibinfo{person}{Gustavo Hern{\'a}ndez-Pe{\~n}aloza},
  \bibinfo{person}{Alberto~Belmonte Hern{\'a}ndez}, \bibinfo{person}{Alicia
  Lozano-Diez}, {and} \bibinfo{person}{Federico {\'A}lvarez}.}
  \bibinfo{year}{2021}\natexlab{}.
\newblock \showarticletitle{A deep learning approach for robust detection of
  bots in twitter using transformers}.
\newblock \bibinfo{journal}{\emph{IEEE Access}}  \bibinfo{volume}{9}
  (\bibinfo{year}{2021}), \bibinfo{pages}{54591--54601}.
\newblock


\bibitem[\protect\citeauthoryear{Mazza, Cresci, Avvenuti, Quattrociocchi, and
  Tesconi}{Mazza et~al\mbox{.}}{2019}]%
        {mazza2019rtbust}
\bibfield{author}{\bibinfo{person}{Michele Mazza}, \bibinfo{person}{Stefano
  Cresci}, \bibinfo{person}{Marco Avvenuti}, \bibinfo{person}{Walter
  Quattrociocchi}, {and} \bibinfo{person}{Maurizio Tesconi}.}
  \bibinfo{year}{2019}\natexlab{}.
\newblock \showarticletitle{Rtbust Exploiting temporal patterns for botnet
  detection on twitter}. In \bibinfo{booktitle}{\emph{Proceedings of the 10th
  ACM conference on web science}}. \bibinfo{pages}{183--192}.
\newblock


\bibitem[\protect\citeauthoryear{Mendoza, Providel, Santos, and
  Valenzuela}{Mendoza et~al\mbox{.}}{2024}]%
        {mendoza2024detection}
\bibfield{author}{\bibinfo{person}{Marcelo Mendoza}, \bibinfo{person}{Eliana
  Providel}, \bibinfo{person}{Marcelo Santos}, {and}
  \bibinfo{person}{Sebasti{\'a}n Valenzuela}.} \bibinfo{year}{2024}\natexlab{}.
\newblock \showarticletitle{Detection and impact estimation of social bots in
  the Chilean Twitter network}.
\newblock \bibinfo{journal}{\emph{Scientific Reports}} \bibinfo{volume}{14},
  \bibinfo{number}{1} (\bibinfo{year}{2024}), \bibinfo{pages}{6525}.
\newblock


\bibitem[\protect\citeauthoryear{Miller, Dickinson, Deitrick, Hu, and
  Wang}{Miller et~al\mbox{.}}{2014}]%
        {miller2014twitter}
\bibfield{author}{\bibinfo{person}{Zachary Miller}, \bibinfo{person}{Brian
  Dickinson}, \bibinfo{person}{William Deitrick}, \bibinfo{person}{Wei Hu},
  {and} \bibinfo{person}{Alex~Hai Wang}.} \bibinfo{year}{2014}\natexlab{}.
\newblock \showarticletitle{Twitter spammer detection using data stream
  clustering}.
\newblock \bibinfo{journal}{\emph{Information Sciences}}  \bibinfo{volume}{260}
  (\bibinfo{year}{2014}), \bibinfo{pages}{64--73}.
\newblock


\bibitem[\protect\citeauthoryear{Murdoch, Singh, Kumbier, Abbasi-Asl, and
  Yu}{Murdoch et~al\mbox{.}}{2019}]%
        {murdoch2019definitions}
\bibfield{author}{\bibinfo{person}{W~James Murdoch}, \bibinfo{person}{Chandan
  Singh}, \bibinfo{person}{Karl Kumbier}, \bibinfo{person}{Reza Abbasi-Asl},
  {and} \bibinfo{person}{Bin Yu}.} \bibinfo{year}{2019}\natexlab{}.
\newblock \showarticletitle{Definitions, methods, and applications in
  interpretable machine learning}.
\newblock \bibinfo{journal}{\emph{Proceedings of the National Academy of
  Sciences}} \bibinfo{volume}{116}, \bibinfo{number}{44}
  (\bibinfo{year}{2019}), \bibinfo{pages}{22071--22080}.
\newblock


\bibitem[\protect\citeauthoryear{Ng and Carley}{Ng and Carley}{2023}]%
        {ng2023botbuster}
\bibfield{author}{\bibinfo{person}{Lynnette Hui~Xian Ng} {and}
  \bibinfo{person}{Kathleen~M Carley}.} \bibinfo{year}{2023}\natexlab{}.
\newblock \showarticletitle{Botbuster: Multi-platform bot detection using a
  mixture of experts}. In \bibinfo{booktitle}{\emph{Proceedings of the
  International AAAI Conference on Web and Social Media}},
  Vol.~\bibinfo{volume}{17}. \bibinfo{pages}{686--697}.
\newblock


\bibitem[\protect\citeauthoryear{Ng, Li, and Ye}{Ng et~al\mbox{.}}{2011}]%
        {ng2011multirank}
\bibfield{author}{\bibinfo{person}{Michaek Kwok-Po Ng}, \bibinfo{person}{Xutao
  Li}, {and} \bibinfo{person}{Yunming Ye}.} \bibinfo{year}{2011}\natexlab{}.
\newblock \showarticletitle{Multirank: co-ranking for objects and relations in
  multi-relational data}. In \bibinfo{booktitle}{\emph{Proceedings of the 17th
  ACM SIGKDD international conference on Knowledge discovery and data mining}}.
  \bibinfo{pages}{1217--1225}.
\newblock


\bibitem[\protect\citeauthoryear{Ou, Cui, Pei, Zhang, and Zhu}{Ou
  et~al\mbox{.}}{2016}]%
        {ou2016asymmetric}
\bibfield{author}{\bibinfo{person}{Mingdong Ou}, \bibinfo{person}{Peng Cui},
  \bibinfo{person}{Jian Pei}, \bibinfo{person}{Ziwei Zhang}, {and}
  \bibinfo{person}{Wenwu Zhu}.} \bibinfo{year}{2016}\natexlab{}.
\newblock \showarticletitle{Asymmetric transitivity preserving graph
  embedding}. In \bibinfo{booktitle}{\emph{Proceedings of the 22nd ACM SIGKDD
  international conference on Knowledge discovery and data mining}}.
  \bibinfo{pages}{1105--1114}.
\newblock


\bibitem[\protect\citeauthoryear{Pan, Wu, Zhu, Zhang, and Wang}{Pan
  et~al\mbox{.}}{2016}]%
        {pan2016tri}
\bibfield{author}{\bibinfo{person}{Shirui Pan}, \bibinfo{person}{Jia Wu},
  \bibinfo{person}{Xingquan Zhu}, \bibinfo{person}{Chengqi Zhang}, {and}
  \bibinfo{person}{Yang Wang}.} \bibinfo{year}{2016}\natexlab{}.
\newblock \showarticletitle{Tri-party deep network representation}. In
  \bibinfo{booktitle}{\emph{Proceedings of the Twenty-Fifth International Joint
  Conference on Artificial Intelligence}}. \bibinfo{pages}{1895--1901}.
\newblock


\bibitem[\protect\citeauthoryear{Pastor-Galindo, M{\'a}rmol, and
  P{\'e}rez}{Pastor-Galindo et~al\mbox{.}}{2022}]%
        {pastor2022profiling}
\bibfield{author}{\bibinfo{person}{Javier Pastor-Galindo},
  \bibinfo{person}{F{\'e}lix~G{\'o}mez M{\'a}rmol}, {and}
  \bibinfo{person}{Gregorio~Mart{\'\i}nez P{\'e}rez}.}
  \bibinfo{year}{2022}\natexlab{}.
\newblock \showarticletitle{Profiling users and bots in Twitter through social
  media analysis}.
\newblock \bibinfo{journal}{\emph{Information Sciences}}  \bibinfo{volume}{613}
  (\bibinfo{year}{2022}), \bibinfo{pages}{161--183}.
\newblock


\bibitem[\protect\citeauthoryear{Peng, Li, Yan, Gong, Wang, Liu, Wang, and
  Ren}{Peng et~al\mbox{.}}{2020}]%
        {peng2020dynamic}
\bibfield{author}{\bibinfo{person}{Hao Peng}, \bibinfo{person}{Jianxin Li},
  \bibinfo{person}{Hao Yan}, \bibinfo{person}{Qiran Gong},
  \bibinfo{person}{Senzhang Wang}, \bibinfo{person}{Lin Liu},
  \bibinfo{person}{Lihong Wang}, {and} \bibinfo{person}{Xiang Ren}.}
  \bibinfo{year}{2020}\natexlab{}.
\newblock \showarticletitle{Dynamic network embedding via incremental skip-gram
  with negative sampling}.
\newblock \bibinfo{journal}{\emph{Science China Information Sciences}}
  \bibinfo{volume}{63} (\bibinfo{year}{2020}), \bibinfo{pages}{1--19}.
\newblock


\bibitem[\protect\citeauthoryear{Peng, Yang, Wang, Li, He, Philip~S., Zomaya,
  and Ranjan}{Peng et~al\mbox{.}}{2021}]%
        {peng2021lime}
\bibfield{author}{\bibinfo{person}{Hao Peng}, \bibinfo{person}{Renyu Yang},
  \bibinfo{person}{Zheng Wang}, \bibinfo{person}{Jianxin Li},
  \bibinfo{person}{Lifang He}, \bibinfo{person}{Yu Philip~S.},
  \bibinfo{person}{Albert~Y Zomaya}, {and} \bibinfo{person}{Rajiv Ranjan}.}
  \bibinfo{year}{2021}\natexlab{}.
\newblock \showarticletitle{Lime: Low-cost and incremental learning for dynamic
  heterogeneous information networks}.
\newblock \bibinfo{journal}{\emph{IEEE Trans. Comput.}} \bibinfo{volume}{71},
  \bibinfo{number}{3} (\bibinfo{year}{2021}), \bibinfo{pages}{628--642}.
\newblock


\bibitem[\protect\citeauthoryear{Perozzi, Al-Rfou, and Skiena}{Perozzi
  et~al\mbox{.}}{2014}]%
        {perozzi2014deepwalk}
\bibfield{author}{\bibinfo{person}{Bryan Perozzi}, \bibinfo{person}{Rami
  Al-Rfou}, {and} \bibinfo{person}{Steven Skiena}.}
  \bibinfo{year}{2014}\natexlab{}.
\newblock \showarticletitle{Deepwalk: Online learning of social
  representations}. In \bibinfo{booktitle}{\emph{Proceedings of the 20th ACM
  SIGKDD international conference on Knowledge discovery and data mining}}.
  \bibinfo{pages}{701--710}.
\newblock


\bibitem[\protect\citeauthoryear{Pham, Nguyen, Vo, and Yun}{Pham
  et~al\mbox{.}}{2022}]%
        {pham2022bot2vec}
\bibfield{author}{\bibinfo{person}{Phu Pham}, \bibinfo{person}{Loan~TT Nguyen},
  \bibinfo{person}{Bay Vo}, {and} \bibinfo{person}{Unil Yun}.}
  \bibinfo{year}{2022}\natexlab{}.
\newblock \showarticletitle{Bot2Vec: A general approach of intra-community
  oriented representation learning for bot detection in different types of
  social networks}.
\newblock \bibinfo{journal}{\emph{Information Systems}}  \bibinfo{volume}{103}
  (\bibinfo{year}{2022}), \bibinfo{pages}{101771}.
\newblock


\bibitem[\protect\citeauthoryear{Pozzar, Hammer, Underhill-Blazey, Wright,
  Tulsky, Hong, Gundersen, Berry, et~al\mbox{.}}{Pozzar et~al\mbox{.}}{2020}]%
        {pozzar2020threats}
\bibfield{author}{\bibinfo{person}{Rachel Pozzar}, \bibinfo{person}{Marilyn~J
  Hammer}, \bibinfo{person}{Meghan Underhill-Blazey}, \bibinfo{person}{Alexi~A
  Wright}, \bibinfo{person}{James~A Tulsky}, \bibinfo{person}{Fangxin Hong},
  \bibinfo{person}{Daniel~A Gundersen}, \bibinfo{person}{Donna~L Berry},
  {et~al\mbox{.}}} \bibinfo{year}{2020}\natexlab{}.
\newblock \showarticletitle{Threats of bots and other bad actors to data
  quality following research participant recruitment through social media:
  cross-sectional questionnaire}.
\newblock \bibinfo{journal}{\emph{Journal of Medical Internet Research}}
  \bibinfo{volume}{22}, \bibinfo{number}{10} (\bibinfo{year}{2020}),
  \bibinfo{pages}{e23021}.
\newblock


\bibitem[\protect\citeauthoryear{Samariya and Thakkar}{Samariya and
  Thakkar}{2023}]%
        {samariya2023comprehensive}
\bibfield{author}{\bibinfo{person}{Durgesh Samariya} {and}
  \bibinfo{person}{Amit Thakkar}.} \bibinfo{year}{2023}\natexlab{}.
\newblock \showarticletitle{A comprehensive survey of anomaly detection
  algorithms}.
\newblock \bibinfo{journal}{\emph{Annals of Data Science}}
  \bibinfo{volume}{10}, \bibinfo{number}{3} (\bibinfo{year}{2023}),
  \bibinfo{pages}{829--850}.
\newblock


\bibitem[\protect\citeauthoryear{Samek, Wiegand, and M{\"u}ller}{Samek
  et~al\mbox{.}}{2017}]%
        {samek2017explainable}
\bibfield{author}{\bibinfo{person}{Wojciech Samek}, \bibinfo{person}{Thomas
  Wiegand}, {and} \bibinfo{person}{Klaus-Robert M{\"u}ller}.}
  \bibinfo{year}{2017}\natexlab{}.
\newblock \showarticletitle{Explainable artificial intelligence: Understanding,
  visualizing and interpreting deep learning models}.
\newblock \bibinfo{journal}{\emph{ITU Journal: ICT Discoveries}}
  (\bibinfo{year}{2017}), \bibinfo{pages}{1--10}.
\newblock


\bibitem[\protect\citeauthoryear{Shao, Ciampaglia, Varol, Yang, Flammini, and
  Menczer}{Shao et~al\mbox{.}}{2018}]%
        {shao2018spread}
\bibfield{author}{\bibinfo{person}{Chengcheng Shao},
  \bibinfo{person}{Giovanni~Luca Ciampaglia}, \bibinfo{person}{Onur Varol},
  \bibinfo{person}{Kai-Cheng Yang}, \bibinfo{person}{Alessandro Flammini},
  {and} \bibinfo{person}{Filippo Menczer}.} \bibinfo{year}{2018}\natexlab{}.
\newblock \showarticletitle{The spread of low-credibility content by social
  bots}.
\newblock \bibinfo{journal}{\emph{Nature communications}} \bibinfo{volume}{9},
  \bibinfo{number}{1} (\bibinfo{year}{2018}), \bibinfo{pages}{1--9}.
\newblock


\bibitem[\protect\citeauthoryear{Singla and Karambir}{Singla and
  Karambir}{2012}]%
        {singla2012comparative}
\bibfield{author}{\bibinfo{person}{Amit Singla} {and} \bibinfo{person}{Mr
  Karambir}.} \bibinfo{year}{2012}\natexlab{}.
\newblock \showarticletitle{Comparative analysis \& evaluation of euclidean
  distance function and manhattan distance function using k-means algorithm}.
\newblock \bibinfo{journal}{\emph{International Journal of Advanced Research in
  Computer Science and Software Engineering (IJARSSE)}} \bibinfo{volume}{2},
  \bibinfo{number}{7} (\bibinfo{year}{2012}), \bibinfo{pages}{298--300}.
\newblock


\bibitem[\protect\citeauthoryear{Stella, Ferrara, and De~Domenico}{Stella
  et~al\mbox{.}}{2018}]%
        {stella2018bots}
\bibfield{author}{\bibinfo{person}{Massimo Stella}, \bibinfo{person}{Emilio
  Ferrara}, {and} \bibinfo{person}{Manlio De~Domenico}.}
  \bibinfo{year}{2018}\natexlab{}.
\newblock \showarticletitle{Bots increase exposure to negative and inflammatory
  content in online social systems}.
\newblock \bibinfo{journal}{\emph{Proceedings of the National Academy of
  Sciences}} \bibinfo{volume}{115}, \bibinfo{number}{49}
  (\bibinfo{year}{2018}), \bibinfo{pages}{12435--12440}.
\newblock


\bibitem[\protect\citeauthoryear{Suarez-Lledo, Alvarez-Galvez,
  et~al\mbox{.}}{Suarez-Lledo et~al\mbox{.}}{2022}]%
        {suarez2022assessing}
\bibfield{author}{\bibinfo{person}{Victor Suarez-Lledo},
  \bibinfo{person}{Javier Alvarez-Galvez}, {et~al\mbox{.}}}
  \bibinfo{year}{2022}\natexlab{}.
\newblock \showarticletitle{Assessing the Role of Social Bots During the
  COVID-19 Pandemic: Infodemic, Disagreement, and Criticism}.
\newblock \bibinfo{journal}{\emph{Journal of Medical Internet Research}}
  \bibinfo{volume}{24}, \bibinfo{number}{8} (\bibinfo{year}{2022}),
  \bibinfo{pages}{e36085}.
\newblock


\bibitem[\protect\citeauthoryear{Tang, Qu, Wang, Zhang, Yan, and Mei}{Tang
  et~al\mbox{.}}{2015}]%
        {tang2015line}
\bibfield{author}{\bibinfo{person}{Jian Tang}, \bibinfo{person}{Meng Qu},
  \bibinfo{person}{Mingzhe Wang}, \bibinfo{person}{Ming Zhang},
  \bibinfo{person}{Jun Yan}, {and} \bibinfo{person}{Qiaozhu Mei}.}
  \bibinfo{year}{2015}\natexlab{}.
\newblock \showarticletitle{Line: Large-scale information network embedding}.
  In \bibinfo{booktitle}{\emph{Proceedings of the 24th international conference
  on world wide web}}. \bibinfo{pages}{1067--1077}.
\newblock


\bibitem[\protect\citeauthoryear{Wang, Jia, Zhang, and Gong}{Wang
  et~al\mbox{.}}{2018}]%
        {wang2018structure}
\bibfield{author}{\bibinfo{person}{Binghui Wang}, \bibinfo{person}{Jinyuan
  Jia}, \bibinfo{person}{Le Zhang}, {and} \bibinfo{person}{Neil~Zhenqiang
  Gong}.} \bibinfo{year}{2018}\natexlab{}.
\newblock \showarticletitle{Structure-based sybil detection in social networks
  via local rule-based propagation}.
\newblock \bibinfo{journal}{\emph{IEEE Transactions on Network Science and
  Engineering}} \bibinfo{volume}{6}, \bibinfo{number}{3}
  (\bibinfo{year}{2018}), \bibinfo{pages}{523--537}.
\newblock


\bibitem[\protect\citeauthoryear{Wang, Cui, and Zhu}{Wang
  et~al\mbox{.}}{2016}]%
        {wang2016structural}
\bibfield{author}{\bibinfo{person}{Daixin Wang}, \bibinfo{person}{Peng Cui},
  {and} \bibinfo{person}{Wenwu Zhu}.} \bibinfo{year}{2016}\natexlab{}.
\newblock \showarticletitle{Structural deep network embedding}. In
  \bibinfo{booktitle}{\emph{Proceedings of the 22nd ACM SIGKDD international
  conference on Knowledge discovery and data mining}}.
  \bibinfo{pages}{1225--1234}.
\newblock


\bibitem[\protect\citeauthoryear{Wang, Wen, Wu, Huang, and Xiong}{Wang
  et~al\mbox{.}}{2019}]%
        {wang2019fdgars}
\bibfield{author}{\bibinfo{person}{Jianyu Wang}, \bibinfo{person}{Rui Wen},
  \bibinfo{person}{Chunming Wu}, \bibinfo{person}{Yu Huang}, {and}
  \bibinfo{person}{Jian Xiong}.} \bibinfo{year}{2019}\natexlab{}.
\newblock \showarticletitle{Fdgars: Fraudster detection via graph convolutional
  networks in online app review system}. In \bibinfo{booktitle}{\emph{Companion
  proceedings of the 2019 World Wide Web conference}}.
  \bibinfo{pages}{310--316}.
\newblock


\bibitem[\protect\citeauthoryear{Wang, Tang, Aggarwal, Chang, and Liu}{Wang
  et~al\mbox{.}}{2017}]%
        {wang2017signed}
\bibfield{author}{\bibinfo{person}{Suhang Wang}, \bibinfo{person}{Jiliang
  Tang}, \bibinfo{person}{Charu Aggarwal}, \bibinfo{person}{Yi Chang}, {and}
  \bibinfo{person}{Huan Liu}.} \bibinfo{year}{2017}\natexlab{}.
\newblock \showarticletitle{Signed network embedding in social media}. In
  \bibinfo{booktitle}{\emph{Proceedings of the 2017 SIAM international
  conference on data mining}}. SIAM, \bibinfo{pages}{327--335}.
\newblock


\bibitem[\protect\citeauthoryear{Wang, Wang, Zhang, Yang, Zhao, and Liu}{Wang
  et~al\mbox{.}}{2023}]%
        {wang2023user}
\bibfield{author}{\bibinfo{person}{Yifei Wang}, \bibinfo{person}{Yupan Wang},
  \bibinfo{person}{Zeyu Zhang}, \bibinfo{person}{Song Yang},
  \bibinfo{person}{Kaiqi Zhao}, {and} \bibinfo{person}{Jiamou Liu}.}
  \bibinfo{year}{2023}\natexlab{}.
\newblock \showarticletitle{User: Unsupervised structural entropy-based robust
  graph neural network}.
\newblock \bibinfo{journal}{\emph{Proceedings of the AAAI Conference on
  Artificial Intelligence}}, \bibinfo{pages}{10235--10243}.
\newblock


\bibitem[\protect\citeauthoryear{Wei and Nguyen}{Wei and Nguyen}{2019}]%
        {wei2019twitter}
\bibfield{author}{\bibinfo{person}{Feng Wei} {and} \bibinfo{person}{Uyen~Trang
  Nguyen}.} \bibinfo{year}{2019}\natexlab{}.
\newblock \showarticletitle{Twitter bot detection using bidirectional long
  short-term memory neural networks and word embeddings}. In
  \bibinfo{booktitle}{\emph{2019 First IEEE International Conference on Trust,
  Privacy and Security in Intelligent Systems and Applications (TPS-ISA)}}.
  IEEE, \bibinfo{pages}{101--109}.
\newblock


\bibitem[\protect\citeauthoryear{Weng and Lin}{Weng and Lin}{2022}]%
        {weng2022public}
\bibfield{author}{\bibinfo{person}{Zixuan Weng} {and} \bibinfo{person}{Aijun
  Lin}.} \bibinfo{year}{2022}\natexlab{}.
\newblock \showarticletitle{Public opinion manipulation on social media: Social
  network analysis of twitter bots during the covid-19 pandemic}.
\newblock \bibinfo{journal}{\emph{International journal of environmental
  research and public health}} \bibinfo{volume}{19}, \bibinfo{number}{24}
  (\bibinfo{year}{2022}), \bibinfo{pages}{1--17}.
\newblock


\bibitem[\protect\citeauthoryear{Wischnewski, Ngo, Bernemann, Jansen, and
  Kr{\"a}mer}{Wischnewski et~al\mbox{.}}{2022}]%
        {wischnewski2022agree}
\bibfield{author}{\bibinfo{person}{Magdalena Wischnewski},
  \bibinfo{person}{Thao Ngo}, \bibinfo{person}{Rebecca Bernemann},
  \bibinfo{person}{Martin Jansen}, {and} \bibinfo{person}{Nicole Kr{\"a}mer}.}
  \bibinfo{year}{2022}\natexlab{}.
\newblock \showarticletitle{“I agree with you, bot!” How users (dis) engage
  with social bots on Twitter}.
\newblock \bibinfo{journal}{\emph{New Media \& Society}}
  (\bibinfo{year}{2022}), \bibinfo{pages}{14614448211072307}.
\newblock


\bibitem[\protect\citeauthoryear{Wu, Liu, Yang, Zheng, and Wang}{Wu
  et~al\mbox{.}}{2020}]%
        {wu2020using}
\bibfield{author}{\bibinfo{person}{Bin Wu}, \bibinfo{person}{Le Liu},
  \bibinfo{person}{Yanqing Yang}, \bibinfo{person}{Kangfeng Zheng}, {and}
  \bibinfo{person}{Xiujuan Wang}.} \bibinfo{year}{2020}\natexlab{}.
\newblock \showarticletitle{Using improved conditional generative adversarial
  networks to detect social bots on Twitter}.
\newblock \bibinfo{journal}{\emph{IEEE Access}}  \bibinfo{volume}{8}
  (\bibinfo{year}{2020}), \bibinfo{pages}{36664--36680}.
\newblock


\bibitem[\protect\citeauthoryear{Wu, Chen, Shi, Li, and Xu}{Wu
  et~al\mbox{.}}{2023a}]%
        {wu2023sega}
\bibfield{author}{\bibinfo{person}{Junran Wu}, \bibinfo{person}{Xueyuan Chen},
  \bibinfo{person}{Bowen Shi}, \bibinfo{person}{Shangzhe Li}, {and}
  \bibinfo{person}{Ke Xu}.} \bibinfo{year}{2023}\natexlab{a}.
\newblock \showarticletitle{SEGA: Structural Entropy Guided Anchor View for
  Graph Contrastive Learning}. In \bibinfo{booktitle}{\emph{Proceedings of the
  International Conference on Machine Learning}}. PMLR, \bibinfo{pages}{1--20}.
\newblock


\bibitem[\protect\citeauthoryear{Wu, Chen, Xu, and Li}{Wu
  et~al\mbox{.}}{2022}]%
        {wu2022structural}
\bibfield{author}{\bibinfo{person}{Junran Wu}, \bibinfo{person}{Xueyuan Chen},
  \bibinfo{person}{Ke Xu}, {and} \bibinfo{person}{Shangzhe Li}.}
  \bibinfo{year}{2022}\natexlab{}.
\newblock \showarticletitle{Structural entropy guided graph hierarchical
  pooling}. In \bibinfo{booktitle}{\emph{Proceedings of the International
  Conference on Machine Learning}}. PMLR, \bibinfo{pages}{24017--24030}.
\newblock


\bibitem[\protect\citeauthoryear{Wu, Ye, and Mou}{Wu et~al\mbox{.}}{2023b}]%
        {wu2023botshape}
\bibfield{author}{\bibinfo{person}{Jun Wu}, \bibinfo{person}{Xuesong Ye}, {and}
  \bibinfo{person}{Chengjie Mou}.} \bibinfo{year}{2023}\natexlab{b}.
\newblock \showarticletitle{Botshape: A Novel Social Bots Detection Approach
  Via Behavioral Patterns}. In \bibinfo{booktitle}{\emph{Proceedings of the
  12th International Conference on Data Mining \& Knowledge Management
  Process}}. \bibinfo{pages}{45--60}.
\newblock


\bibitem[\protect\citeauthoryear{Yang, Harkreader, and Gu}{Yang
  et~al\mbox{.}}{2013}]%
        {yang2013empirical}
\bibfield{author}{\bibinfo{person}{Chao Yang}, \bibinfo{person}{Robert
  Harkreader}, {and} \bibinfo{person}{Guofei Gu}.}
  \bibinfo{year}{2013}\natexlab{}.
\newblock \showarticletitle{Empirical evaluation and new design for fighting
  evolving twitter spammers}.
\newblock \bibinfo{journal}{\emph{IEEE Transactions on Information Forensics
  and Security}} \bibinfo{volume}{8}, \bibinfo{number}{8}
  (\bibinfo{year}{2013}), \bibinfo{pages}{1280--1293}.
\newblock


\bibitem[\protect\citeauthoryear{Yang, Zhou, Li, and Liu}{Yang
  et~al\mbox{.}}{2021}]%
        {yang2021generalized}
\bibfield{author}{\bibinfo{person}{Jingkang Yang}, \bibinfo{person}{Kaiyang
  Zhou}, \bibinfo{person}{Yixuan Li}, {and} \bibinfo{person}{Ziwei Liu}.}
  \bibinfo{year}{2021}\natexlab{}.
\newblock \showarticletitle{Generalized out-of-distribution detection: A
  survey}.
\newblock \bibinfo{journal}{\emph{arXiv preprint arXiv:2110.11334}}
  (\bibinfo{year}{2021}).
\newblock


\bibitem[\protect\citeauthoryear{Yang, Varol, Davis, Ferrara, Flammini, and
  Menczer}{Yang et~al\mbox{.}}{2019}]%
        {yang2019arming}
\bibfield{author}{\bibinfo{person}{Kai-Cheng Yang}, \bibinfo{person}{Onur
  Varol}, \bibinfo{person}{Clayton~A Davis}, \bibinfo{person}{Emilio Ferrara},
  \bibinfo{person}{Alessandro Flammini}, {and} \bibinfo{person}{Filippo
  Menczer}.} \bibinfo{year}{2019}\natexlab{}.
\newblock \showarticletitle{Arming the public with artificial intelligence to
  counter social bots}.
\newblock \bibinfo{journal}{\emph{Human Behavior and Emerging Technologies}}
  \bibinfo{volume}{1}, \bibinfo{number}{1} (\bibinfo{year}{2019}),
  \bibinfo{pages}{48--61}.
\newblock


\bibitem[\protect\citeauthoryear{Yang, Varol, Hui, and Menczer}{Yang
  et~al\mbox{.}}{2020}]%
        {yang2020scalable}
\bibfield{author}{\bibinfo{person}{Kai-Cheng Yang}, \bibinfo{person}{Onur
  Varol}, \bibinfo{person}{Pik-Mai Hui}, {and} \bibinfo{person}{Filippo
  Menczer}.} \bibinfo{year}{2020}\natexlab{}.
\newblock \showarticletitle{Scalable and generalizable social bot detection
  through data selection}. In \bibinfo{booktitle}{\emph{Proceedings of the AAAI
  conference on artificial intelligence}}, Vol.~\bibinfo{volume}{34}.
  \bibinfo{pages}{1096--1103}.
\newblock


\bibitem[\protect\citeauthoryear{Yang, Yang, Li, Cui, Yang, Wang, Xu, and
  Xie}{Yang et~al\mbox{.}}{2023a}]%
        {yang2022rosgas}
\bibfield{author}{\bibinfo{person}{Yingguang Yang}, \bibinfo{person}{Renyu
  Yang}, \bibinfo{person}{Yangyang Li}, \bibinfo{person}{Kai Cui},
  \bibinfo{person}{Zhiqin Yang}, \bibinfo{person}{Yue Wang},
  \bibinfo{person}{Jie Xu}, {and} \bibinfo{person}{Haiyong Xie}.}
  \bibinfo{year}{2023}\natexlab{a}.
\newblock \showarticletitle{RoSGAS: Adaptive Social Bot Detection with
  Reinforced Self-Supervised GNN Architecture Search}.
\newblock \bibinfo{journal}{\emph{ACM Transactions on the Web}}
  \bibinfo{volume}{17}, \bibinfo{number}{3} (\bibinfo{year}{2023}),
  \bibinfo{pages}{1--31}.
\newblock


\bibitem[\protect\citeauthoryear{Yang, Yang, Peng, Li, Li, Liao, and Zhou}{Yang
  et~al\mbox{.}}{2023b}]%
        {yang2023fedack}
\bibfield{author}{\bibinfo{person}{Yingguang Yang}, \bibinfo{person}{Renyu
  Yang}, \bibinfo{person}{Hao Peng}, \bibinfo{person}{Yangyang Li},
  \bibinfo{person}{Tong Li}, \bibinfo{person}{Yong Liao}, {and}
  \bibinfo{person}{Pengyuan Zhou}.} \bibinfo{year}{2023}\natexlab{b}.
\newblock \showarticletitle{FedACK: Federated Adversarial Contrastive Knowledge
  Distillation for Cross-Lingual and Cross-Model Social Bot Detection}. In
  \bibinfo{booktitle}{\emph{Proceedings of the Web conference}}.
  \bibinfo{pages}{1314--1323}.
\newblock


\bibitem[\protect\citeauthoryear{Yang, Zhang, Wu, Yang, Sheng, Peng, Li, Xue,
  and Su}{Yang et~al\mbox{.}}{2023c}]%
        {yang2023minimum}
\bibfield{author}{\bibinfo{person}{Zhenyu Yang}, \bibinfo{person}{Ge Zhang},
  \bibinfo{person}{Jia Wu}, \bibinfo{person}{Jian Yang},
  \bibinfo{person}{Quan~Z Sheng}, \bibinfo{person}{Hao Peng},
  \bibinfo{person}{Angsheng Li}, \bibinfo{person}{Shan Xue}, {and}
  \bibinfo{person}{Jianlin Su}.} \bibinfo{year}{2023}\natexlab{c}.
\newblock \showarticletitle{Minimum entropy principle guided graph neural
  networks}. In \bibinfo{booktitle}{\emph{Proceedings of the Sixteenth ACM
  International Conference on Web Search and Data Mining}}.
  \bibinfo{pages}{114--122}.
\newblock


\bibitem[\protect\citeauthoryear{Zeng, Peng, Li, Liu, Liu, Philip~S., and
  He}{Zeng et~al\mbox{.}}{2023c}]%
        {zeng2023unsupervised}
\bibfield{author}{\bibinfo{person}{Guangjie Zeng}, \bibinfo{person}{Hao Peng},
  \bibinfo{person}{Angsheng Li}, \bibinfo{person}{Zhiwei Liu},
  \bibinfo{person}{Chunyang Liu}, \bibinfo{person}{Yu Philip~S.}, {and}
  \bibinfo{person}{Lifang He}.} \bibinfo{year}{2023}\natexlab{c}.
\newblock \showarticletitle{Unsupervised Skin Lesion Segmentation via
  Structural Entropy Minimization on Multi-Scale Superpixel Graphs}. In
  \bibinfo{booktitle}{\emph{Proceedings of the IEEE International Conference on
  Data Mining (ICDM)}}. \bibinfo{pages}{768--777}.
\newblock


\bibitem[\protect\citeauthoryear{Zeng, Peng, Li, Liu, Yang, Liu, and He}{Zeng
  et~al\mbox{.}}{2024b}]%
        {zeng2024semi}
\bibfield{author}{\bibinfo{person}{Guangjie Zeng}, \bibinfo{person}{Hao Peng},
  \bibinfo{person}{Angsheng Li}, \bibinfo{person}{Zhiwei Liu},
  \bibinfo{person}{Runze Yang}, \bibinfo{person}{Chunyang Liu}, {and}
  \bibinfo{person}{Lifang He}.} \bibinfo{year}{2024}\natexlab{b}.
\newblock \showarticletitle{Semi-Supervised Clustering via Structural Entropy
  with Different Constraints}. In \bibinfo{booktitle}{\emph{Proceedings of the
  2024 SIAM International Conference on Data Mining (SDM)}}. SIAM,
  \bibinfo{pages}{208--216}.
\newblock


\bibitem[\protect\citeauthoryear{Zeng, Peng, and Li}{Zeng
  et~al\mbox{.}}{2023a}]%
        {zeng2023effective}
\bibfield{author}{\bibinfo{person}{Xianghua Zeng}, \bibinfo{person}{Hao Peng},
  {and} \bibinfo{person}{Angsheng Li}.} \bibinfo{year}{2023}\natexlab{a}.
\newblock \showarticletitle{Effective and stable role-based multi-agent
  collaboration by structural information principles}. In
  \bibinfo{booktitle}{\emph{Proceedings of the Thirty-Seventh AAAI Conference
  on Artificial Intelligence}}. \bibinfo{publisher}{AAAI Press},
  \bibinfo{pages}{11772--11780}.
\newblock


\bibitem[\protect\citeauthoryear{Zeng, Peng, and Li}{Zeng
  et~al\mbox{.}}{2024a}]%
        {zeng2024adversarial}
\bibfield{author}{\bibinfo{person}{Xianghua Zeng}, \bibinfo{person}{Hao Peng},
  {and} \bibinfo{person}{Angsheng Li}.} \bibinfo{year}{2024}\natexlab{a}.
\newblock \showarticletitle{Adversarial socialbots modeling based on structural
  information principles}. In \bibinfo{booktitle}{\emph{Proceedings of the AAAI
  Conference on Artificial Intelligence}}, Vol.~\bibinfo{volume}{38}.
  \bibinfo{pages}{392--400}.
\newblock


\bibitem[\protect\citeauthoryear{Zeng, Peng, Li, Liu, He, and Yu}{Zeng
  et~al\mbox{.}}{2023b}]%
        {zeng2023hierarchical}
\bibfield{author}{\bibinfo{person}{Xianghua Zeng}, \bibinfo{person}{Hao Peng},
  \bibinfo{person}{Angsheng Li}, \bibinfo{person}{Chunyang Liu},
  \bibinfo{person}{Lifang He}, {and} \bibinfo{person}{Philip~S. Yu}.}
  \bibinfo{year}{2023}\natexlab{b}.
\newblock \showarticletitle{Hierarchical State Abstraction based on Structural
  Information Principles}. In \bibinfo{booktitle}{\emph{Proceedings of the
  Thirty-Second International Joint Conference on Artificial Intelligence,
  {IJCAI-23}}}, \bibfield{editor}{\bibinfo{person}{Edith Elkind}} (Ed.).
  \bibinfo{publisher}{International Joint Conferences on Artificial
  Intelligence Organization}, \bibinfo{pages}{4549--4557}.
\newblock


\bibitem[\protect\citeauthoryear{Zhang, Zhang, Zheng, Qiao, Li, Zhang, Dam,
  Thwal, Tun, Huy, et~al\mbox{.}}{Zhang et~al\mbox{.}}{2023}]%
        {zhang2023complete}
\bibfield{author}{\bibinfo{person}{Chaoning Zhang}, \bibinfo{person}{Chenshuang
  Zhang}, \bibinfo{person}{Sheng Zheng}, \bibinfo{person}{Yu Qiao},
  \bibinfo{person}{Chenghao Li}, \bibinfo{person}{Mengchun Zhang},
  \bibinfo{person}{Sumit~Kumar Dam}, \bibinfo{person}{Chu~Myaet Thwal},
  \bibinfo{person}{Ye~Lin Tun}, \bibinfo{person}{Le~Luang Huy},
  {et~al\mbox{.}}} \bibinfo{year}{2023}\natexlab{}.
\newblock \showarticletitle{A Complete Survey on Generative AI (AIGC): Is
  ChatGPT from GPT-4 to GPT-5 All You Need?}
\newblock \bibinfo{journal}{\emph{arXiv preprint arXiv:2303.11717}}
  (\bibinfo{year}{2023}).
\newblock


\bibitem[\protect\citeauthoryear{Zhang, Cui, and Zhu}{Zhang
  et~al\mbox{.}}{2020}]%
        {zhang2020deep}
\bibfield{author}{\bibinfo{person}{Ziwei Zhang}, \bibinfo{person}{Peng Cui},
  {and} \bibinfo{person}{Wenwu Zhu}.} \bibinfo{year}{2020}\natexlab{}.
\newblock \showarticletitle{Deep learning on graphs: A survey}.
\newblock \bibinfo{journal}{\emph{IEEE Transactions on Knowledge and Data
  Engineering}} \bibinfo{volume}{34}, \bibinfo{number}{1}
  (\bibinfo{year}{2020}), \bibinfo{pages}{249--270}.
\newblock


\bibitem[\protect\citeauthoryear{Zhao, Liu, Yan, Li, Shao, and Peng}{Zhao
  et~al\mbox{.}}{2020}]%
        {zhao2020multi}
\bibfield{author}{\bibinfo{person}{Jun Zhao}, \bibinfo{person}{Xudong Liu},
  \bibinfo{person}{Qiben Yan}, \bibinfo{person}{Bo Li},
  \bibinfo{person}{Minglai Shao}, {and} \bibinfo{person}{Hao Peng}.}
  \bibinfo{year}{2020}\natexlab{}.
\newblock \showarticletitle{Multi-attributed heterogeneous graph convolutional
  network for bot detection}.
\newblock \bibinfo{journal}{\emph{Information Sciences}}  \bibinfo{volume}{537}
  (\bibinfo{year}{2020}), \bibinfo{pages}{380--393}.
\newblock


\bibitem[\protect\citeauthoryear{Zhou, Li, Li, Yu, Liu, Wang, Zhang, Ji, Yan,
  He, et~al\mbox{.}}{Zhou et~al\mbox{.}}{2023}]%
        {zhou2023comprehensive}
\bibfield{author}{\bibinfo{person}{Ce Zhou}, \bibinfo{person}{Qian Li},
  \bibinfo{person}{Chen Li}, \bibinfo{person}{Jun Yu}, \bibinfo{person}{Yixin
  Liu}, \bibinfo{person}{Guangjing Wang}, \bibinfo{person}{Kai Zhang},
  \bibinfo{person}{Cheng Ji}, \bibinfo{person}{Qiben Yan},
  \bibinfo{person}{Lifang He}, {et~al\mbox{.}}}
  \bibinfo{year}{2023}\natexlab{}.
\newblock \showarticletitle{A comprehensive survey on pretrained foundation
  models: A history from bert to chatgpt}.
\newblock \bibinfo{journal}{\emph{arXiv preprint arXiv:2302.09419}}
  (\bibinfo{year}{2023}).
\newblock


\bibitem[\protect\citeauthoryear{Zou, Peng, Huang, Yang, Li, Wu, Liu, and
  Yu}{Zou et~al\mbox{.}}{2023}]%
        {zou2023se}
\bibfield{author}{\bibinfo{person}{Dongcheng Zou}, \bibinfo{person}{Hao Peng},
  \bibinfo{person}{Xiang Huang}, \bibinfo{person}{Renyu Yang},
  \bibinfo{person}{Jianxin Li}, \bibinfo{person}{Jia Wu},
  \bibinfo{person}{Chunyang Liu}, {and} \bibinfo{person}{Philip~S. Yu}.}
  \bibinfo{year}{2023}\natexlab{}.
\newblock \showarticletitle{SE-GSL: A General and Effective Graph Structure
  Learning Framework through Structural Entropy Optimization}. In
  \bibinfo{booktitle}{\emph{Proceedings of the ACM Web Conference 2023}}.
  \bibinfo{pages}{499–510}.
\newblock


\bibitem[\protect\citeauthoryear{Zou, Wang, Li, Peng, Wang, Liu, Sheng, and
  Zhang}{Zou et~al\mbox{.}}{2024}]%
        {zou2024multispans}
\bibfield{author}{\bibinfo{person}{Dongcheng Zou}, \bibinfo{person}{Senzhang
  Wang}, \bibinfo{person}{Xuefeng Li}, \bibinfo{person}{Hao Peng},
  \bibinfo{person}{Yuandong Wang}, \bibinfo{person}{Chunyang Liu},
  \bibinfo{person}{Kehua Sheng}, {and} \bibinfo{person}{Bo Zhang}.}
  \bibinfo{year}{2024}\natexlab{}.
\newblock \showarticletitle{Multispans: A multi-range spatial-temporal
  transformer network for traffic forecast via structural entropy
  optimization}. In \bibinfo{booktitle}{\emph{Proceedings of the 17th ACM
  International Conference on Web Search and Data Mining}}.
  \bibinfo{pages}{1032--1041}.
\newblock


\end{thebibliography}

\end{document}